%% file: main.tex
\documentclass[twocolumn]{aastex61}

\usepackage{amsmath,amstext}
\usepackage[T1]{fontenc}
\usepackage{apjfonts}
\usepackage[figure,figure*]{hypcap}

\newcommand{\CIV}{\ion{C}{4}}
\newcommand{\HI}{\ion{H}{1}}
\newcommand{\OVI}{\ion{O}{6}}
\newcommand{\SiIII}{\ion{Si}{3}}
\newcommand{\SiIV}{\ion{Si}{4}}
\newcommand{\CIII}{\ion{C}{3}}

\newcommand{\kms}{\ensuremath{\mathrm{km\,s}^{-1}}}
\newcommand{\hst}{{\sl HST}}
\newcommand{\lya}{\ensuremath{\mathrm{Ly}\alpha}}
\newcommand{\Lya}{\ensuremath{\mathrm{Ly}\alpha}}
\newcommand{\lyb}{\ensuremath{\mathrm{Ly}\beta}}
\newcommand{\Lyb}{\ensuremath{\mathrm{Ly}\beta}}
\newcommand{\Rgrp}{\ensuremath{R_{\rm grp}}}
\newcommand{\Rvir}{\ensuremath{R_{\rm vir}}}

\shorttitle{Gas in Galaxy Groups}
\shortauthors{Stocke et~al.}

\turnoffedit

\begin{document}

\title{The Ultraviolet Detection of Diffuse Gas in Galaxy Groups}

\author{John T. Stocke}
\affiliation{Center for Astrophysics and Space Astronomy, Department of Astrophysical and Planetary Sciences, University of Colorado, 389 UCB, Boulder, CO 80309, USA}
\author{Brian A. Keeney}
\affiliation{Center for Astrophysics and Space Astronomy, Department of Astrophysical and Planetary Sciences, University of Colorado, 389 UCB, Boulder, CO 80309, USA}
\author{Charles W. Danforth}
\affiliation{Center for Astrophysics and Space Astronomy, Department of Astrophysical and Planetary Sciences, University of Colorado, 389 UCB, Boulder, CO 80309, USA}
\author{Benjamin D. Oppenheimer}
\affiliation{Center for Astrophysics and Space Astronomy, Department of Astrophysical and Planetary Sciences, University of Colorado, 389 UCB, Boulder, CO 80309, USA}
\author{Cameron T. Pratt}
\affiliation{Center for Astrophysics and Space Astronomy, Department of Astrophysical and Planetary Sciences, University of Colorado, 389 UCB, Boulder, CO 80309, USA}
\author{Andreas A. Berlind}
\affiliation{Department of Physics and Astronomy, Vanderbilt University, PMB 401807, 2401 Vanderbilt Place, Nashville, TN 37240-1807, USA}
\author {Chris Impey}
\affiliation{Steward Observatory, University of Arizona, 933 N Cherry Avenue, Tucson, AZ 85721, USA}
\author {Buell Jannuzi}
\affiliation{Steward Observatory, University of Arizona, 933 N Cherry Avenue, Tucson, AZ 85721, USA}

\begin{abstract}
A small survey of the UV-absorbing gas in 12~low-$z$ galaxy groups has been conducted using the Cosmic Origins Spectrograph (COS) on-board the {\sl Hubble Space Telescope} (\hst). Targets were selected from a large, homogeneously-selected sample of groups found in the Sloan Digital Sky Survey (SDSS). A critical selection criterion excluded sight lines that pass close ($<1.5$ virial radii) to a group galaxy, to ensure absorber association with the group as a whole. \edit1{Deeper galaxy redshift observations are used  both to search for closer galaxies and also} to characterize these $10^{13.5}$ to $10^{14.5}~M_{\odot}$ groups, the most massive of which are highly-virialized with numerous early-type galaxies (ETGs). This sample also includes two spiral-rich groups, not yet fully-virialized. At group-centric impact parameters of 0.3-2~Mpc, these $\mathrm{S/N}=15$-30 spectra detected \HI\ absorption in 7 of 12~groups; high (\OVI) and low (\SiIII) ion metal lines are present in 2/3 of the absorption components. None of the three most highly-virialized, ETG-dominated groups are detected in absorption. Covering fractions $\gtrsim50$\% are seen at all impact parameters probed, but do not require large filling factors despite an enormous extent. Unlike halo clouds in individual galaxies, group absorbers have radial velocities which are too low to escape the group potential well without doubt. This suggests that these groups are ``closed boxes'' for galactic evolution in the current epoch. Evidence is presented that the cool and warm group absorbers are not a pervasive intra-group medium (IGrM), requiring a hotter ($T\sim10^6$ to $10^7$~K) IGrM to be present to close the baryon accounting. 
\end{abstract} 


\section{Introduction}
\label{sec:intro}

In the past decade, the {\sl Hubble Space Telescope} (\hst) and its UV spectrographs have brought a tremendous increase in our knowledge of the relationship between galaxies and the diffuse circum- and inter-galactic media (CGM \& IGM) out of which galaxies are still forming. But while QSO absorption-line probes show that diffuse gas in the IGM accounts for the vast majority of baryons at all epochs \citep*{rauch98,shull12}, the ``baryon census'' is still not complete at low-$z$, either for individual spiral galaxies \citep[][although see \citealp{prochaska17} for a different view]{klypin01,stocke13} or on cosmological scales \citep{shull12}. With the IGM becoming more complex with time \citep[e.g.,][]{dave99,cen99}, the effects of galaxy feedback and non-equilibrium cooling complicate both an observational search and a theoretical prediction for diffuse gas associated with individual galaxies \citep{tumlinson11,oppenheimer12,stocke13,werk13,prochaska17}, galaxy groups and clusters \citep{yoon12,yoon13,yoon17}, and large-scale galaxy structures \citep*{wakker15, tejos18}. 

Of particular interest in this regard are highly-ionized metal (\OVI\ \& \ion{Ne}{8}) lines and thermally-broadened \HI\ absorptions, which show that there is a sizable fraction of warm gas \citep[following the naming convention in][since sometimes gas in this temperature range is called the warm-hot IGM, or WHIM]{savage14} locally at $T > 10^{5}$~K. The WHIM is suspected to harbor the bulk of the ``missing baryons'' at low-$z$, and may also be an important gas reservoir in the recycling processes of galactic winds and in-fall that continue to fuel on-going star formation in spiral galaxies \citep[e.g.,][]{binney87}. Surveys for metal ion absorbers \citep [e.g.,][]{steidel10,danforth08,tumlinson11,burchett16} show that gas enriched by star formation has been transported quite far from galaxies \citep*{chen01,adelberger05,stocke06,stocke13,prochaska11,johnson15,pratt18}, up to $\sim600$~kpc or more in the current epoch, much larger than the inferred virial radius of even the most massive galaxies, but somewhat smaller than the size of most galaxy groups.

The science program of the Cosmic Origins Spectrograph (COS) Guaranteed Time Observers \citep[GTOs;][]{stocke13,keeney17} and the various \hst\ guest observer (GO) Projects, including COS-Halos, COS-Dwarfs and COS-GASS \citep{tumlinson11,werk13,bordoloi14,borthakur13}, and the Andromeda Galaxy AMIGA project \citep*{lehner15}, among others, have gone a long way towards defining the relationship between individual galaxies, their massive photo-ionized CGMs ($>10^{10}~M_{\odot}$ at $T \sim 10^4$~K for $L \geq L^*$, late-type galaxies) and the shocked interfaces with hotter gas seen as \OVI\ absorptions \citep{savage10,savage12,tumlinson11,tumlinson13,prochaska11,thom12,borthakur13,stocke13,stocke14,werk13,werk14,bordoloi14,lehner14}. However dramatic the CGM detections have been, these studies still leave $\lesssim50$\% of spiral baryons unaccounted for in the current epoch (but see \citealp{prochaska17} for an accounting in which the CGM baryon content is dominated by warm gas). \edit1{It is possible that most, if not all, of the discrepancy in the baryon census found by these studies is due to differing assumptions involving the value for the low-$z$ extragalactic ionizing flux and the conversion between stellar mass and halo mass for massive spiral galaxies \citep{keeney17}.} If there are as-yet undetected baryons in the CGM, this uncertainty leaves important questions unanswered, including what are the physical conditions of these baryons and are these baryons still bound to the galaxies or blown out into the IGM? 

\citet[][Paper~1 hereafter]{savage14} used \hst/COS spectra with $\mathrm{S/N}=20$-50 to detect and analyze 54~\OVI\ absorbers at low-$z$, including at least 14~systems which are demonstrably collisionally ionized gas at warm temperatures ($T>10^5$~K) too hot to be cool, photo-ionized gas. The remaining \OVI\ systems are either cooler than this limit, suggesting photo-ionized gas, or have velocity misalignments that preclude conclusive temperature analysis (see Paper~1 for detailed methodology). The broad, shallow absorptions seen in many of these \OVI\ plus \lya\ systems are not obviously associated with individual galaxies, since many of these absorbers are found well outside the virial radius of the nearest galaxy \citep{keeney17,pratt18}.

In \citet[][Paper~2 hereafter]{stocke14}, the galaxy environments of nearby ($z\leq0.15$) photo-ionized and collisionally ionized \OVI\ absorbers from Paper~1 were investigated. Based on deep and wide galaxy surveys around these sight lines, groups of galaxies were found in all cases. Using a subset of these groups for which at least eight members were identified, weak correlations ($\sim95$\% confidence level) were found between the gas temperatures as inferred from the \OVI\ and \HI\ $b$-values and the velocity dispersions and total luminosities of the groups. While an association between some of these absorbers and the nearest group galaxy to the sight line is plausible in several cases, Paper~2 puts forward the hypothesis that most of these warm absorbers are associated with an entire galaxy group.

If this is the case, a simple argument suggests that these warm absorbers are very large, and massive enough to account for the remainder of the missing baryons in late-type galaxies; viz. using the warm absorber line density of $d\mathcal{N}/dz \geq 4$ per unit redshift (Paper~2) in conjunction with the local space density of galaxy groups \citep*{pisani03,berlind06} requires that these absorbers have radii $\geq1$~Mpc at high covering factor (larger still, if ``patchy''), and therefore quite massive ($\lesssim10^{12}~M_{\odot}$; Paper~2). This is comparable to the mass of the X-ray emitting intra-cluster medium (ICM) in elliptical-dominated groups and poor clusters, where it is the dominant baryon reservoir at halo masses of $\geq10^{14.5}~M_{\odot}$ \citep{mulchaey00}.
  
However, little is known about the intra-group medium (IGrM) in the lower-mass halos ($10^{12}$ to $10^{14}~M_{\odot}$) typical of small groups of galaxies, the most numerous ensembles of galaxies in the Universe. If a massive IGrM is present at the expected temperatures \citep[$\log{[T/\mathrm{K}]} = 6.2$-6.5;][]{mulchaey96}, it would be difficult to detect --- too cool to emit significantly at keV energies, and too hot to create narrow, high column density \HI\ \lya\ and metal absorption lines. But at the very least, a rudimentary (i.e., patchy and multi-temperature) IGrM is expected because disk galaxies merge to form the central elliptical in elliptical-dominated groups, which do exhibit massive, X-ray emitting IGrMs.
  
If an IGrM at warm temperatures or slightly higher is volume-filling, its mass is comparable to the amount needed to bring groups up to the cosmic mean baryon-to-dark matter ratio \citep[e.g.][]{klypin99, mcgaugh00}. This would make groups of galaxies, like the more massive clusters, ``closed boxes'' for galactic evolution. Recent observational work by \citet{pratt18} confirms earlier studies which showed that high metallicity gas ($Z \geq 0.1\,Z_{\odot}$ as traced by \OVI\ absorption) does not extend farther than $\approx600$~kpc from its source inside star-forming galaxies; this maximum distance that metal absorption (specifically \OVI) extends from galaxies is insufficient for the gas to escape from the group. On the other hand, current numerical simulations differ by factors of 3-5 on the percentage of baryons retained by groups in this critical mass range ($\log{[M/M_{\odot}]} = 13$-14.5); e.g., COSMO-Owls \citep{lebrun14}, EAGLE \citep{schaye15} and ILLUSTRIS \citep{nelson18} find divergent values in this halo mass range due to differing feedback prescriptions.

Reconciling the observational results with the numerical simulations is important. If all the baryons expelled from individual galaxies are retained inside galaxy groups, this would have important consequences for galactic evolution; e.g., some of the observed width in the mass-metallicity relationship \citep{tremonti04} could be due to differing chemical histories between groups of galaxies. Additionally, the continuing high star formation rate in spirals and the so-called ``G-dwarf problem'' \citep{binney87,pagel09} must be reconciled within individual galaxy groups. Given these important consequences, the hypotheses put forward above, which are based on the \hst/COS results of Papers~1 \& 2, need to be tested. This is one of the purposes of the current paper.

This paper is organized as follows: \autoref{sec:design} describes the experimental design for this investigation and the selection of the sample of low-$z$ galaxy groups with QSO probes. \autoref{sec:cos} presents the \hst/COS UV spectra and the basic analysis of the absorption systems detected at the redshifts of the foreground groups, including an analysis of the few possible broad \Lya\ absorbers (BLAs) in these groups. \autoref{sec:groups} presents the analysis of the galaxy redshift surveys conducted around each sight line based on the data found in \citet{keeney18}, and describes the group-finding algorithm and derived group parameters (e.g., membership, sky position, redshift and velocity dispersion) for each group. \autoref{sec:disc} discusses the results of this survey of gas in galaxy groups, including whether the absorption is associated with an individual group galaxy or the entire group. Discussions of gas covering fractions, absorber kinematics and absorber/group correlations are also included in this Section. \autoref{sec:conc} summarizes the basic observational properties of this work and describes the important conclusions derived from this survey. Detailed descriptions of the \hst/COS spectroscopy are presented in \autoref{sec:abs}, including \autoref{tab:abs}, which provides measurements for each absorption line detected that is associated with these groups. \autoref{sec:indiv} presents detailed notes on each galaxy group. Throughout this paper we assume a WMAP9 cosmology \citep[i.e., $H_0 = 69.7~\mathrm{km\,s^{-1}\,Mpc^{-1}}$, $\Omega_{\Lambda}=0.718$ and $\Omega_{\rm m}=0.282$;][]{hinshaw13}.

\section{Experimental Design and Sample Selection}
\label{sec:design}

\input{./tab_cos.tex}

This project investigates whether galaxy groups possess a cool and/or warm component to the IGrM by performing the reverse experiment to that of Papers~1 and 2. In those papers, the \hst/COS spectra were used to find and identify WHIM absorbers, then the foreground galaxy distribution at the absorber distances were examined; groups of galaxies were found to be present in virtually all cases. However, groups of galaxies are a very common phenomenon in the low-$z$ Universe \citep[$3\times10^{-3}~\mathrm{Mpc}^{-3}$ at $M_{\rm grp} \geq 10^{13}~M_{\odot}$;][Paper~3 hereafter]{berlind06}. Moreover, these warm gas-selected groups in Paper 2 have rather heterogeneous properties, ranging from systems with total luminosities of $\sim50\,L^*$ and $\sigma_{\rm grp} \approx300$-500~\kms\ to much smaller groups similar to the M51\,/\,M101 group. This diversity casts some doubt on the associations proposed in Paper~2; in some cases, the absorption could be due to individual galaxy halos, not to gas associated with the entire group. Since galaxies and groups of galaxies are not independent phenomena, it is difficult to ascribe association of gas to one and not the other using the method of Papers~1 and 2.

Here we reverse this experiment. We start by pre-selecting a homogeneous set of low-$z$ galaxy groups defined by a uniform, objective selection method. By using a homogenously selected sample of groups, we avoid the problem of assigning association to a very heterogeneous population for which a potential chance coincidence between absorbers and groups is possible. A large catalog of galaxy groups ($>7500$) chosen systematically from the Sloan Digital Sky Survey (SDSS) using the algorithm of Paper~3 was used to identify appropriate groups for study.

Restricting the groups searched to redshifts of $z=0.1$-0.2: (1) allows good SDSS galaxy group membership selection and characterization (e.g., estimated total luminosities, sizes and velocity dispersions); (2) facilitates excellent follow-up observations using multi-object galaxy spectroscopy \citep[MOS already in-hand;][]{keeney18} and (3) allows both the \OVI\ doublet and \lya\ (as well as \Lyb, \SiIII, and other lines of potential interest) to be observed with COS in the highest throughput G130M mode in a minimum number of \hst\ orbits. Much of this wavelength range is blueward of the \lya\ rest wavelength, making an \OVI\ identification much easier and more secure, even if only one line of the doublet is detected \citep{stocke17}.

Next, we cross-correlated the redshift-restricted group catalog with a list of bright ($V \lesssim 17.5$) background QSO targets, which allows high-S/N spectra to be obtained in just a few orbits. We required that the AGN sight lines intersect these groups at a range of impact parameters ($\rho$): $0.25 \leq \rho \leq 1.5$ group virial radii (symbolized \Rgrp\ herein to distinguish this scaling parameter from an individual galaxy's virial radius, \Rvir). Some of these impact parameters are modified by our more detailed examination of group membership and properties described in \autoref{sec:groups}. While sight lines at larger impact parameters were available, association between absorbers and indvidual groups become less secure at larger radii. Importantly, we rejected sight lines which passed within $1.5\,\Rvir$ of a group galaxy to make sure we were observing group gas, not gas associated with individual galaxies \citep[see e.g.,][]{prochaska11,stocke13,keeney17}. \edit1{An initial evaluation of sight line ``isolation'' using the SDSS spectroscopic survey finds no proximate galaxies at $L \geq 0.9$-$2.5\,L^*$ (limit depends on the specific group redshift). Sight line isolation is reevaluated in \autoref{sec:groups} using a considerably deeper galaxy redshift survey conducted after the \hst/COS observations were secured. This reevaluation finds two groups (25124 \& 32123) with galaxies close to the sight line, which removes one or more of the detected absorption systems from consideration as group gas.}

The resulting QSO target list along with basic observational details is shown in \autoref{tab:cos}. Despite having thousands of groups in the catalog from which to choose, the restrictions mentioned above leave only about two dozen which have bright QSOs projected nearby. Only six of these targets had been previously observed by COS and all but one of these six have S/N too low for detecting broad, shallow WHIM absorbers.

However, FBQS\,1010$+$3003 has medium-quality archival COS data \citep[$\mathrm{S/N}=15$ (20) at \OVI\ (\lya);][]{stocke17} with a broad ($b \approx 100$~\kms), shallow, $4\sigma$ detection of \OVI\ 1032~\AA\ at approximately the redshift of a group in our catalog. While \lya\ shows several narrower, photo-ionized CGM components most likely associated with individual group galaxies, there is no \lya\ absorber that can be associated with the \OVI. This requires $T>10^6$~K as in \citet{savage10}. There is no other plausible identification for the \OVI\ line since it is found blueward of the \lya\ rest wavelength.

Based on an analysis of five SDSS galaxies and 15~from WIYN/HYDRA \citep{keeney18} located at $\rho<2$~Mpc and $|\Delta v_{\rm abs}|<2000~\kms$ from the \OVI\ absorber, this group has $\sigma_{\rm grp} \approx 200$~\kms\ \citep{stocke17}, similar to what was found by Paper~2 for some absorber-selected galaxy groups. In this case, the group centroid on the sky and its mean recession velocity made a group association with the \OVI\ absorber unlikely. Instead, \citet{stocke17} concluded that the warm gas detection most likely was associated with the single, nearest galaxy to the absorber. This test case is a cautionary tale for this project, highlighting both the importance of choosing targets well away from bright group galaxies, and also the difficulty in both characterizing groups of galaxies and determining if the observed UV absorption system is related primarily to an individual galaxy or to the entire group. 

With the absence of sufficient high-S/N UV spectra in the \hst\ archive to conduct this experiment, we used the SDSS group catalog to identify ten new, bright targets which span a range of impact parameters through low-$z$ galaxy groups. Of the ten targets selected, one (CSO\,1022) was observed by Program~13444 (PI: Wakker) during the previous cycle. These ten QSO sight lines plus the FBQS\,1010$+$3003 sight line probe a well-defined sample of 13~pre-selected galaxy groups at a variety of impact parameters (two sight lines probe two distinct groups at two different redshifts; see \autoref{tab:cos}). In conjunction with the ``serendipitous'' sample of group WHIM detections from Papers~1 and 2, these detections either will affirm or deny the proposed model for these warm absorbers as an IGrM.

\input{./tab_berlind.tex}

\autoref{tab:berlind} lists the properties of the groups defined by the SDSS analysis of Paper~3. The group identifier is listed in column~1, followed by the number of group members in column~2. The group centroid on the sky is listed in columns~3 and 4, its redshift in column~5, and its estimated mass ($M_0$) and velocity dispersion ($\sigma_0$) in columns~6 \& 7. Finally, the coordinates of the QSO sight line that probes the group (\autoref{tab:cos}) are listed in columns~8 \& 9. Note that despite the modest numbers of group members ($N_0 = 3$-10), these groups have $M_0>10^{14}~M_{\odot}$ in most cases.

\section{\hst/COS Spectroscopy}
\label{sec:cos}

This section details the UV absorption systems associated with the 13~SDSS galaxy groups. Systems of absorption are considered within $\pm2.5\,\sigma_0$ of the mean redshift of the group (see \autoref{tab:berlind}). This broad window is necessary both because individual components of the system may be related either to an individual galaxy or the entire group and also because the original group mean position and recession velocity were based on just the few brightest SDSS group galaxies. We revise the original estimate of group parameters using SDSS combined with the MOS presented in \autoref{sec:groups}, and will revisit the velocity, velocity dispersion and sky offsets of each group as it pertains to the detected absorbers.

For each of the nine sight lines in our \hst\ observing program (\#14277: Stocke, PI) a single COS G130M spectrum was obtained over several orbits in \hst\ Cycle~23 (see \autoref{tab:cos} for details). The G130M spectrum of CSO\,1022 was taken by program \#13444 in the previous \hst\ cycle and obtained $\mathrm{S/N} \approx 16$ at 1200~\AA. The G130M spectrum of FBQS\,1010$+$3003 exceeds this criteria at \lya\ and is close to this S/N ratio at \OVI\ (see above).

All other sight lines have a S/N ratio at this level or greater, allowing the $4\sigma$ detection of an unresolved \OVI\ 1032~\AA\ line with an equivalent width $W_{\lambda}\ga 40$~m\AA, corresponding to an \OVI\ column density of $N_{\rm O\,VI} \ga 10^{13.5}~\mathrm{cm}^{-2}$ in the co-added spectra. However, one spectrum obtained by this program, SDSS\,J1540$-$0205, has a higher redshift Lyman limit system (LLS) that obscures the wavelength where the \OVI\ absorption would appear in group~16803 ($z_0 = 0.14839$). Because \SiIII\ absorption is not detected in a high-S/N portion of this same spectrum at the group redshift, we conservatively identify this sight line as an \HI-only detection, even though the potential \OVI\ wavelength does not meet the above criteria.  

Spectra were processed using the techniques employed in \citet{danforth16}, including automated line-identification procedures based on the Faint Object Spectrograph procedure developed by \citet{schneider93}, with statistical significance for each line determined by the method developed specifically for the COS data by \citet{keeney12}. In our first analysis of each spectrum, we searched for systems of absorbers \citep[see detailed discussion of systems versus components in][]{danforth16} and then refit each spectrum to determine the component(s) associated with the \OVI\ absorption, if present.

\input{./tab_abssummary.tex}

We present a top-level summary of the absorption-line findings in \autoref{tab:abssummary}. Four sight lines (five groups) show strong \HI\ absorption and significant (though not necessarily strong) \OVI\ absorption in at least one component.  Four sight lines show \HI\ absorption, but no \OVI; this includes the SDSSJ\,15403--0205 sight line, where a weak \Lya\ feature is seen $\sim900$~\kms\ from the SDSS center of group~16803, but the corresponding \OVI\ is obscured by an intervening LLS. Finally, two sight lines (three groups) show no significant absorption of any kind within the established velocity range of the group (see \autoref{sec:disc:fcov}).

\begin{figure}
  \epsscale{1.2}
  \centering\plotone{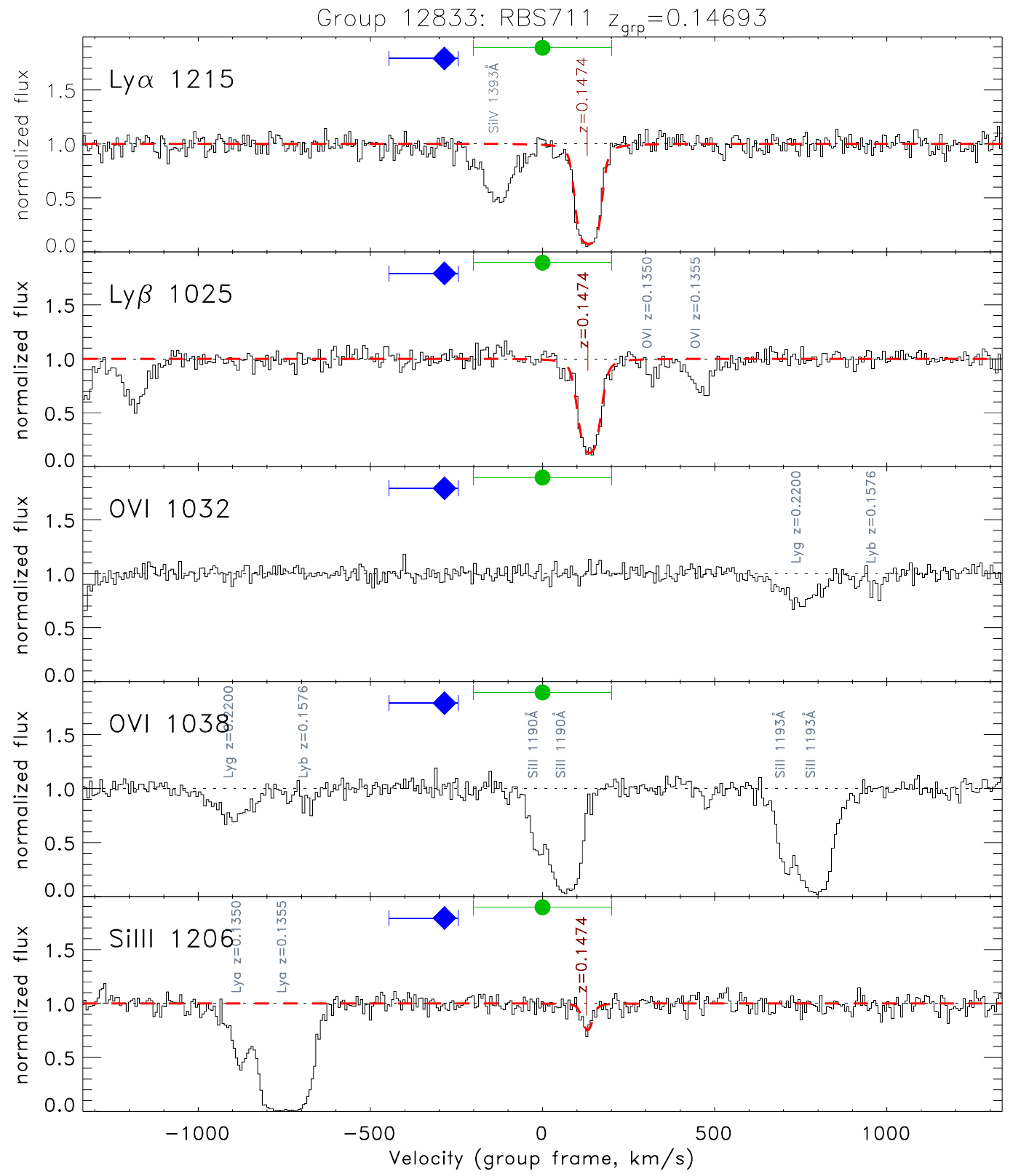}
  \caption{\HI, \OVI, and \SiIII\ absorption within $\pm2.5\,\sigma_0$ of group~12833. The panels are centered on the SDSS group redshifts (green circles), and use the SDSS velocity dispersion, $\sigma_0$, from \autoref{tab:berlind}. Blue diamonds show the group redshift and its uncertainty derived from our updated Monte Carlo analysis (\autoref{tab:params}). Strong, narrow \HI\ is seen at $\Delta v_{\rm abs}=+129$~\kms\ with no corresponding \OVI.  A weak \SiIII\ feature aligned with the \HI\ absorption shows the existence of enriched, photo-ionized gas.  Absorption features unrelated to the galaxy group are identified in gray with either wavelengths (for Galactic features) or redshifts (for unrelated IGM lines).
  \label{fig:abs}}
\end{figure}

\autoref{fig:abs} shows all absorption within $\pm2.5\,\sigma_0$ of the SDSS group redshifts in \autoref{tab:berlind}, indicated by the green circles, whose error bars indicate the SDSS redshift uncertainty. The blue diamonds with error bars show the revised group redshifts and uncertainties from \autoref{tab:params} (see \autoref{sec:groups:algorithm} for details). Red dashed lines and labels indicate absorption near the group redshift, while gray labels indicate absorption at other redshifts including Galactic lines. The galaxy symbol indicates the velocity of the nearest galaxies within $1.5\,\Rvir$ of the sight line (3~cases only). 

Discussions of the absorption toward each sight line are presented in \autoref{sec:abs}, including \autoref{tab:abs} which details line measurements of all \HI\ and \OVI\ lines detected within $\pm2.5\,\sigma_0$ of the SDSS group centers. In many cases, particularly for absorbers which fall at large velocity offsets from the group, these systems may be coincidental and not associated with the galaxy groups.  We list these line parameters for completeness. The limiting factor on the velocity accuracy for individual absorbers, especially \lya\ absorbers, is often set by the ability to de-convolve complex line profiles into individual components.  A complete discussion of the line identification and measurement techniques is found in \citet{danforth16}.

\subsection{Metal-line Absorbers}
\label{sec:cos:metals}

As listed in \autoref{tab:abs} (see \autoref{sec:abs}) and shown in \autoref{fig:abs}, metal lines are detected in two-thirds (12 out of 18) of the absorption components. \OVI\ is detected in seven of the metal-bearing systems as expected \citep{pratt18} given the large impact parameters from individual galaxies designed into this study. The remaining absorbers contain significant low-ion absorption, usually \SiIII\ 1206~\AA, similar to the photo-ionized absorbers in the CGM of individual galaxies. In the following discussion, the individual absorption components will be labeled by: [group\,/\,\Lya\ redshift].
 
Because this study was able to observe these targets only with COS/G130M, the higher ionization transitions normally seen in photo-ionized gas (\SiIV\ and \CIV) are not available, making a robust ionization model impossible for these absorbers. In lieu of that preferred solution, we assume that the five absorbers with only low ions (usually \SiIII) are close to photo-ionization equilibrium (PIE), with a characteristic gas temperature of $\sim20,000$~K as a very rough estimate. These absorbers are 12833\,/\,0.14743; 25124\,/\,0.18528, 32123\,/\,0.15874, 36001\,/\,0.18585 (\CIII\ is the only metal line detected) and 44739\,/\,0.11706. Three absorbers are detected in \HI\ and \OVI\ only, and have \OVI\ and \HI\ line widths consistent with warm gas; their characteristic temperatures are derived in \autoref{sec:cos:BLAs}.

This leaves four absorbers with both \SiIII\ and \OVI\ detected. For three (32123\,/\,0.16081; 36001\,/\,0.18527 and 44739\,/\,0.11537) of these four composite systems, both the \OVI\ and \SiIII\ are too weak to provide any definitive information, so the temperatures of these absorbers are poorly estimated. By our automated measurement the first two of these three absorbers have $b_{\rm O\,VI} > b_{\rm H\,I}$, an unphysical situation if the \HI\ and \OVI\ reside in the same gas. Although the error bars on these measurements do overlap, the more straightforward solution is that the narrow \HI\ ($b\approx25$~\kms) is not co-located with the broader \OVI\ ($b\approx 30$~\kms), but is associated with the \SiIII\ absorption. 

We suspect that a BLA is present in the saturated \lya\ profile that matches the \OVI\ line widths in these two cases but the \OVI\ profiles are too weak to provide tight enough constraints to specify the BLA parameters. Instead, we can set only very loose upper limits on the temperature of these two absorbers by assuming that the \OVI\ line widths are entirely thermal: $T< 8\times10^5$~K and $T<1\times10^6$~K, respectively. This assumption leaves the observed \SiIII\ lines and the narrow \HI\ lines as low temperature, photo-ionized absorbers.

For the third system in this list (44739\,/\,0.11537), $b_{\rm H\,I}>b_{\rm O\,VI}$ and \SiIII\ is much stronger than the \OVI. In this case, a photo-ionization model seems plausible and no BLA is required. In the fourth absorber (44739\,/\,0.11766) the \OVI\ is very broad and symmetrical suggesting that a BLA is present and the absorber temperature is $>10^5$ K (see \autoref{sec:cos:BLAs}).

For the metal-free absorbers, no definitive ionization-state or temperature determination is possible, except for absorber 32123\,/\,0.16142, which has a very broad and symmetrical \lya\ profile with a line width suggesting a temperature in CIE just barely exceeding $10^5$~K. As discussed previously \citep{stocke13, werk14, keeney17}, absorbers with no metal lines detectable in these spectra means only that the metallicities are low, typically $\lesssim 10$\% Solar metallicity, so that some metals may well be present in these six ``metal-free'' absorbers as well.

\subsection{Presence of Broad \lya\ Absorbers (BLAs)}
\label{sec:cos:BLAs}

The spectra shown in \autoref{fig:abs} and absorptions described in \autoref{tab:abs} use the automated line-finding and measurement program described in \citet{danforth16}. However, that routine is not well-suited to the detection of broad, shallow components in complex \lya\ profiles, which can indicate the presence of warm-hot gas in these groups. Since Paper~2 found evidence linking systems containing broad BLAs as well as broad, shallow \OVI\ absorption with galaxy groups, it is important to establish how many absorbers found here are consistent with possessing BLAs and thus with warm-hot gas in these groups. 

\input{./tab_bla.tex}

To investigate the possibility that BLAs are present in these spectra, the three \lya\ lines were refit for cases in which the \OVI\ absorption suggests the presence of warm gas (see procedure developed in Papers~1 and 2 using the \HI\ and \OVI\ absorptions only). Briefly, the velocity and $b$-value of the \OVI\ lines are used to constrain the wavelength and width of a potential BLA. The high-S/N ratio of these spectra is a necessity for this procedure.

\begin{figure}
  \epsscale{1.2}
  \centering\plotone{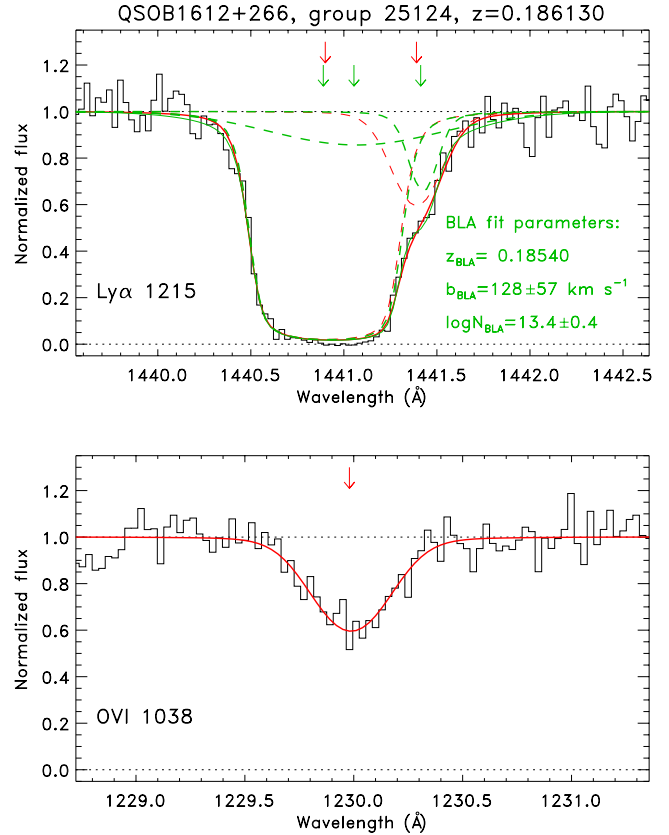}
  \caption{A \edit1{probable} BLA in the absorption associated with group~25124. Nominal fits are shown in solid red lines for \Lya\ and \OVI\ 1038~\AA\ (the \OVI\ 1032~\AA\ line is blended with Ly10 at a different redshift as discussed in the text) with red arrows marking the velocities. Individual components are shown as dashed lines. The green line shows the \Lya\ profile refitted with a third component centered at the same velocity as the broad, symmetrical \OVI\ absorber; individual velocity components for this second fit are shown as dashed lines. The green arrows indicate the positions of the new \lya\ components.  The BLA has best-fit parameters shown in the plot, with a large $b$-value implying warm gas temperatures ($T\approx 8\times10^5$~K, but with considerable uncertainty).
  \label{fig:bla}}
\end{figure}

\paragraph{Group~25124} The presence of a BLA is suggested in this case by an offset in the velocity centroid between the strong, broad, symmetrical \OVI\ 1038 line compared to the two \HI\ components and low-ionization metal ions. The stronger \OVI\ line is blended with a Ly10 absorption at $z$=0.330 (see \autoref{fig:abs}). Using the location and width of the \OVI\ 1038~\AA\ line, the \Lya\ profile was refit and a BLA matching the \OVI\ can be present (see \autoref{fig:bla} and \autoref{tab:bla}). The \ion{N}{2}, \ion{Si}{2} and \SiIII\ lines follow the narrow \HI\ absorption, as does the blue component of the \CIII\ line; the redward \CIII\ line and the \ion{N}{3} line match the location of the BLA and \OVI. The narrower blue component likely has a low temperature consistent with PIE, and is closely associated with a nearby $2\,L^*$ galaxy 168~kpc ($0.8\,\Rvir$) away. Using the measured $b$-values for the BLA and \OVI\ in this case suggests a gas temperature in CIE of $T\approx 8 \times10^5$~K. This \edit1{probable} BLA may be associated either with the nearby galaxy or the entire galaxy group, since the velocity difference between the broad and narrow components suggest different physical locations for these absorbers.

\paragraph{Group 32123} In this sight line there are two possibilities for BLAs. One at $z=0.16081$ contains broad \OVI\ absorption seen in both lines of the doublet but at low-S/N. A BLA corresponding to this \OVI\ can be fit within the \Lya\ absorption complex, but is not well-constrained; i.e., we can set only a very loose upper limit on temperature ($T<8\times10^5$~K) by assuming that the observed \OVI\ line width is entirely thermal. More interesting is a possible, metal-free BLA absorber at $z=0.16142$, whose symmetrical line width suggests a temperature of $1.5\times10^5$~K, \edit1{which would make it a collisionally ionized absorber. However, the limited S/N of this weak feature means that it can also be a blend of two narrower components.}   

\paragraph{Group 36001} In this spectrum at $z=0.18527$ there is a weak, broad \OVI\ 1032~\AA\ absorber, too weak to be detected in the 1038~\AA\ line. While a BLA \edit1{\textit{could}} be present, its parameters are not well-constrained. As with the possible warm absorber in group~32123, we set a very loose upper limit of $T< 1\times10^6$ K by assuming that the OVI line width is entirely thermal. Two possibilities seem equally plausible; a warm absorber with a BLA and broad \OVI\ and a cool absorber with narrow \HI\ and \SiIII\ are both present, or there is a single cool absorber which contains both photo-ionized \SiIII\ and \OVI. There is also another \HI\ plus \OVI\ absorber possibly associated with group~36001 at a relative velocity of $-906~\kms$, $>2.5\,\sigma_{\rm grp}$ from the group mean velocity and thus outside the velocity bounds we have set for association (see \autoref{sec:disc:fcov}). In this case, the \HI\ and \OVI\ have nearly the same line widths suggesting a low temperature and a turbulent medium, but a BLA associated with the \OVI\ could be present in the saturated \lya\ profile. Assuming that the \OVI\ width is entirely thermal sets a firm upper limit of $T<1\times 10^6$~K. 

\paragraph{Group 44739} In this case, a very broad, shallow \OVI\ 1032~\AA\ line is well-aligned with a narrow \HI\ line that also has moderately strong \SiIII\ (see \autoref{fig:abs} and \autoref{tab:abs}) at $z=0.11766$. While a BLA could be present in this case, and is suggested by the breadth of the symmetrical \OVI\ doublet lines, the BLA is predicted to have $b_{\rm H\,I}\approx200$~\kms\ assuming minimal gas turbulence or velocity components in the \OVI. This line width is so broad that the predicted BLA would be undetectable, and suggests $T>10^6$~K. In this scenario, a cool, photo-ionized absorber would also be present at nearly the same velocity, based on the narrow \HI\ and \SiIII. \edit1{But the presence of this BLA is very uncertain.}

\paragraph{Group 44858} In this case, the original line fit listed in \autoref{tab:abs} for this $z=0.12567$ absorber does not change by applying the method used above. The somewhat broad \HI\ and \OVI\ lines indicate a gas temperature just slightly in excess of $10^5$ K; \edit1{however, given the uncertain amount of non-thermal motion contributing to the line widths we cannot prove that this absorber is collisionally-ionized warm gas.} The nearest galaxy is $7\,\Rvir$ away, making a group association with this gas most likely.

\bigskip
\edit1{In summary, there are two probable BLAs in this sample of absorbers, one of which is possibly associated with an individual galaxy (25124\,/\,0.18540). One other system (44858\,/\,0.12567) with \HI\ and \OVI\ only, has line widths close to but not defintively within the collisionally-ionized regime. Additionally, there are four possible BLAs, including three systems with \HI, \OVI, and \SiIII\ (with line-width-inferred temperature limits in excess of $10^5$~K), and one \HI-only BLA (32123\,/\,0.16142) with a broad, symmetrical profile yielding a temperature estimate of $1.5\times10^5$~K assuming the line width is entirely thermal. One other system (36001\,/\,0.18429) has a poorly measured upper limit on temperature of $<10^6$~K, but is outside the velocity bounds established for association with its group. A separate absorber in this same group (36001\,/\,0.18527) has a possible BLA that sets a very similar limit on temperature.} 

\edit1{Overall, this survey detected eight absorbers in six groups with metal lines consistent with being in PIE at a fiducial temperature of $\sim 2\times10^4$~K. Two probable BLA absorbers have temperature estimates at $T \gtrsim 10^5$~K and four other possible BLAs are present with temperature estimates or limits in the collisionally-ionized regime. However, unlike in Paper 2 there is no strong evidence for warm gas ubiquitously associated with the galaxy groups in this sample.}

\newpage
\section{Characterizing the SDSS Groups Probed by these Sight Lines}
\label{sec:groups}

\subsection{MOS Observations}
\label{sec:groups:mos}

While the acquisition and analysis of the galaxy redshift and magnitude data is presented and described in detail in \citet{keeney18}, here we summarize the basic parameters of those observations as they apply to the group membership analysis. For all of the target sight line fields, the SDSS spectroscopic survey data is used for the brighter galaxies ($r \leq 17.8$), while new multi-object spectroscopy (MOS) on fainter galaxies was obtained specifically for this program at the MMT Observatory 6m using the Hectospec MOS \citep{fabricant05} and at the WIYN 3.5m telescope using the HYDRA MOS.

These groups have modest numbers of SDSS galaxies ($N_0 = 3$-10), so we designed the MOS survey to increase the number of group members to $N\geq20$ per group to better constrain basic group parameters, including velocity dispersion, centroid position and extent on the sky, which can be used to infer, e.g., the group halo mass and group virial radius. Our goal for these observations was not very high completeness \edit1{(i.e., $>$90\%)}, but rather a fair sampling of potential group members so that the group is well-characterized. The overall completeness of these observations is $\geq 60$\% for these target fields, where the overall completeness is defined to be the fraction of all targets with $g < 20$ within $20\arcmin$ of the sight line for which redshifts are available from this survey or other sources \citep[][see their Table~8 for individual field completeness values and other related quantities]{keeney18}. These limits correspond to group galaxies at $L\geq 0.2\,L^*$, although only a small percentage of those galaxies with redshifts in-hand are actually at the redshifts of these groups; it is likely that a similarly small percentage of unobserved galaxies are also at the group redshift. 

\edit1{The targeting of galaxies in these fields was blind with respect to galaxy color, type or clustering but rather relied entirely on the apparent magnitude, proximity to the sight line and SDSS photometric redshift estimate (galaxies with $z_{phot} > 0.25$ were targeted at lower priority). See \citealp{keeney18} for more details on the targeting process. The brightest galaxies ($g < 18$ or $L \geq L^*$ at the redshifts of these groups) and the closest galaxies ($\rho < 5\arcmin$ or $\approx700$~kpc at the redshifts of these groups) were targeted with highest priority. This results in completeness fractions of $\gtrsim 75$\% for the closest galaxies down to very faint magnitudes ($g < 20$ or $L \gtrsim 0.2\,L^*$ at the group redshifts) and for the brightest galaxies ($g \leq 18$ or $\sim L^*$ at the group redshifts) out to $\sim2.5$~Mpc from the sight lines (see Appendix~A in \citealp{keeney18} for statistics by sight line). These completeness fractions are for \textbf{all} galaxies with these parameters, not just the group galaxies. The high completeness fractions down to faint limits close to these sight lines means that it is quite unlikely that we have missed a nearest galaxy to these absorbers (see next subsection). The high completeness for bright galaxies out to large radii means that the full extent and constituency of these groups has been very well-sampled.}   

At WIYN/HYDRA we used the 600@10.1 grating centered at 5200~\AA\ ($\mathcal{R} \approx 1200$ from 3800-6600~\AA); the S/N of these spectra is $9^{+5}_{-3}$ per pixel when the galaxy $g$-band magnitude range is $19.4^{+0.5}_{-0.7}$. At MMT/Hectospec, we used the 270gpm grating to achieve $\mathcal{R} \approx 1000$ from 3700-9100~\AA; the S/N of these spectra is $7^{+6}_{-3}$ per pixel when the galaxy $g$-band magnitude range is $19.8^{+0.7}_{-0.8}$. We chose our wavelength coverage to ensure good sensitivity to \ion{Ca}{2} H \& K absorption at $z\approx0$ to maximize the return of accurate redshifts for absorption-line galaxies in these low-$z$ galaxy groups.

The design criterion to obtain redshifts for a representative sample of $>20$ group members is achieved in all but two cases, group~36001 (6~group galaxies; see \autoref{tab:groups}) and group~44726 (8~group members; \autoref{tab:groups}). Group~36001 has only 57\% overall completeness, but Group~44726 easily exceeds the design completeness level (60\%) for this survey. Since the MOS on these two fields did not differ significantly in completeness from the others, these are truly sparse groups, not fields with high incompleteness.

\subsection{Nearest Galaxies to the Absorbers}
\label{sec:groups:ng}

For all of our groups, we have located the galaxy with $|\Delta v_{\rm abs}| \leq 1000$~\kms\ nearest to the QSO sight line to determine whether an individual galaxy is more likely associated with the observed absorption than the group as a whole. While our original criteria for choosing sight lines excluded QSOs whose impact parameter to a bright group galaxy was $\leq 1.5\,\Rvir$, deeper MOS found two cases where fainter galaxies were found close to the sight line at the redshift of the absorber.

\input{./tab_ng.tex}

\autoref{tab:ng} lists the properties of the nearest galaxy to each absorber detected in our \hst/COS spectra at or near the group redshift. Column~9 lists the rest-frame velocity difference between the absorber and galaxy; i.e., $\Delta v_{\rm abs} = c\,(z_{\rm gal} - z_{\rm abs})/(1 + z_{\rm abs})$. The final column lists the galaxy's three-dimensional distance from the absorber, $D$, normalized by the galaxy's virial radius, \Rvir. The virial radius is estimated from the galaxy's rest-frame $g$-band luminosity, $L_{\rm gal}$\footnote{As in \citet{keeney17}, we adopt $M_g^* = -20.3$ \citep[corrected to $H_0=70$~\kms]{montero-dorta09} with $K$-corrections from \citet*{chilingarian10} and \citet{chilingarian12}.}, as detailed in \citet{stocke13}. The three-dimensional absorber-galaxy distance is calculated using:
\begin{equation}
\label{eqn:D}
D^2 = \rho^2 + D_z^2
\end{equation}
where $\rho$ is the galaxy's impact parameter with respect to the QSO sight line and $D_z$ is the distance along the line of sight in Mpc. The line-of-sight distance is calculated using a reduced Hubble flow model such that
\begin{align}
\label{eqn:Dz}
D_z = \left\{ \begin{array}{lc}
              0                             & \hspace{5mm} |\Delta v_{\rm abs}| \leq v_{\rm red} \\
              (|\Delta v_{\rm abs}| - v_{\rm red})/H_0 & \hspace{5mm} |\Delta v_{\rm abs}| > v_{\rm red} \\
              \end{array} \right.
\end{align}

We assume $v_{\rm red}=400$~\kms\ for consistency with previous studies of galaxy-absorber associations \citep[see, e.g., Paper~2,][]{prochaska11,stocke13,keeney17,keeney18,pratt18}. This convention is used to account for possible peculiar velocities between galaxies and absorbers (e.g., due to outflow/infall); the reduced Hubble flow cutoff value was chosen to match the maximum rotation speeds of the most massive galaxies.

There are only two cases for which a galaxy is close to the sight line in units of virial radii: group~25124 ($0.8\,\Rvir$ from the sight line) and group~32123 ($0.6\,\Rvir$ away). Based on the observed distributions of nearest and next-nearest neighbors to a large sample of absorbers, \citet{keeney17} find that low-$z$ absorption can be attributed with confidence to the CGM of individual galaxies only when $D/\Rvir \lesssim 1.4$. This suggests that some of the absorption in groups~25124 and 32123  is likely associated with individual galaxies in \autoref{tab:ng} rather than the group itself.

To investigate this uncertainty, the sight line towards FBQS\,1010+3003 was used as a test case for the current study. This sight line was selected in the same way as the current study, except that a bright SDSS galaxy at the group redshift was known to be present close to the sight line prior to deeper MOS. A relatively high-S/N (15-20) COS spectrum was in-hand for FBQS\,1010+3003 from the science program of the COS-GTOs \citep{stocke13, keeney17} and a deep galaxy survey of the region is available from \citet{keeney18}. A detailed examination in \citet{stocke17} tried to distinguish between an absorber association with an individual $2.7\,L^*$ galaxy located $1.0\,\Rvir$ from the absorber and a small group of nearby galaxies, of which the individual galaxy is not a member. For an \OVI-only absorber in the FUV spectrum of FBQS\,1010+3003, a better match both in sky position and velocity was found for the individual galaxy association. For comparison, the nearest galaxy data for this case is reproduced at the bottom of \autoref{tab:ng} and labeled group~49980. The \OVI-only absorber is at $z=0.11351$; the other absorbers are detected in \lya, some of which likely are associated with the galaxy group (see \citealp{stocke17} for details).

The situation is not so clear for groups~25124 and 32123. For the $z=0.18529$, 0.18540 (the BLA plus \OVI\ absorber), 0.18569 \Lya\ complex in group~25124, the high and low-ion components cannot be at the same location given their velocity difference. All three components can be associated with the single galaxy in \autoref{tab:ng}, or the BLA plus broad \OVI\ can be associated with the galaxy group. To be conservative, we associate all three with the galaxy and count group~25124 as a non-detection. The case of group~32123 is somewhat clearer in that a total of four absorbers are present; the two at lower velocity match the velocity of the nearest galaxy (see \autoref{tab:ng}), which is not a group member. The other two absorbers are too far away in velocity to be plausibly associated with this same galaxy. For these two absorbers, which include an \HI\ and \OVI\ absorber and a BLA (see \autoref{fig:abs}), the closest galaxy is $1.8\,\Rvir$ away, so we identify the absorption with the galaxy group. 

The individual galaxies associated with absorbers near the redshifts of groups~25124 and 32123 are both early types, with specific star formation rates consistent with being passive galaxies ($\mathrm{sSFR} < 10^{−11}~\mathrm{yr}^{−1}$). Additionally, the nearest galaxy in the FBQS\,1010+3003 sight line is an Sa galaxy whose sSFR suggests that it is also passive. We detect \OVI\ absorption in all three of these cases as well as a few other cases in previous investigations \citep[Paper~2 and][]{keeney17}, but the COS-Halos collaboration found very few \OVI\ absorptions associated with passive galaxies in their sample \citep{tumlinson11}. The reported absence of \OVI\ is almost certainly due to the COS-Halos spectra having lower S/N than the present data and the data obtained in the COS-GTO program\edit1{, resulting in a rather high column density limit ($\log{N_{\rm O\,VI}} > 14.3$) for their detections}. While \OVI\ is undoubtedly present around most passive galaxies, the larger impact parameters of those absorbers suggests an association with the entire group to which the galaxy is a member \citep[Paper~2 and][]{keeney17}. The rather constant covering fraction of 50-70\% for absorption-line galaxies as a function of impact parameter out to several virial radii ($4\,\Rvir$ or $\sim1$~Mpc) suggests that the gaseous structure creating the absorption around passive galaxies is much larger than an individual galaxy CGM \citep[see Figures~7 and 8 in][]{keeney18}.

\begin{figure}
  \epsscale{1.2}
  \centering\plotone{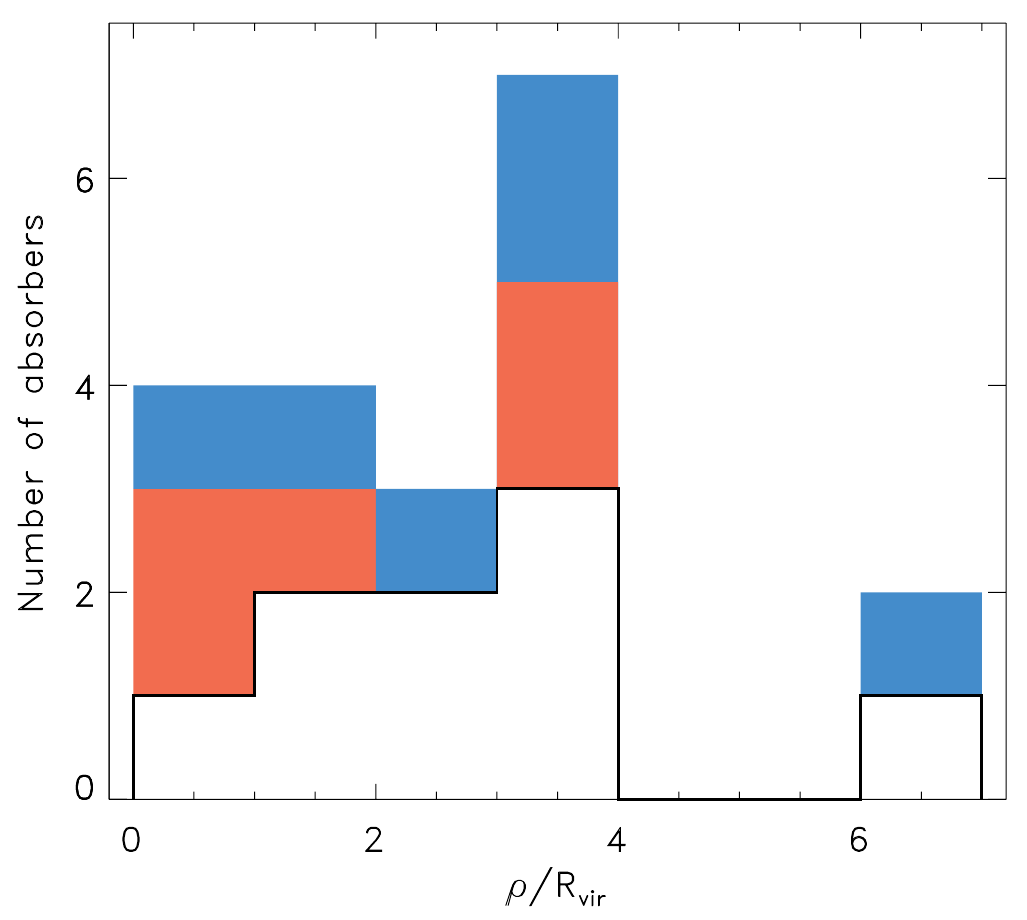}
  \caption{Impact parameters of nearest galaxies to the absorbers in this survey. Open boxes are metal-free absorbers; blue boxes indicate the presence of \OVI\ while the red boxes are absorbers with only low ions. The FBQS\,1010+3003 field, which has a close galaxy at $\rho/\Rvir < 1$, is included in this compilation despite being chosen somewhat differently from the rest of the sample (see text).
  \label{fig:ng}}
\end{figure}

For the other groups, despite the much deeper MOS obtained by us, there are no galaxies close to the sight lines, making the claim that these absorbers are associated with the entire galaxy group strong. The full range of nearest galaxy impact parameters is shown as a histogram in \autoref{fig:ng}. The bins are in units of virial radii, and open symbols indicate \HI-only absorption, red indicates low-ion without \OVI, and blue indicates \OVI\ absorption with or without low ions. The FBQS\,1010+3003 field listed at the bottom as group~49980 is included for completeness, despite having a previously known galaxy close to the sight line (see \citealp{stocke17}).

In summary, while all the absorptions listed in \autoref{tab:ng} could be ascribed to the foreground groups targeted by this study, we conservatively associate the absorption complex in group~25124 and two of four absorbers near group~32123 to the individual closest galaxy listed in \autoref{tab:ng}. We associate the other two absorbers near group~32123 (those not close to an individual galaxy) with the entire group. That is, we count group~25124 as a non-detection, while group~32123 is counted as a detection because only one of the two absorber systems detected is associated with an individual galaxy.

\subsection{Group-Finding Algorithm and Parameters}
\label{sec:groups:algorithm}

The group-finding algorithm employed here differs somewhat from that presented in our previous group studies. Paper~2 and \citet{stocke17} used a friends-of-friends (FoF) algorithm to create an initial guess at group membership, which was subsequently refined by a ``virialization'' step. Here, we remove the initial FoF step altogether and use Monte Carlo simulations to explore the systematic effects of initial parameters in the virialization procedure.

Our group-finding algorithm assumes a group mass-to-light ratio of $\Upsilon_{\rm grp} \equiv M_{\rm grp}/L_{\rm grp}$, which allows us to estimate the group's virial radius, \Rgrp, and velocity dispersion, $\sigma_{\rm grp}$, based on the total luminosity of the group members:
\begin{align}
\label{eqn:Lgrp}
L_{\rm grp} &= \sum_{i=1}^{N_{\rm grp}} L_i \\
\label{eqn:Mgrp} 
M_{\rm grp} &= \Upsilon_{\rm grp} \left(\frac{L_{\rm grp}}{L^*}\right) \times 10^{10}~M_{\Sun} \\
\label{eqn:Rgrp}
\Rgrp &= \left(\frac{3M_{\rm grp}}{4\pi\Delta_{\rm crit}\rho_{\rm crit}}\right)^{1/3} \nonumber \\
&= 957 \left(\frac{M_{\rm grp}}{10^{14}~M_{\Sun}}\right)^{1/3}~\mathrm{kpc} \\
\label{eqn:sgrp}
\sigma_{\rm grp} &= \sqrt{\frac{GM_{\rm grp}}{3\Rgrp}} \nonumber \\
&= 387 \left(\frac{M_{\rm grp}}{10^{14}~M_{\Sun}}\right)^{1/3}~\mathrm{km\,s}^{-1}
\end{align}
We assume $\Delta_{\rm crit}=200$ and $\rho_{\rm crit} = 9.205\times10^{-30}\,h_{70}^2~\mathrm{g\,cm}^{-3}$ \citep*{shull12}, and ignore any evolution with redshift since our groups are all located at $z<0.2$.
  
We start with an initial group position, redshift, and mass close to those in \autoref{tab:berlind}, then search for galaxies that are located within a specified distance of the group center on the plane of the sky and in velocity space  (see \autoref{sec:groups:algorithm:mc} for details). All galaxies in this volume are assumed to be group members, and new group positions, redshifts, and masses are estimated based on the revised membership. These updated group properties are used as the starting point for the next iteration of the group finder, and the algorithm iterates until it converges (i.e., the identities of group members do not change from one iteration to the next), which is typically achieved in five or fewer iterations.

As mentioned above, all galaxies within our search volume are considered to be group members, so it is important to choose the boundaries carefully. For the boundary on the plane of the sky, we adopt the group splashback radius\footnote{The ``splashback radius'' is the apocenter of satellites during their first orbit after falling into a dark matter halo, and has been suggested as a physically motivated definition of a halo boundary \citep*{adhikari14, diemer14, more15}.}, $R_{\rm sp} \sim \Omega_{\rm m}^{-1/3}\,\Rgrp \approx 1.5\,\Rgrp$ \citep{diemer17}. For the boundary in velocity space, we adopt $\pm2\,\sigma_{\rm grp}$ from the group redshift. This value was chosen by analyzing a large mock catalog, where galaxies were placed inside dark matter halos in an $N$-body simulation and given velocities that trace the velocities of dark matter particles; this analysis suggests that $>99$\% of all satellite galaxies are located within $2\,\sigma_{\rm grp}$ of the parent halo's velocity.

\subsubsection{Monte Carlo Analysis}
\label{sec:groups:algorithm:mc}

The basic group properties that we wish to derive are its position on the sky, redshift, and membership. Unfortunately, these properties depend sensitively on nuisance parameters\footnote{We use the term ``nuisance parameter'' in the statistical sense; i.e., a model parameter that is not of physical interest but must be accounted for to constrain parameters that are physically interesting.} whose values are unknown or ill-constrained. We employ Monte Carlo simulations to marginalize over these parameters by allowing them to vary over a range of values.

Our initial estimates of the group location, redshift, and mass come from the SDSS analysis in \autoref{tab:berlind}. The uncertainties in the group position and redshift are the uncertainties in the mean value for the SDSS group members (i.e., they are proportional to $N_0^{-1/2}$). We assume Gaussian priors for these values, with mean and standard deviation from \autoref{tab:berlind}.

The uncertainties in the group mass were estimated by running the SDSS FoF algorithm of Paper~3 on mock galaxy catalogs where the true halo masses were known. Our prior on $\log{M_0}$ assumes a Gaussian with mean and standard deviation listed in \autoref{tab:berlind} that is then weighted by a halo mass function (HMF) to account for the expectation that low-mass halos are more common than high-mass halos. The HMF was evaluated at $z=0.15$ with $\Delta_{\rm crit}=200$, cosmological parameters from {\sl Planck} \citep{planck14}, and fitting functions from \citet{tinker08} using the online HMF calculator of \citet*{murray13}.

We adopt a Gaussian prior on $\log{\Upsilon_{\rm grp}}$ with a mean of 2.5 and a standard deviation of 0.3~dex, which is a good characterization of the $g$-band values from \citet{proctor15}. While it is typically assumed that redder optical magnitudes (i.e., SDSS $r$- or $i$-band) correlate more strongly with an individual galaxy's stellar mass, $\Upsilon_{\rm grp}$ shows no difference in correlation for the SDSS $g$-, $r$-, and $i$-bands \citep[i.e., the Gaussian distribution of $\log{\Upsilon_{\rm grp}}$ has a width of 0.3~dex in all three bands;][]{proctor15}. Thus, we retain the $g$-band as our standard of reference for all galaxy luminosities for consistency with past studies \citep[Paper~2,][]{stocke13,stocke17,keeney17,keeney18}.

For each SDSS galaxy group in \autoref{tab:berlind}, we perform 10,000 Monte Carlo realizations, allowing the nuisance parameters to vary as described above. For each realization, we record the values of the nuisance parameters for that trial and the group members identified by our algorithm. From these, physically-meaningful values such as the group position, redshift, and luminosity, $M_{\rm grp}$, \Rgrp, $\sigma_{\rm grp}$ are derived.\footnote{We caution the reader that we are using different notation than Paper~2 and \citet{stocke17}. We now call the group velocity dispersion expected based on its mass $\sigma_{\rm grp}$ instead of $\sigma_{\rm vir}$, and the observed velocity dispersion $\sigma_{\rm obs}$ instead of $\sigma_{\rm grp}$. We believe that this new nomenclature is more transparent and internally consistent, and apologize for any confusion.} The observed size ($R_{\rm obs}$) and velocity dispersion ($\sigma_{\rm obs}$) of the group members, and the impact parameter between the group center and the QSO sight line ($\rho_\star$) also are derived. The ensemble of all realizations are then analyzed to determine the range of the derived quantities for each galaxy group.

As in Paper~2 and \citet{stocke17}, we have used robust estimators of the group properties, as initially defined by \citet*{beers90}. The position of the group on the sky and its redshift are calculated with the bi-weight location estimator, and the observed velocity dispersion of the group ($\sigma_{\rm obs}$) is calculated using the ``gapper'' scale estimator as illustrated in Equations~3-4 of Paper~2. Additionally, we estimate the observed size of the group ($R_{\rm obs}$) by calculating each group member's impact parameter with respect to the group center and adopting the 68th percentile value as $R_{\rm obs}$ (i.e., 68\% of the identified group members are within a projected distance of $R_{\rm obs}$ from the group center). Unlike $\sigma_{\rm obs}$, we do not assign any physical meaning to $R_{\rm obs}$; it is merely an empirical means of assessing whether the identified group is approximately the size that we would expect given its inferred mass (i.e., $R_{\rm obs} \sim \Rgrp$).

\autoref{tab:params} summarizes our Monte Carlo realizations. Column~1 lists the group identifier, columns~2 \& 3 the sky position in degrees, and column~4 the redshift ($z_{\rm grp}$). Column~5 lists the number of group members ($N_{\rm grp}$), column~6 the total group luminosity ($L_{\rm grp}$; \autoref{eqn:Lgrp}), and column~7 the total group mass ($M_{\rm grp}$; \autoref{eqn:Mgrp}). Columns~8 and 9 are the observed group radius ($R_{\rm obs}$) and velocity dispersion ($\sigma_{\rm obs}$). Column~11 lists the ratio $R_{\rm obs}/\Rgrp$, and column~12 the ratio $\sigma_{\rm obs}/\sigma_{\rm grp}$, where \Rgrp\ and $\sigma_{\rm grp}$ are calculated using \autoref{eqn:Rgrp} and \autoref{eqn:sgrp}, respectively. Columns~10 and 13 list the sight line impact parameter from the group centroid in kpc and as a fraction of \Rgrp. Details of individual groups are discussed in \autoref{sec:indiv}.

\clearpage
\input{./tab_params.tex}

\begin{figure*}
\epsscale{0.96}
\centering\plotone{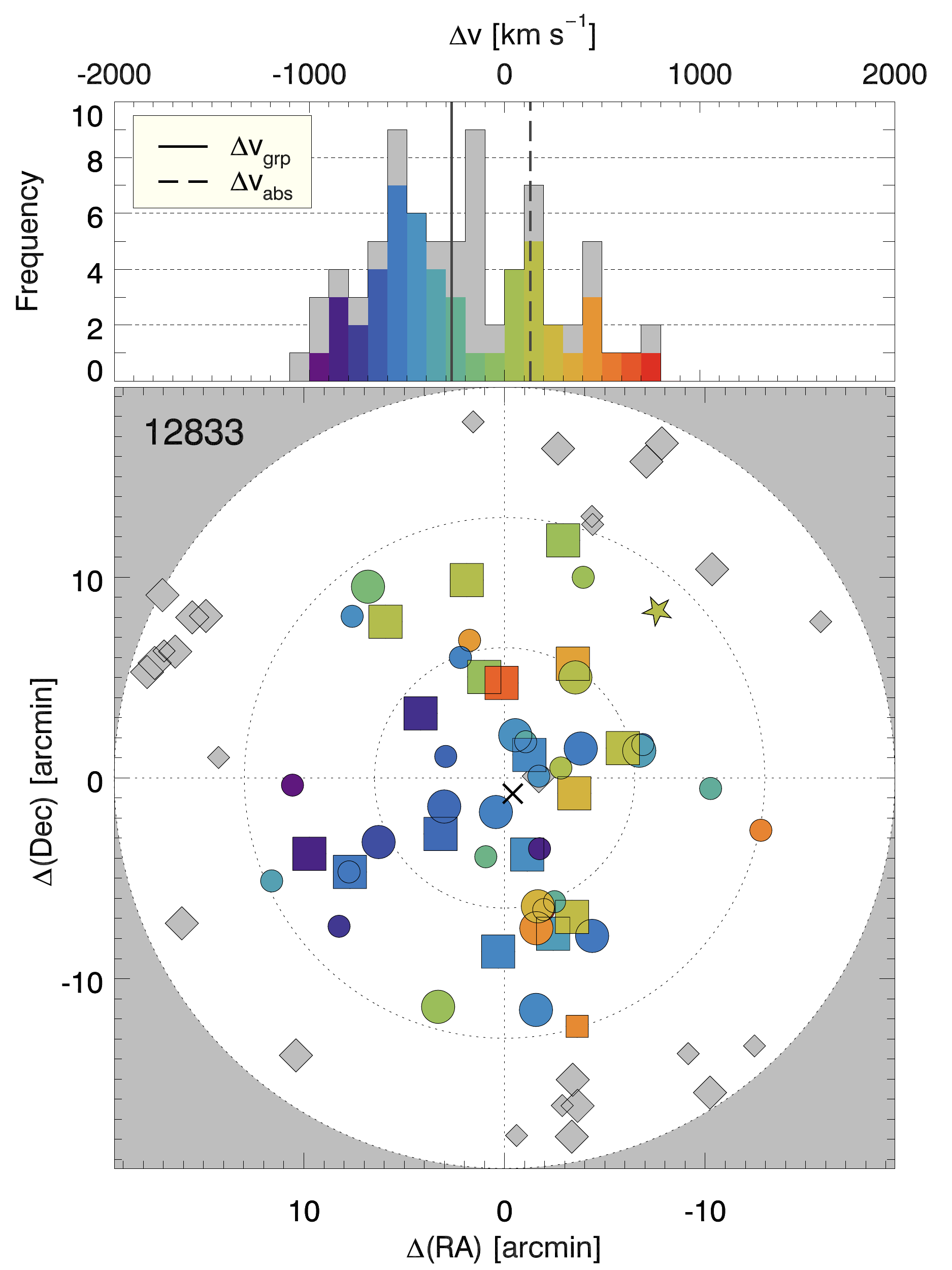}
\caption{\textit{Top:} Velocity histogram of galaxies near group 12833. Group members are color coded from purple to red for the lowest and highest redshift group members, respectively. Field galaxies are gray. The solid vertical line shows the group velocity centroid and dashed vertical lines show the positions of absorption in the \hst/COS spectra. All velocities are calculated with respect to the initial estimate of the group redshift from the SDSS analysis in \autoref{tab:berlind}. \textit{Bottom:} Distribution on the sky of galaxies near group 12833. Group members are colored as in the top panel. Gray diamonds represent field galaxies, squares represent SDSS galaxies, and circles represent new galaxy redshifts from our MOS campaign. \edit1{Symbol size is indicative of galaxy luminosity with the largest symbols indicating $L>L^*$, intermediate size symbols indicating $0.5\,L^* <  L  < L^*$ while the smallest symbols indicate $L < 0.5\,L^*$.} The position of the group centroid on the sky is marked with an ``$\times$'', and the position of the \hst/COS sight line is marked with a star. Circles are shown at 1~Mpc intervals, centered on the initial estimate of the group position from the SDSS analysis in \autoref{tab:berlind}.
\label{fig:12833}}
\end{figure*}

\clearpage
\input{./tab_12833.tex}

\input{./tab_groups.tex}



\subsection{Adopted Group Membership}
\label{sec:groups:members}

The Monte Carlo analysis described in \autoref{sec:groups:algorithm} allows us to assess which galaxies are group members probabilistically. Regardless of the uncertainties of this process it is essential that a final group membership is adopted so that its overall properties (mean recession velocity, mass, etc) and relationship to any absorption can be assessed, as well as whether the nearest galaxy to the QSO sight line is a group member \citep[e.g., the passive galaxy closest to one of the absorbers in Group~32123 is not a member of the group identified in this field;][]{stocke17}.

As a first step, we determine each galaxy's membership fraction, $f_{\rm mem}$; i.e., the fraction of the Monte Carlo trials in which a particular galaxy was identified as a group member. While it is tempting to define the final group as the set of all galaxies with $f_{\rm mem} > 50$\%, there is no guarantee that this set of galaxies is self-consistent (i.e., all located within $1.5\,\Rgrp$ of the group center and $2\,\sigma_{\rm grp}$ of the group redshift, to the exclusion of all other galaxies).

Instead, we calculate several group parameters (sky position, $z_{\rm grp}$, $N_{\rm grp}$, $M_{\rm grp}$) for each Monte Carlo realization, and compare them with the distribution of values over all realizations (i.e., marginal likelihoods\footnote{These are the histograms in \autoref{fig:corner} (see \autoref{sec:indiv}).}) to determine whether the value for a particular realization is a common outcome. This comparison uses an adaptive kernel density estimate \citep{terrell92} of the marginal likelihoods to avoid artifacts from arbitrary choices of histogram bin sizes and centers.

Each realization is then assigned a joint likelihood by multiplying the marginal likelihood values described above over all parameters. We adopt the membership of the realization with the highest joint likelihood as the final group configuration. This procedure ensures that we only adopt group configurations that are representative and self-consistent.

\autoref{fig:12833} shows the adopted membership of individual groups, with group members displayed in color and the gradation in color indicating the recession velocity gradient for group members. Gray symbols are non-group galaxies in the field-of-view and velocity range. Zero velocity in the top histogram and the center in the bottom plot are the SDSS group velocity and sky position in \autoref{tab:berlind}, respectively. The solid vertical line in the top panel, and the ``$\times$''symbol in the bottom panel, mark the adopted group velocity and sky centroids, respectively. The star marks the AGN sight line location in the bottom panel, while the absorber velocity or velocities are shown as dashed vertical lines in the top panel.

\autoref{tab:12833} lists the adopted members of individual galaxy groups, with galaxies listed in order of decreasing membership fraction, $f_{\rm mem}$. This table includes by column: (1) a galaxy identifier, which is either a SDSS ``ObjID'' or, for galaxies without SDSS redshifts, a string in the style of \citet{keeney18}; (2 \& 3) galaxy sky position in degrees; (4) galaxy redshift; (5) rest-frame $g$-band luminosity; (6) galaxy spectral type;  (7 \& 8) galaxy offset on the sky (in kpc) and in velocity (in \kms) from the SDSS group centroids in \autoref{tab:berlind}; and (9) $f_{\rm mem}$. A galaxy's spectral type is determined visually as described in \citet{keeney18}, and abbreviated ``E'' for emission-line galaxies, ``A'' for absorption-line galaxies, and ``C'' for composite galaxies that show both emission and absorption.

The properties of our adopted groups are summarized in \autoref{tab:groups}, where all values are calculated using only galaxies that are adopted as group members by the procedure described above. The columns list largely the same information as those of \autoref{tab:params}, except for columns~12 and 13, which list the offsets between the adopted sky position and redshift and the SDSS values (\autoref{tab:berlind}), and column~14, which lists the group's spiral fraction. We estimate a group's spiral fraction using the spectral types of the members, where all emission-line and composite galaxies are considered ``spirals''. These percentages use only those galaxies at $L>0.6\,L^*$ where the MMTO galaxy survey work is largely complete \citep{keeney18}.

The centroids for 9 of the 12~groups did not vary significantly between the SDSS analysis and the results found here using much deeper data. The deeper MOS data find that groups 44564 and 44565 in the CSO\,1022 sight line are part of a larger group with centroids intermediate between the two, and the new analysis finds group~12833 in the RBS\,711 sight line to be $1.5\sigma$ ($\sim300~\kms$) lower in redshift than in \autoref{tab:berlind}. The remaining 9 groups showed little change in their 3D locations between \autoref{tab:berlind} and \autoref{tab:groups}. Derived halo masses range from $10^{13.5}$ to $10^{14.6} M_{\odot}$, with a corresponding range of virial radii of $\Rgrp=650$-1500~kpc and velocity dispersions of $\sigma_{\rm grp}=270$-610~\kms\ (see \autoref{tab:groups}). The range of impact parameters in this study is 0.36-$2\,\Rgrp$, with few sight lines $<\Rgrp$ due to the criterion that these sight lines could not intercept an SDSS galaxy halo at $<1.5\,\Rvir$, naturally excluding sight lines in the dense, inner regions of groups.

There is little difference between the total halo masses computed by Paper~3 using just the SDSS spectroscopic survey galaxies and the total halo masses computed using the deeper MMTO redshift survey (see \autoref{tab:groups}). The revised values are only marginally lower (statistically less by 0.05~dex or 12\%) than the SDSS values, much smaller than the factor 2-3 combined errors on individual mass estimates (see \autoref{tab:berlind} and \autoref{tab:params}). Therefore, despite the modest number of group members in the SDSS analysis, the group halo masses derived by Paper~3 appear to be quite accurate and can be used as viable indicators of group mass for comparisons with other group properties, including those of the absorbers associated with groups at higher and lower estimated halo masses (see \autoref{sec:groups:properties}, \ref{sec:disc:fcov} and \ref{sec:disc:kinematics}).

\begin{figure}
  \epsscale{1.2}
  \centering\plotone{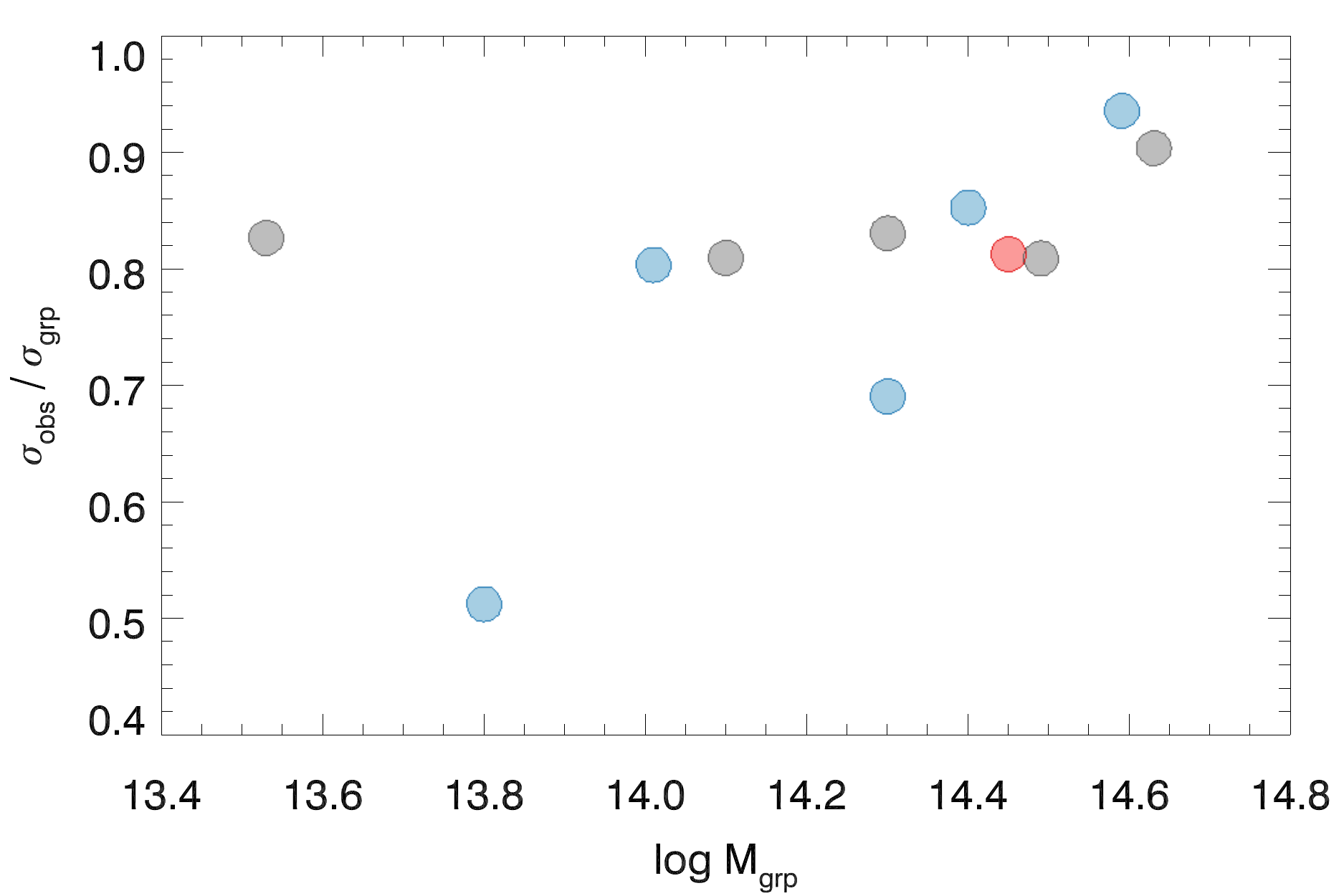}
  \caption{The calculated group mass ($M_{\rm grp}$) versus $\sigma_{\rm obs}/\sigma_{\rm grp}$, which is a measure of the virialization of these systems; e.g., $\sigma_{\rm obs}/\sigma_{\rm grp}$ = 1 suggests a highly-virialized system.
    \edit1{Gray symbols are groups with only metal-free absorbers or non-detections in \HI, while red symbols are groups with absorbers having only low ions present.} \OVI-bearing groups are blue. All but two groups have $\sigma_{\rm obs}/\sigma_{\rm grp} \gtrsim 0.8$; interestingly, the two less-virialized groups have rich associated UV absorption systems and high spiral fractions (see below).
  \label{fig:deviation}}
\end{figure}
    
\begin{figure}
  \epsscale{1.2}
  \centering\plotone{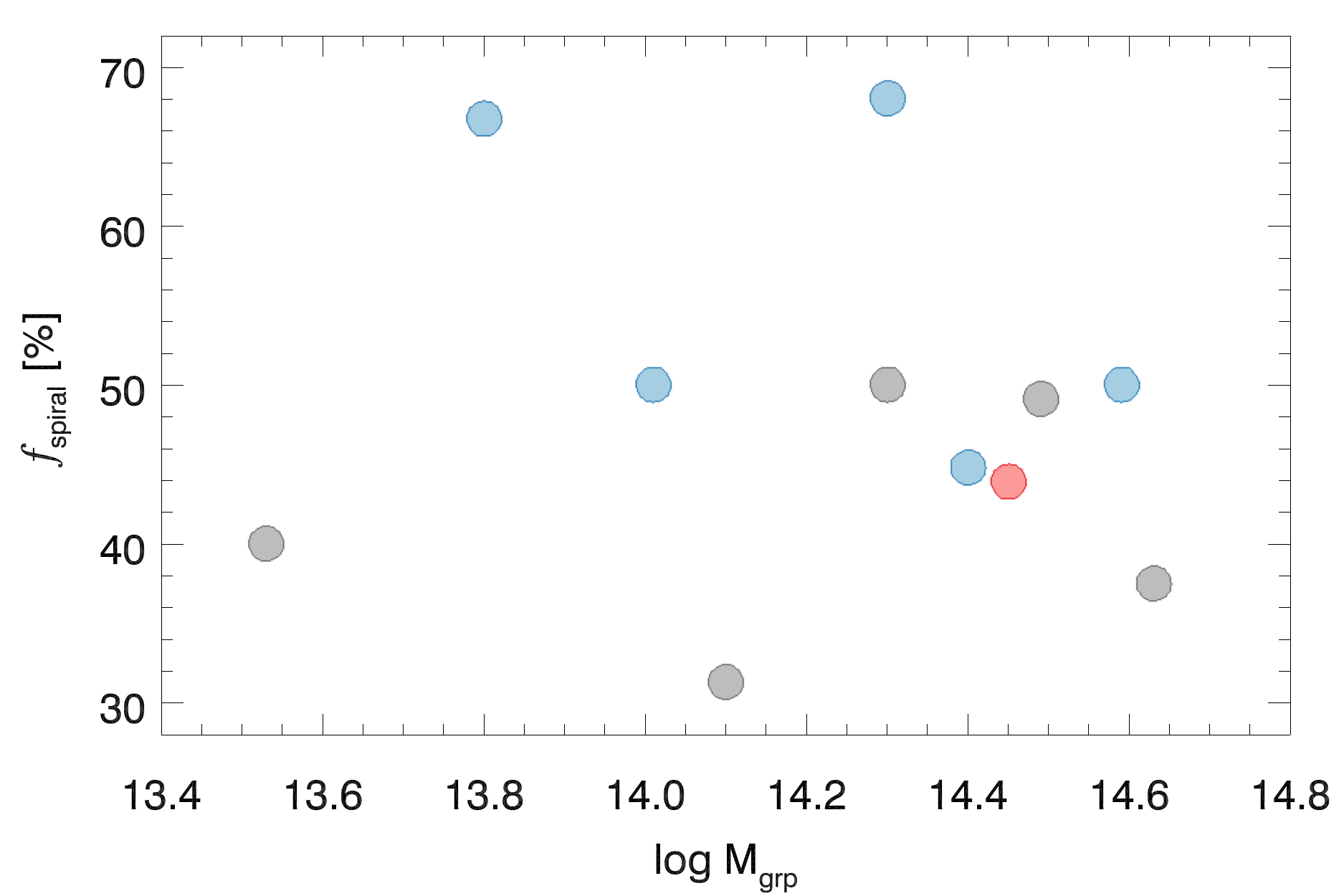}
  \caption{The calculated group mass ($M_{\rm grp}$) versus the fraction of $L>0.6\,L^*$ group galaxies that are star-forming ($f_{\rm spiral}$). For the star-forming/spiral classification we have used the spectroscopic proxy of emission lines; i.e., all galaxy spectra with either emission lines only or emission lines plus absorption lines are classified as spirals for this purpose \citep{keeney18}.  \edit1{Gray symbols are groups with only metal-free absorbers or non-detections in \HI, while red symbols are groups with absorbers having only low ions present.} \OVI-bearing groups are blue. 
  \label{fig:Mfspiral}}
\end{figure}
    
\begin{figure}
  \epsscale{1.2}
  \centering\plotone{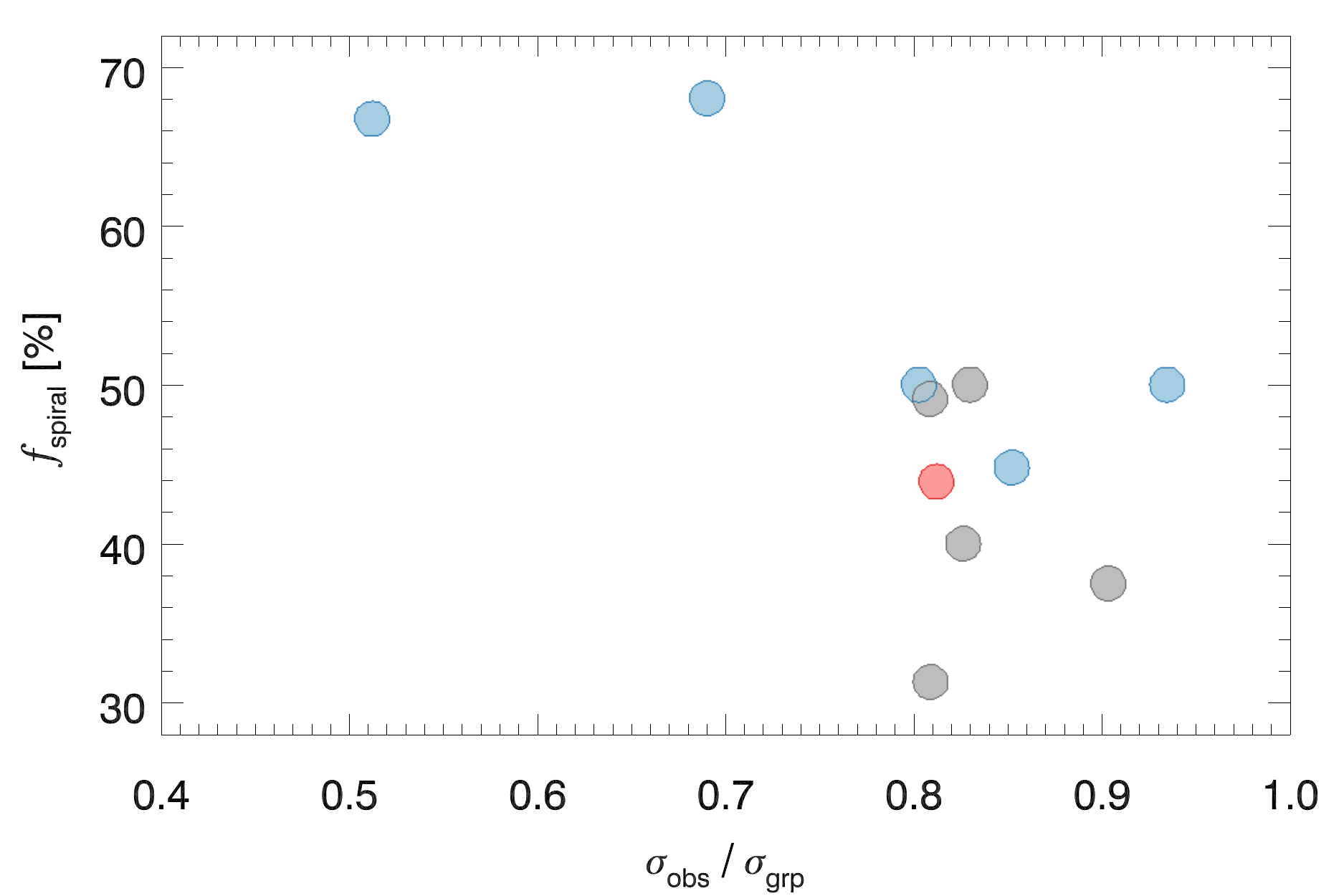}
  \caption{The virialization parameter  ($\sigma_{\rm obs}/\sigma_{\rm grp}$) versus the fraction of $L>0.6\,L^*$ group galaxies that are star-forming ($f_{\rm spiral}$). For the star-forming/spiral classification we have used the spectroscopic proxy of emission lines; i.e., all galaxy spectra with either emission lines only or emission lines plus absorption lines are classified as spirals for this purpose \citep{keeney18}. \edit1{Gray symbols are groups with only metal-free absorbers or non-detections in \HI, while red symbols are groups with absorbers having only low ions present.} \OVI-bearing groups are blue.  In this plot the strong correlation between ETGs and virialization is apparent, although more examples of spiral-rich, unvirialized groups are required to make this conclusion robust. 
  \label{fig:Sigmafspiral}}
\end{figure}

\subsection{Basic Group Properties}
\label{sec:groups:properties}

In this section the basic properties of these groups are explored in order to place the UV absorptions into their proper context. While the number of group galaxies ($N_{\rm grp}$), the total group luminosity ($L_{\rm grp}$), total group mass ($M_{\rm grp}$), group virial radius (\Rgrp), and velocity dispersion ($\sigma_{\rm grp}$) are all directly related by \autoref{eqn:Lgrp}-\ref{eqn:sgrp}, the observed velocity dispersion ($\sigma_{\rm obs}$) and the fraction of spiral or star-forming galaxies are independent of these derived group properties. The observed velocity dispersion differs from the calculated velocity dispersion ($\sigma_{\rm grp}$) because the group galaxies may not be in virial equilibrium.

In order to further explore the kinematic signature of virialization in these groups, \autoref{fig:deviation} shows the observed relationship between calculated group mass and the ratio between the observed and calculated velocity dispersion, $\sigma_{\rm obs}/\sigma_{\rm grp}$. At $M_{\rm grp} > 10^{14}~M_{\odot}$, all but one of these groups have $\sigma_{\rm obs}/\sigma_{\rm grp} \geq 0.8$, suggesting that these massive groups are mostly virialized. The sample also includes a couple of groups (36001 and 44739) which are not close to being virialized by this same measure. These two groups are among the few in this sample with very rich absorption systems (e.g., group~44739 has 4~absorbers; one \HI-only; one \HI\ + \SiIII; and two \HI\ + \OVI; see \autoref{sec:disc:kinematics}). Thus, this small sample appears to bridge the halo mass region where groups are becoming fully virialized in the current epoch.

\autoref{fig:Mfspiral} and \ref{fig:Sigmafspiral}, show the spiral (i.e., star-forming) fraction of galaxies in these groups as a function of halo mass and of virialization, respectively. Because we do not possess good morphological information about these group galaxies, we have used the spectral classification of \citet{keeney18} as a proxy; viz., emission-line and composite (emission-line plus absorption-line) galaxies are classified as star-forming/spirals for this purpose. The spiral fraction is computed using all group members with $L>0.6\,L^*$. The MMTO MOS group galaxy survey used for these classifications \citep{keeney18} is $>90$\% complete to this limit out to $\rho_\star \leq 1$~Mpc for all groups except 25124, which is complete only to a limit slightly deeper than the SDSS ($\sim 3\,L^*$ out to $\rho_\star \leq 1$~Mpc). If group~25124, which is a massive ($10^{14.6}~M_{\odot}$), highly-virialized ($\sigma_{\rm obs}/\sigma_{\rm grp}=0.8$) group with $f_{\rm spiral}=0.45$, were excluded from this discussion it would not alter these plots noticeably. 

It is clear from \autoref{fig:deviation} and \ref{fig:Mfspiral} that the mass range of this survey spans a range of group properties from nearly fully-virialized massive groups with large fractions of early-type galaxies (ETGs) to partially-virialized groups with large spiral fractions. \autoref{fig:Sigmafspiral} highlights that the spiral fraction and partial virialization go together, although this conclusion certainly requires a much larger group sample, particularly at lower halo mass and lesser virialization, to be robust.  Nevertheless, in the most massive, highly-virialized groups the spiral fraction is $<50$\%, so these groups are dominated by ETGs.

Among the nine groups with high virialization factors, the three groups with the lowest spiral fraction have no detected UV absorption lines. These groups are neither highly centrally-concentrated nor dominated by a single ETG like the massive X-ray-detected groups of \citet{mulchaey96}, but may have similar gas properties. Deep X-ray continuum observations are appropriate to test whether a hot IGrM is present in the ETG-dominated groups in this sample. 

And while the least-massive groups in the sample are highly spiral-dominated and not always highly-virialized, they are still an order of magnitude more massive than the small, spiral-dominated groups like the Local Group where most of the star-formation in the current Universe takes place.  New UV spectroscopy of spiral-rich groups at $M_{\rm grp} \leq 10^{13.5}~M_{\odot}$ is required to investigate typical star-forming sites in the local Universe.

\section{Discussion of Results}
\label{sec:disc}

\subsection{Covering Fraction}
\label{sec:disc:fcov}

Despite the wide range of impact parameters targeted, \HI\ absorption was detected toward 7 of the 12~groups probed through the full range of impact parameters (group~25124 is counted as a non-detection because its single absorber system is at least partly associated with an individual galaxy; see \autoref{tab:ng}). \autoref{tab:abs} (see \autoref{sec:abs}) lists all absorbers within $\pm2.5\,\sigma_0$ (typically, $\pm1000$-1500~\kms) of the group velocity centroid using the initial SDSS analysis (\autoref{tab:berlind}), but for the important choice of the absorber velocity range for group association, we adopt a range corresponding to $\pm2.5\,\sigma_{\rm grp}$, based on the adopted group parameters listed in \autoref{tab:groups}.

While this choice is somewhat arbitrary, it is a good match to the criterion we used for including galaxies as group members; i.e., based on mock catalogs produced from $N$-body simulations, $>99$\% of all satellite halos are found within $\pm2\,\sigma_{\rm grp}$ of the primary halo's velocity (see \autoref{sec:groups:algorithm}). Increasing this range from $\pm2\,\sigma_{\rm grp}$ to $\pm2.5\,\sigma_{\rm grp}$ provides a small allowance for both the statistical sampling of the potential well using the modest number of galaxies available (i.e., uncertainties in mean velocity and velocity dispersion), and also for the peculiar velocity of gas which may have been expelled from, or is falling into, any individual galaxy in the group.

Enforcing these velocity bounds removes only one absorber system in \autoref{tab:abs} from group association: 36001\,/ 0.18429 at $-906$ \kms\ (a velocity difference of $2.8\,\sigma_{\rm grp}$). While there are other absorbers that remain in the sample at larger velocity differences, this particular absorber is excluded because group~36001 has the smallest $\sigma_{\rm grp}$ in the sample; however, there are two other absorber systems associated with group~36001, so it is retained as a group detection. Decreasing this limit to $2\,\sigma_{\rm grp}$ eliminates only an additional 2~systems of the 18 total. There are no viable individual galaxy associations associated with any of the systems with large $|\Delta v|$ (see \autoref{fig:abs} and \autoref{tab:ng}).

While there are 10~sight lines through 12 galaxy groups as defined by our original set of observations, two modifications are required to correctly assess the covering fraction. First, our deeper MOS and subsequent analysis finds evidence that groups~44564 and 44565 in the CSO~1022 sight line are, in fact, one considerably more massive group; this leaves 11~groups probed. Second, the single absorber system in group~25124 was determined to be associated with a single bright galaxy, not the group as a whole; however, a BLA present within the \lya\ absorption profile (see \autoref{sec:cos:metals}) is similar to other galaxy group absorptions investigated by Paper~2. Due to this uncertainty, we eliminate group~25124 from the covering fraction discussion rather than count it as a non-detection, leaving ten groups only.

\input{./tab_cf.tex}

The resulting \lya\ covering fraction of the 10~groups of galaxies is shown in \autoref{tab:cf} as a function of normalized impact parameter, $\rho_\star/\Rgrp$. Despite the small number statistics, \autoref{tab:cf} shows that the covering fraction is substantial ($\gtrsim 50$\%) out to at least $2\,\Rgrp$ ($\sim2$~Mpc, given the masses of the groups probed). This result is consistent with the rather constant distribution of covering fractions with radius ($\sim50$\% out to several virial radii, or $\sim1$~Mpc) seen for absorption-line galaxies in \citet{keeney18}. It appears that the CGM of absorption-line galaxies joins seamlessly with group gas to which those galaxies typically belong.

\begin{figure}
  \epsscale{1.2}
  \centering\plotone{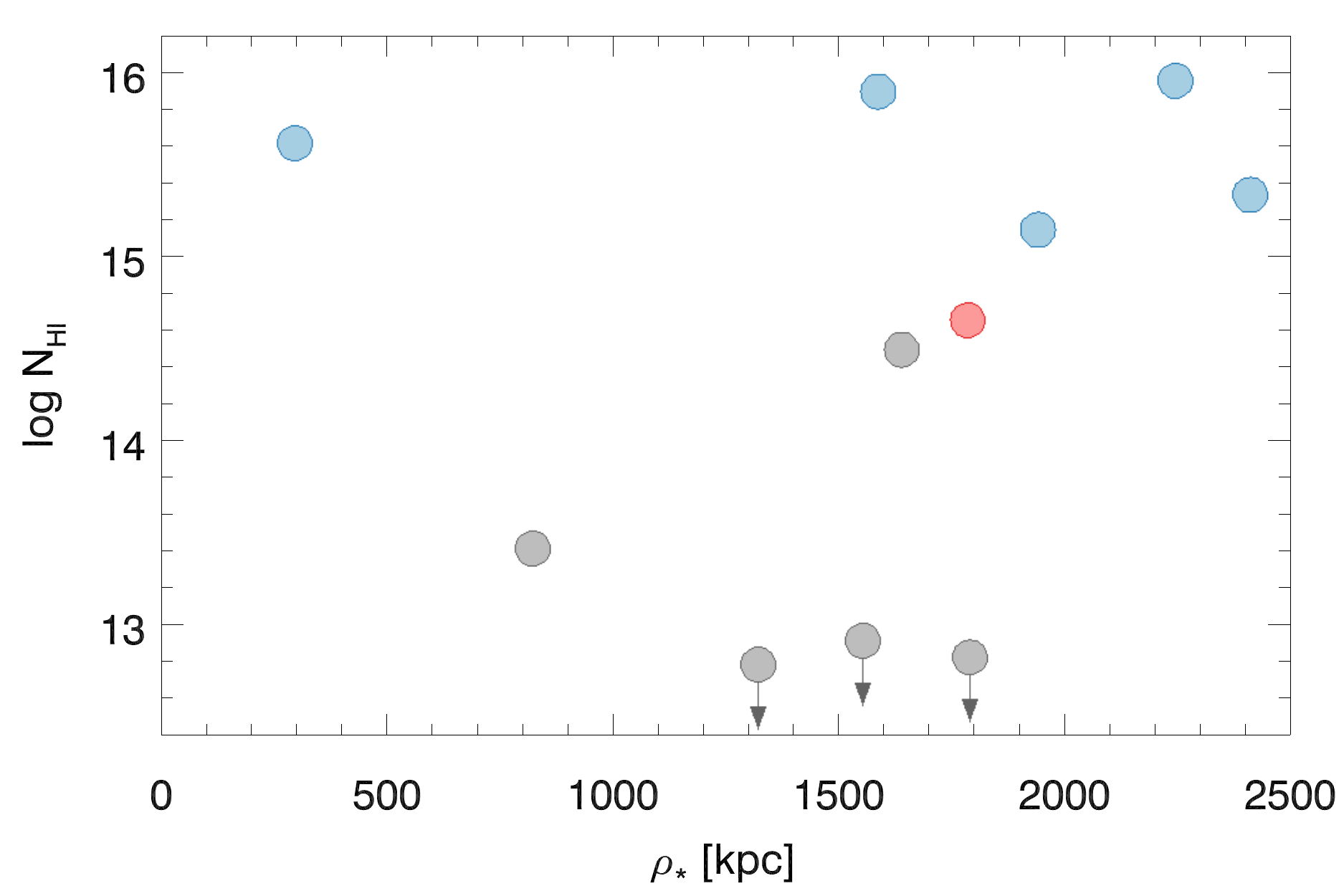}
  \caption{The total $N_{\rm H\,I}$ in all absorption components associated with the group as a function of the physical impact parameter $\rho_\star$. \edit1{Gray symbols are groups with only metal-free absorbers or non-detections in \HI, while red symbols are groups with absorbers having only low ions present.} \OVI-bearing groups are blue. The \HI\ column densities do not show any obvious trend with the group-centric impact parameter out to 2~Mpc, or $\sim2\,\Rgrp$.
  \label{fig:NHIrho}}
\end{figure}

Similarly, in \autoref{fig:NHIrho} the total \HI\ column density within $2.5\,\sigma_{\rm grp}$ of $z_{\rm grp}$ along these group sight lines is shown as a function of sight-line impact parameter from the group centroid on the sky. The obvious conclusion from this plot is that cool and/or warm gas remains abundant out to at least 2~Mpc (or $\sim 2\,\Rgrp$) for these groups. While the high, but $<100$\%, covering fraction found here may be due to a patchy distribution of UV-absorbing gas in all these groups, it is also possible that the presence or absence of this gas can be due to other factors.

\begin{figure}
  \epsscale{1.2}
  \centering\plotone{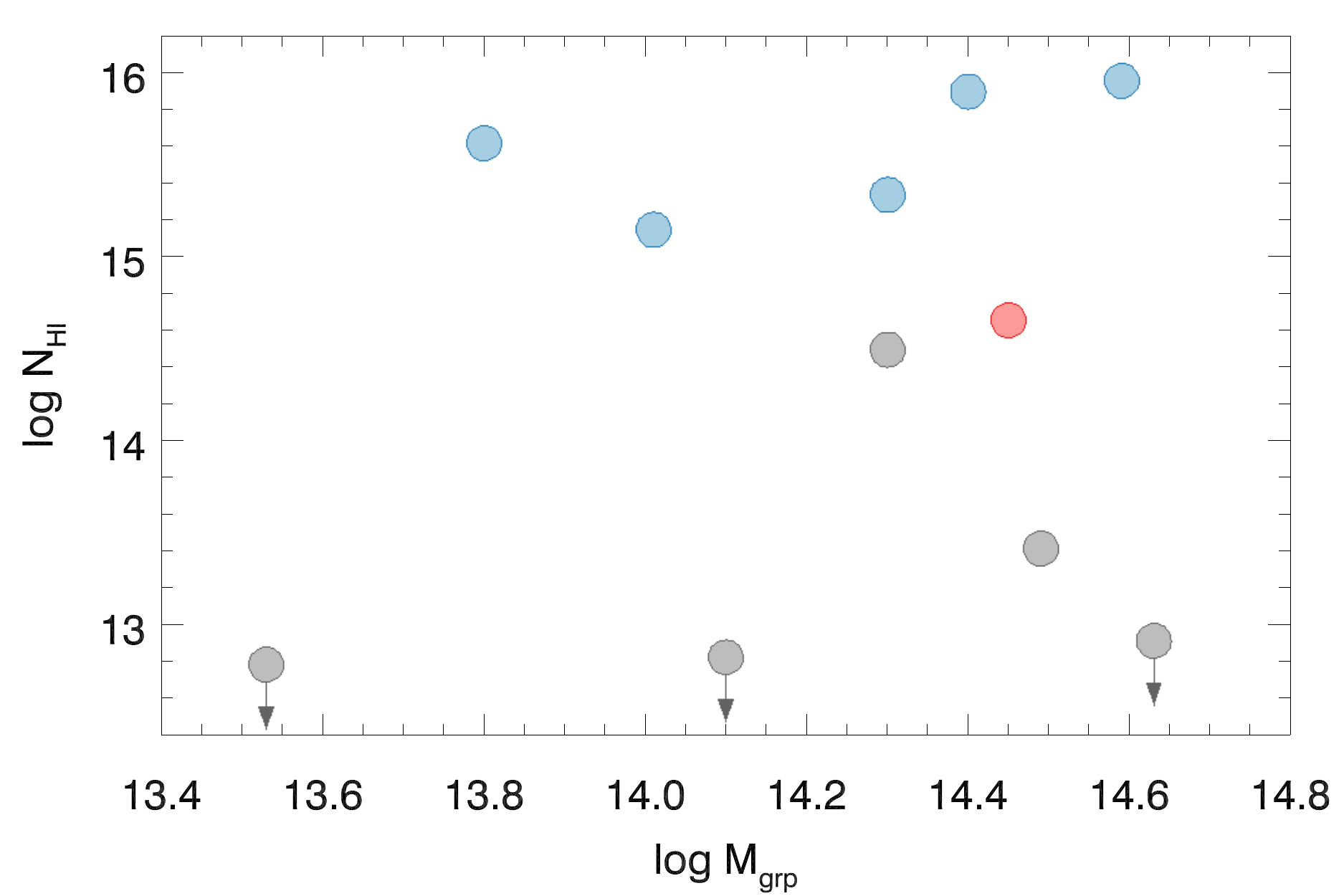}
  \caption{\edit1{The total $N_{\rm H\,I}$ in all absorption components associated with the group versus the calculated group mass ($M_{\rm grp}$). Gray symbols are groups with only metal-free absorbers or non-detections in \HI, while red symbols are groups with absorbers having only low ions present.} \OVI-bearing groups are blue. Despite including both virialized and unvirialized groups, there is no obvious trend between the presence of absorbers and total group halo mass in this sample. Neither the \HI\ detections/non-detections nor the \HI\ column densities show any obvious trend with the estimated mass of the group.   
  \label{fig:MNHI}}
\end{figure}

In Paper~2, an argument was presented that the area covered by warm group gas is large, $\sim1$~Mpc in radius per group if unity covering factor is assumed. This estimate is based on the number density per unit redshift ($d\mathcal{N}/dz= 4$) of warm absorbers ($T\geq 10^5$~K) seen at low-$z$ and the space density of groups of galaxies ($\approx 3 \times 10^{-4}~\mathrm{Mpc^{-3}}$; Paper~3). The $\gtrsim50$\% covering fraction out to $\rho \geq 2$~Mpc found by this study supports the previous conclusion that a large gaseous medium exists in galaxy groups.

However, the huge area covered by \HI\ absorption does not require a huge mass because the filling factor of this gas is unknown and could be small; e.g., if the cool gas is in small clouds like in the CGM and the warm gas is in filamentary structures scattered throughout the volume of the group, the filling factor of cool plus warm group gas could be similar to the filling factor (3-5\%) of cool gas in the CGM, or even smaller. In this case, the UV-detected group gas can have a total gas mass \textit{per galaxy} of only 2-4\% of the full baryon budget. This makes it unlikely that the UV-detected group gas solves the baryon problem. But, when viewed as transition gas, the UV absorbing gas may be cospatial with even hotter gas that constitutes a volume-filling IGrM. Further, in this sample there is no strong evidence for warm or warm-hot gas found associated with these groups (see \autoref{sec:cos:BLAs}). 

\autoref{fig:MNHI} shows the total \HI\ column density of all absorption components in these groups as a function of group mass. Despite their lower mass, the two relatively un-virialized groups, 36001 ($\log{M_{\rm grp}}=13.80$) and 44739 ($\log{M_{\rm grp}}=14.30$) both have \HI-rich absorption systems. There is no obvious trend between group mass and either detections/non-detections or the total $N_{\rm H\,I}$ in this mass range.

However, it is notable that the three groups that lack \HI\ absorption (\autoref{tab:abs} in \autoref{sec:abs}) are the three most ETG-rich groups (\autoref{tab:groups}), and among the most highly virialized (\autoref{fig:Sigmafspiral}). While a larger sample is certainly needed, this result suggests that groups which are spiral-rich and not completely virialized will be detected strongly and multiply in \HI\ and metals. This result predicts that less-virialized groups at lower halo mass will be found to have large reservoirs of cool plus warm UV-detectable gas which is not obviously associated with a specific group galaxy.  

\edit1{Different from the results found in Paper~2, there are very few (2~probable and 4~possible) warm absorbers found in these groups. Despite the results of Paper~2 this suggests that the warm gas in groups is rather patchy and with potentially low covering fraction, i.e., $\lesssim10$\%.}

\subsection{Absorber Kinematics}
\label{sec:disc:kinematics}

\begin{figure}
  \epsscale{1.2}
  \centering\plotone{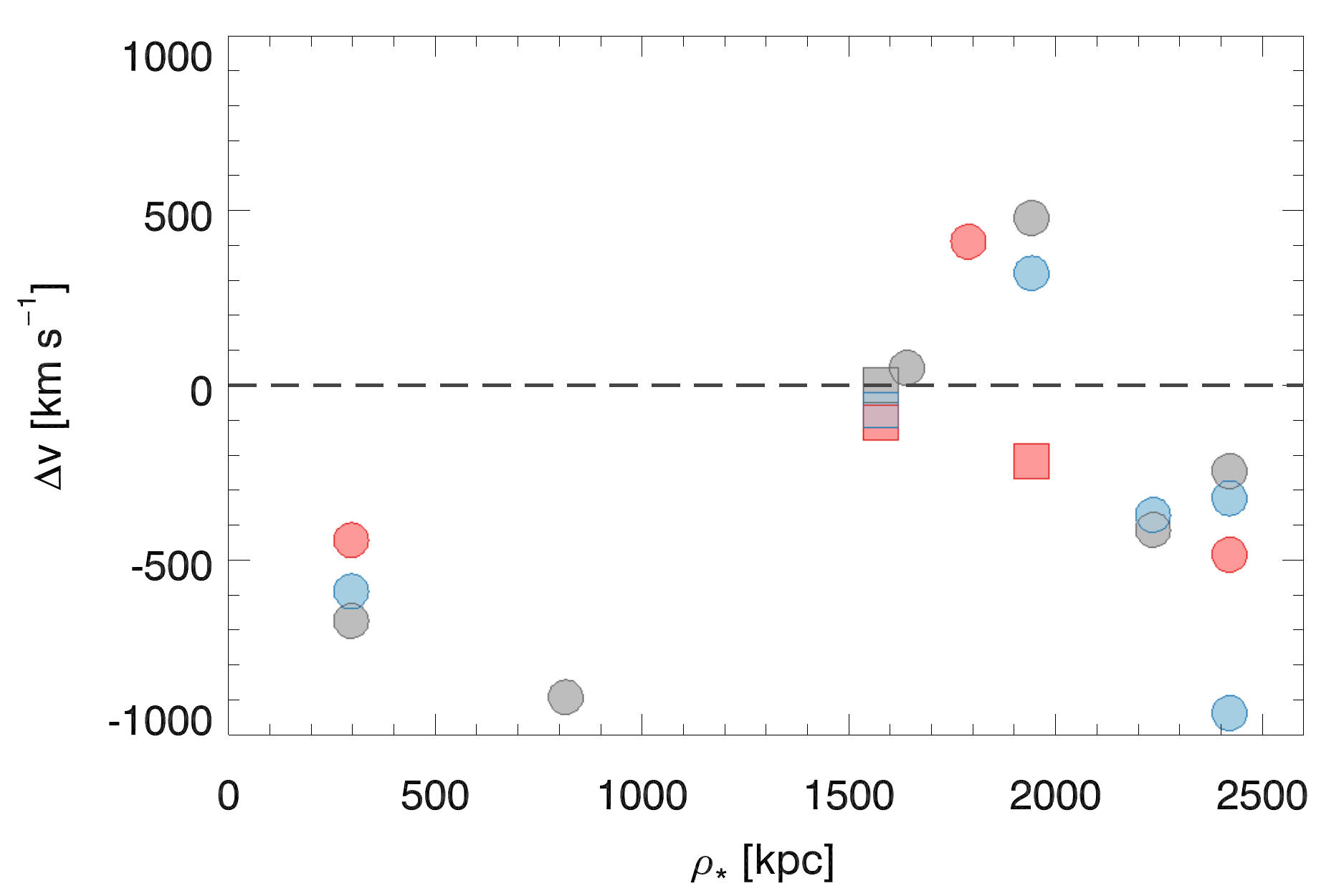}
  \caption{The physical impact parameter in kpc is plotted against the (absorber-group) velocity difference in \kms. Colors are as in \autoref{fig:MNHI}, and squares are absorbers which could be associated with individual galaxies.
  \label{fig:dv_rho}}
\end{figure}

Limited by our available measurement tools, the kinematic information we have for these absorbers consists of: (1) a two-dimensional location on the sky with respect to the sky centroid of the group, and (2) the radial component of the velocity relative to the group centroid velocity. In \autoref{fig:dv_rho}, these basic quantities are displayed in a plot of impact parameter, $\rho_\star$, versus the one dimensional velocity difference, $\Delta v = c(z_{\rm abs} - z_{\rm grp})/(1+z_{\rm grp})$. There is no strong concentration around $\Delta v = 0$, which would be expected if the detected gas is a massive IGrM as seen in X-ray bremsstralung emission for more massive systems \citep{mulchaey00}. The relative dearth of absorbers at $\rho_\star \leq 1.5$~Mpc is due to our imposed criterion that no sight line pass within $1.5\,\Rvir$ of an SDSS group galaxy, thus avoiding the dense central regions of groups.

There is no obvious kinematic difference between the metal-bearing absorbers (red symbols for only low ions present and blue symbols for \OVI\ present) and the metal-free absorbers (gray symbols); nor is there any difference between the locations of the high and low ions. The square symbols show the location of the absorbers possibly associated with individual bright galaxies in groups~25124 and 32123. If any or all of those absorbers are associated with the entire group rather than individual galaxies, the basic results of this plot are not altered. 

\begin{figure}
  \epsscale{1.2}
  \centering\plotone{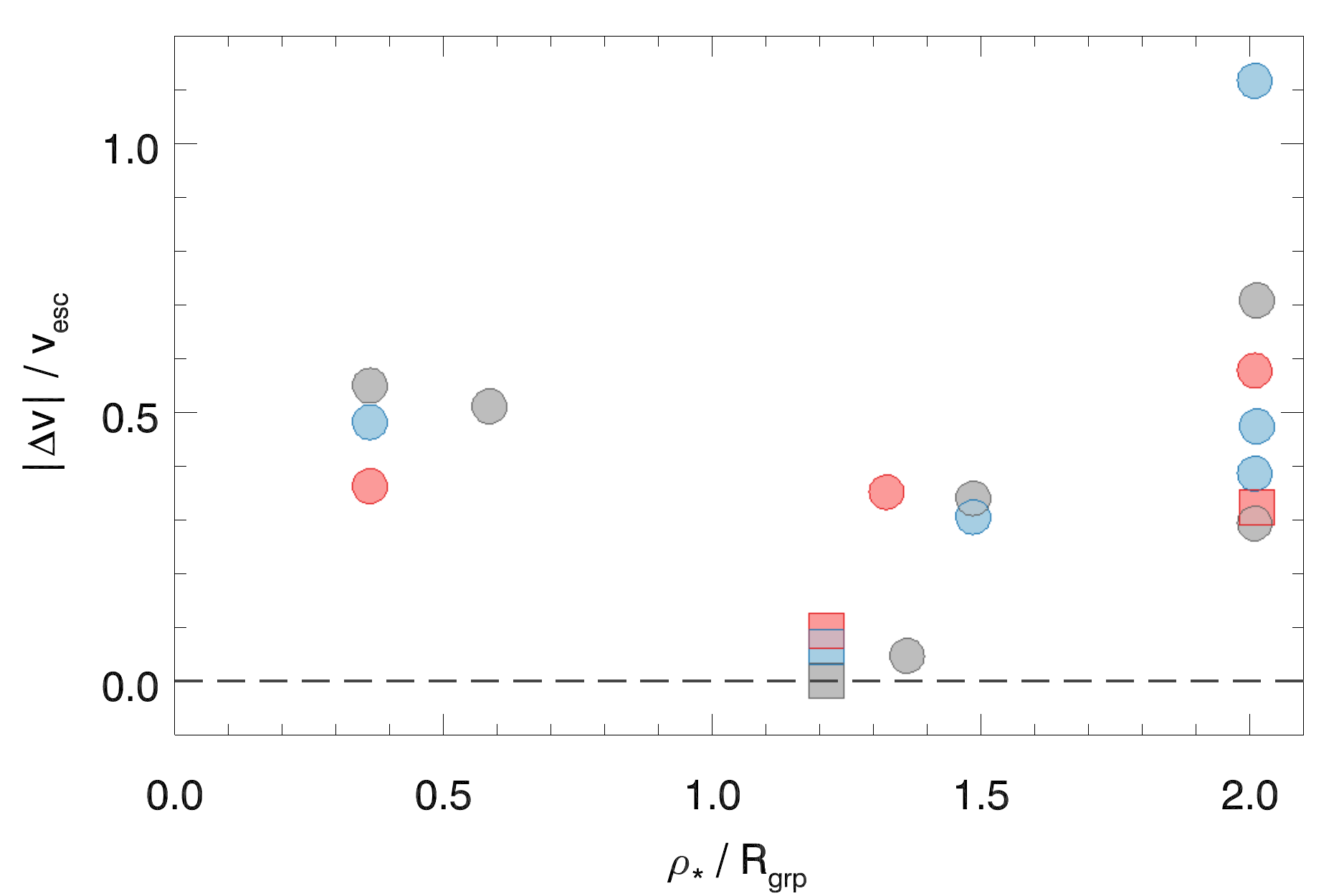}
  \caption{Symbols are similar to their depictions in \autoref{fig:dv_rho}, but in this case the impact parameter ($\rho_\star$) is displayed in units of the group virial radius (\Rgrp), while the velocity difference is in units of the escape velocity ($v_{\rm esc}$) at the radius of the impact parameter.
  \label{fig:dv_norm}}
\end{figure}

More revealing is \autoref{fig:dv_norm}, which shows these same observables normalized by the theoretical quantities of group virial radius (i.e., $\rho_\star/\Rgrp$) on the horizontal axis and escape velocity (i.e., $|\Delta v|/v_{\rm esc}$) on the vertical axis, where the escape velocity is calculated at a radius equal to the observed impact parameter. What is striking in this plot is that only one of the absorbers has sufficient radial velocity to escape the group gravitational potential well.

This result is similar to the results found the COS-Halos study \citep{tumlinson11} of the inner half of the CGM for which absorbers do not possess sufficient radial velocity to escape their home galaxy's gravitational potential. While \citet{stocke13} and \citet{keeney17} found a similar result to COS-Halos in the inner CGM, when the outer CGM ($\rho>0.5\,\Rvir$) is probed the results are quite different. At $\rho > 0.5\,\Rvir$, the one-dimensional velocity difference between absorbers and their associated galaxies cluster around the value for the escape velocity (see Figure~8 in \citealp{stocke13} and Figure~13 in \citealp{keeney17}). For individual galaxies, this suggests that many of these CGM absorbers will escape if their direction of motion is out-going, although some few still are unlikely to escape \citep[see e.g.,][]{tumlinson99, keeney05, stocke10, tumlinson11, stocke13}. Presumably the absorbers discovered in this small survey are examples of gas which has escaped its home galaxy CGM and is now bound to its home galaxy group.

This result does not appear to be caused by the limit placed on absorber velocity, by which absorbers are excluded for association if $|\Delta v| > 2.5\,\sigma_{\rm grp}$ from the group mean velocity. The one absorber excluded by this velocity cut (36001\,/\,0.18429) has a radial velocity relative to this group of only $0.7\,v_{\rm esc}$. On the other hand, the absorber whose relative velocity exceeds the escape speed (44739\,/\,0.11537, a mixed \HI, \OVI, and \SiIII\ absorber) has a relative velocity of $< 2\,\sigma_{\rm grp}$ from the mean group velocity and so would have been included even if the limit were lowered.

To ensure that any fast-moving absorber that could have escaped from these groups has not been excluded from consideration by the assumed cutoff in velocity that was imposed, a search for such absorbers was conducted around all the sight lines in this survey. Increasing the relative velocity bounds to include all absorbers at $|\Delta v| \leq 5\,\sigma_{\rm grp}$ (roughly $\pm2500$~\kms) from the group redshift, an additional dozen absorbers were found; however, all but three of these are metal-free and not relevant to these concerns. Of the remaining three, two show absorption in \HI\ and \OVI, and are located only $0.83\,\Rvir$ from the same $0.5\,L^*$ galaxy, an individual galaxy association by our earlier discussion. The third absorber is $2.3\,\Rvir$ from a $2\,L^*$ galaxy near group~19670, so this absorber could be associated with a lower luminosity galaxy not observed in our program, or it could be associated with another group of galaxies to which this nearest galaxy is a member, or it has been ejected from group~19670 at $\sim1800$~\kms, or $1.5\,v_{\rm esc}$. Even if the latter is correct, only 2 of 13 metal-bearing absorbers in these groups could have relative radial velocities exceeding the escape velocity for the groups they probe, thus any group mass lost by this process to the IGM is small.

\begin{figure}
  \epsscale{1.2}
  \centering\plotone{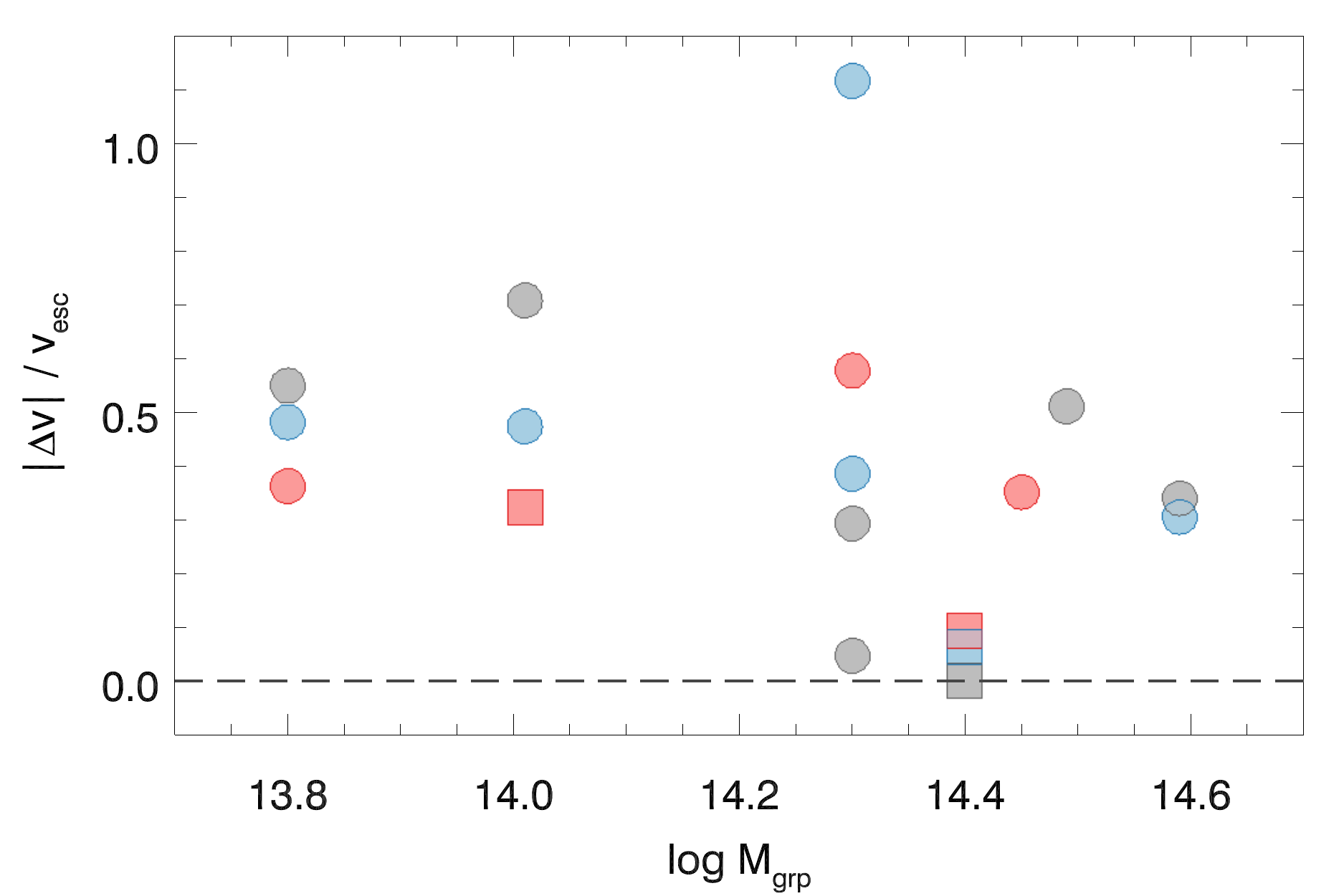}
  \caption{A plot of the calculated group mass ($M_{\rm grp}$) versus the  absorber-group radial velocity difference, where the velocity difference is in units of the escape velocity ($v_{\rm esc}$) at the radius of the impact parameter.
  \label{fig:dv_M}}
\end{figure}

Reinforcing the results shown in \autoref{fig:dv_rho} and \ref{fig:dv_norm} is \autoref{fig:dv_M}, which shows that there is little trend between the normalized absorber-group velocity difference and the group mass. Except for the one absorber mentioned above which exceeds the escape speed in radial velocity difference, there is little evidence that these absorption systems will escape from their parent groups. 

For galaxy groups, \autoref{fig:dv_norm} shows that most absorbers may \textbf{not escape} from the group potential unless their direction of motion is mostly tangential on the sky or inward. This result can have significant implications for galaxy evolution; e.g., these groups are close to being ``closed boxes'' for galactic evolution in the current epoch. While some gas can be accreted onto the group from the intergalactic medium, any metals currently produced by group galaxies may not leave the group. Any baryons and metals which left the region of this group must have done so at much earlier cosmic times.

The BAHAMAS cosmological hydrodynamic simulations \citep{mccarthy17}, which have been tuned to observational X-ray and Sunyaev-Zeldovich Effect constraints for groups and clusters, predict that about 1/3 of the baryons associated with a $10^{13.8}~M_{\odot}$ halo mass are ejected beyond the virial radius, while a $10^{14.6}~M_{\odot}$ halo should retain almost all of its bayons and be considered essentially baryonically closed. Almost all of the groups here are at estimated halo masses in excess of $10^{14}~M_{\odot}$, in the regime where the BAHAMAS simulations suggest that most of the baryons should be retained by the group suggesting that a hot, massive IGrM is present in these groups. 

\subsection{Gas versus Group Properties}
\label{sec:disc:temp}

In our initial study of gas in galaxy groups (Paper~2) we used previously detected \HI\ plus \OVI\ absorbers from Paper~1 as markers to search for groups at the absorber redshift. Groups of varying richness (total group luminosities ranging from a few $L^*$ to nearly $100\,L^*$) were found near these absorbers. In that study, weak correlations were seen between gas and group properties, which supported the identification of those absorbers with the groups distinct from gas associated with individual galaxies in the groups.

Herein, the selection criteria excluded sight lines near ($\rho \leq 1.5\,\Rvir$) bright galaxies in these groups, strongly supporting a group association for these absorbers. Still, many (9) absorbers were found within these groups with low ions like \SiIII\ typical of cool, photo-ionized CGM absorbers. Only two of these apparently photo-ionized absorbers are at $\rho < 1.5\,\Rvir$ from the nearest galaxy; seven others are at larger distances including three at $\rho> 3\,\Rvir$ away.

While it is possible that there are sub-$L^*$ galaxies closer to some of these absorbers that were not observed by our galaxy survey \citep{keeney18}, more likely these metal-bearing clouds have escaped their home galaxy's potential well and are now bound within the confines of their home galaxy group. The cool clouds in these groups that are well-away from galaxies could be CGM clouds that have escaped their home galaxy as predicted by \citet{stocke13} and \citet{keeney17} based on CGM cloud kinematics. 

\begin{figure}
  \epsscale{1.2}
  \centering\plotone{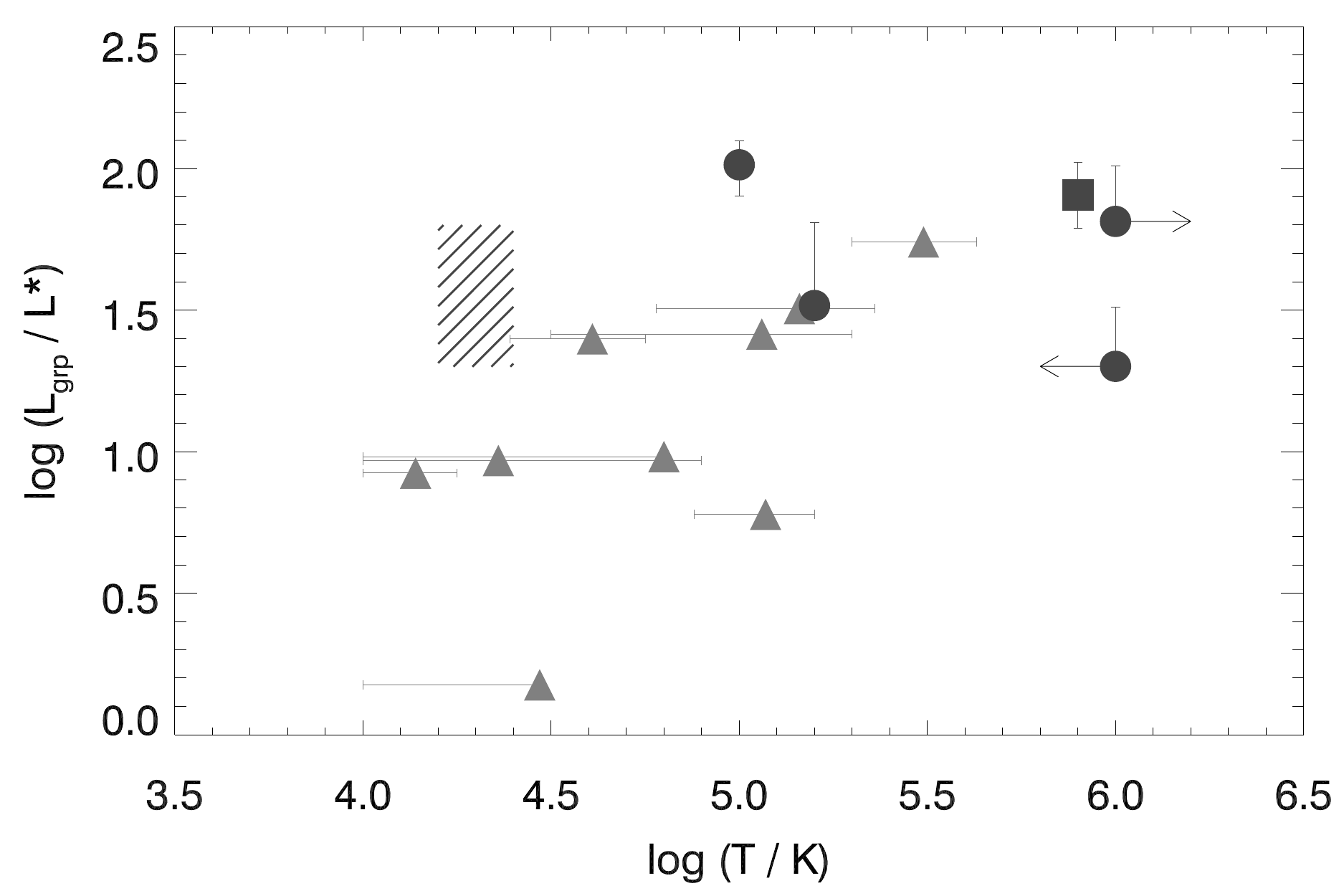}
  \caption{Group absorber temperatures versus total group luminosity, reproduced from Figure~10 of Paper~2. The gray points are from Paper~2, while the black points are warm absorbers from this study. The current sample temperatures are all quite uncertain and should be viewed as estimates. The hashed area is the approximate location for four photo-ionized group absorbers from this study. With the addition of the new data there is no obvious correlation found between these parameters.
  \label{fig:Tlum}}
\end{figure}

\begin{figure}
  \epsscale{1.2}
  \centering\plotone{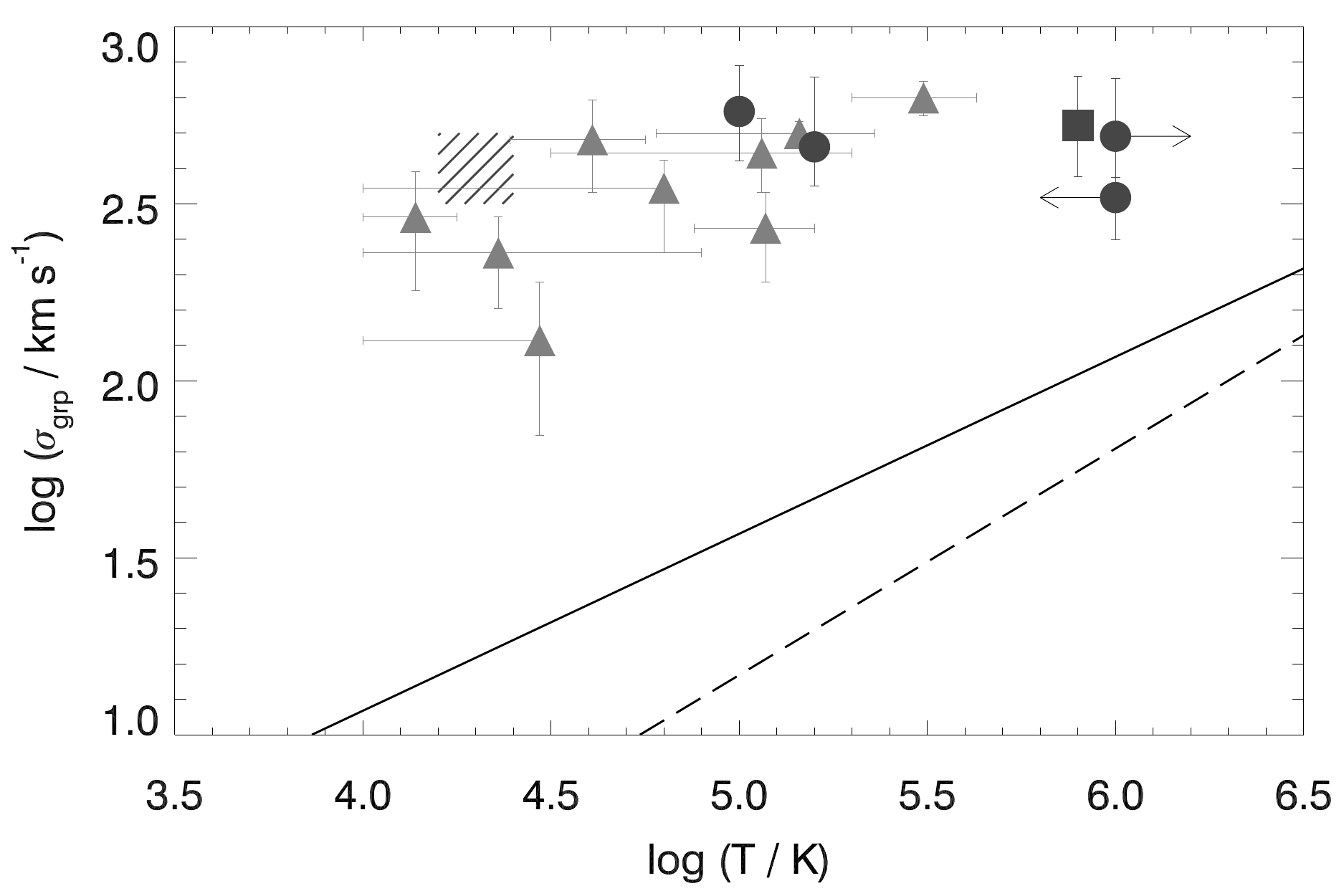}
  \caption{Group absorber temperatures vs. group velocity dispersion, reproduced from Figure~11 of Paper~2. The gray points are from Paper~2, while the black points are warm absorbers from this study. The current sample temperatures are all quite uncertain and should be viewed as estimates. The hashed area is the approximate location for four photo-ionized group absorbers from this study. With the addition of the new data there is no obvious correlation found between these parameters. The solid line is the extrapolation of $\beta_{\rm spec}= 1$ (energy equipartition between galaxies and gas) from the clusters and groups data found in \citet[]{osmond04}; see their Figure~16. The dashed line is an extrapolation of the best-fit power law for clusters and groups from \citet{wu99}. Notice that the points and limits for the warm absorbers in groups are far above either of these extrapolations from richer galaxy systems, illustrating that this gas is too cool by one order of magnitude or more from the expected volume-filling intra-group gas conditions.
  \label{fig:Tsig}}
\end{figure}

Two plots from Paper~2 suggested weak correlations between absorber temperature and total group luminosity (\autoref{fig:Tlum}) and between absorber temperature and group velocity dispersion (\autoref{fig:Tsig}). These are reproduced here with the addition of a few points from this study selected in the same way; i.e., \OVI\ absorbers in groups with either warm or cool absorbers depending on \HI, \OVI, and \SiIII\ properties (see \autoref{sec:cos:BLAs}). \edit1{For these Figures we have used the possible warm absorber temperatures derived in \autoref{sec:cos:BLAs} to provide the maximum possibility for showing correlations in these plots.} These Figures show that any correlation between these quantities is very weak or nonexistent. In these two figures, five points from the current study are added at $T> 10^5$~K (group~32123 has two possible warm absorbers; the one with a measured value is plotted, not the upper limit). The cross-hatched region shows the range of cool absorber properties where $T \approx 20,000$~K is assumed. \edit1{If no warm absorbers are present in this sample, as appears likely based on the results presented in \autoref{sec:cos:BLAs}, the possibility for correlations in these two plots is diminished further.}

While the groups studied here are comparable to only the most luminous and massive groups found by Paper~2, these groups possess gas at the full range of temperatures from 20,000 K to $>10^6$~K, diminishing the already marginal correlations found by the earlier work. Further, the absence of correlations in \autoref{fig:Tlum} and \autoref{fig:Tsig} show that the gas temperature is not directly related to the mass of the group (i.e., its potential well depth). Nor are the absorbers detected at velocities close to the group centroid velocities. These are other indicators that the UV-detected gas is not volume filling but is, instead, that gas most easily detectable through UV absorption. 

Since the possibly warm clouds have temperatures too hot to be photo-ionized by the UV background, either these absorbers are heated by interaction with a hotter medium or are condensing out of it. In either case, a hot intra-group medium is suggested to be present that could be connected more intimately to the gravitational potential well depth. In richer systems like clusters and elliptical-dominated groups, this hotter gas is detected in X-ray bremsstrahlung emission.

Based on extrapolations from the richer systems shown in \autoref{fig:Tsig}, the groups studied here are expected to have an intra-group medium  temperature of $\log{(T/\mathrm{K})} = 6$-7, much hotter than the warm clouds we detect as BLAs and broad \OVI\ absorbers. The solid line extrapolation assumes energy equipartition between gas and galaxies \citep[i.e., $\beta_{\rm spec} = 1$;][]{osmond04}, while the dashed line is the best-fit power-law extrapolation based on the \citet{wu99} compilation of clusters and groups. Our data points in \autoref{fig:Tsig} are clearly too cool to fall near these extrapolations, nor do the current data possess the slope appropriate for an intra-group medium based on these extrapolations.

\section{Summary and Conclusions}
\label{sec:conc}

This study uses UV absorption lines detected in COS/ G130M spectra of 10~AGN targets that probe 12 galaxy groups selected from a homogeneous catalog culled from the SDSS spectroscopic survey in the range $z_0=0.10$-0.19, which facilitates the detection of both the \OVI\ doublet and \Lya\ in the single G130M exposures allocated. These sight lines occur at group-centered impact parameters ranging from 0.3-2 group virial radii with absorber-group radial velocity differences as large as $\pm1000$~\kms.

In order to ensure that these spectra probe gas associated with the foreground groups distinct from individual galaxies within the groups, an important sample selection criterion is that no sight line passes within 1.5 virial radii of a bright SDSS group galaxy. Earlier work \citep{keeney17} showed that an association between an absorber and a single galaxy is plausible only if the impact parameter $\rho \leq 1.4\,\Rvir$. But this selection criterion also has the effect of minimizing the number of sight lines close to group centers where the galaxy density is highest, minimizing sample statistics close to group centers.

Excepting for SDSS J1540$-$0205 (where a Lyman limit system at $z=0.33$ obscures the \OVI\ region in group~16803), $\mathrm{S/N} = 15$-30 was achieved for these targets in the wavelength regions of \Lya\ and the \OVI\ doublet at the group redshift. Absorption associated with these groups was detected for 7 of the 12 groups probed, including group~16803 where only \lya\ was detected. Metal lines, usually the \OVI\ doublet and/or \SiIII\ 1206~\AA, are seen in 12 of the 18 absorber components detected. See \autoref{fig:abs} for the COS spectra; all line parameters are listed in \autoref{tab:abs}.

These \hst/COS observations were supported by a deep and wide galaxy redshift survey around each of these sight lines (see \citealp{keeney18} for details of these observations and subsequent spectral analysis). While the SDSS spectroscopic survey provides complete galaxy redshift information down to $L \gtrsim L^*$, the MOS obtained by \citet{keeney18} extends to $L>0.2\,L^*$ at a completeness level over a $20\arcmin$-radius field-of-view of $\gtrsim60$\%. \edit1{Completeness levels for this galaxy redshift survey are substantially better ($\gtrsim75$\%) both for faint galaxies ($g \leq 20$) close ($\lesssim700$~kpc) to the sight line and for luminous group galaxies out to large radii ($<2.5$~Mpc).} 

These MOS observations were designed to obtain $\geq20$ group members so that each group could be correctly characterized from the galaxy data. This observing goal was accomplished in all but three cases (groups~36001, 49980 and 44726), which have few member galaxies in toto, not because they were poorly observed. The galaxy positions and redshifts provide ample data to characterize these groups using a unique Monte-Carlo approach that is described in detail in \autoref{sec:groups}. All MOS galaxy redshifts used in this study can be found at MAST via \dataset[doi:10.17909/T9XH52]{http://dx.doi.org/10.17909/T9XH52}\footnote{\url{https://archive.stsci.edu/prepds/igm-gal/}}.

The resulting group parameters from the Monte Carlo process are listed in \autoref{tab:params} and \autoref{tab:groups}, and are generally very similar to the parameters found by the algorithm of Paper~3 using the SDSS data alone \autoref{tab:berlind}. These groups have varying richness with total luminosities $L=10$-$130\,L^*$, virial radii of $\Rgrp = 680$-1540~Mpc, velocity dispersions of $\sigma_{\rm grp}=270$-630~\kms, and estimated halo masses of $\log{(M_{\rm grp}/M_{\odot})} = 13.5$-14.5. At these masses, the groups in this sample have a mixture of early- and late-type galaxies and are more massive than most nearby groups in the Local Supercluster, including the Local Group. Additionally, this sample spans the range from almost fully virialized groups doimnated by early-type, passive galaxies to partially-virialized, spiral rich groups \edit1{as determined by new analysis tools developed for this study and described in \autoref{sec:groups:algorithm}-\ref{sec:groups:properties}}. These groups are similar to the groups in which Paper~2 found warm absorbers.

Based on the information derived from these two datasets, the basic results from this study include:
\begin{enumerate}
\item The covering fraction of \Lya\ absorption due to these groups is $\gtrsim50$\% at impact parameters $\rho=0.4$-$2\,\Rgrp$ \edit1{(see \autoref{tab:cf} in \autoref{sec:disc:fcov} showing the small number statistics for this result)}; $\Rgrp\approx1$~Mpc for the masses of these groups. Additionally, there is no obvious trend between the observed $N_{\rm H\,I}$  and impact parameter out to $\approx2$~Mpc. This is quite different from the rapidly declining CGM $N_{\rm H\,I}$ around individual star-forming galaxies. While large covering fractions over such a huge area around these groups is consistent with the number density of warm absorbers ($d\mathcal{N}/dz = 4$ per unit redshift; Paper~2), high covering fractions do not lead necessarily to large volume-filling factors and very large masses of cool and/or warm gas (see below). \edit1{However, unlike the results found by Paper 2, in this sample there are few demonstrably warm absorbers suggesting that warm gas specifically could have a rather small covering fraction in massive galaxy groups.} Given the unstable condition of warm, collisionally-ionized gas, it is likely that this gas exists in filaments that fill only a few percent of the group's volume and account for only a few percent of the total baryon budget of the group.  

\item While the \HI\ covering fraction of these groups is high over a very large surface area, it is possible that some few massive groups always possess detectable warm and/or cool \HI\ absorption and some do not. In this sample, the three \HI\ non-detections are all quite massive, highly-virialized groups with spiral galaxy fractions of $\sim40$\% (see discussion in \autoref{sec:disc:fcov}). This result leads to the prediction that lower mass (approximately one order of magnitude less massive that those studied here), spiral-rich groups will have \HI\ \lya\ detections at close to 100\% covering fraction out to a couple of virial radii in impact parameter. Since these lower halo mass groups, which are similar to the Local Group in mass and kinematics, are the sites where most of the current-epoch star formation occurs, probing these groups for UV absorption is vital to understand the full context of star formation in the local Universe.
  
\item Different from the circumstance for single-galaxy CGMs, the absorbers associated with groups do \textbf{not} possess sufficient relative radial velocities to escape from these groups; i.e., $|\Delta v|/v_{\rm esc} \leq 0.6$ except for one absorber with $|\Delta v| = 1.1\,v_{\rm esc}$ (see \autoref{sec:disc:kinematics}). Since seven \HI\ and low-metal-ion (usually \SiIII) absorbers are found in these groups well away from individual galaxies, cool CGM clouds do escape from their home galaxy \citep[as suggested by CGM cloud radial velocities close to $v_{\rm esc}$; see][]{stocke13,keeney17} but do not escape from their home galaxy group. This has the important consequence for massive galaxy groups in the mass range studied here that little or no gas escapes from galaxy groups into the IGM in the current epoch. If baryons and metals have left the region of this group, that process must have occurred at earlier cosmic times. 

\item There are several observational results from this study which strongly suggest that the cool and/or warm gas detected in groups is neither volume filling nor massive.  These observational results include: (1) the cool or warm gas detected is not close to either the sky centroid of these groups nor to their radial velocity centroid (see \autoref{fig:dv_rho}); (2) the temperature of this gas does not correlate with two different measurements of group halo mass, its total luminosity (see \autoref{fig:Tlum}) and the velocity dispersion of its galaxies (see \autoref{fig:Tsig}); and (3) the warm gas temperatures, both those possible detections found herein and those found in \citet{stocke14}, are too cool by on order of magnitude to be a volume-filling IGrM based on extrapolations from richer galaxy systems (see \autoref{fig:Tsig}). 

\end{enumerate}

While plentiful warm and cool gas has been found associated with rich groups of galaxies, a massive IGrM that completes the baryon inventory in star-forming and passive galaxies has yet to be discovered. Given its predicted temperature range of $10^6$ to $10^7$~K, new soft X-ray instrumentation and telescopes will be required to discover it. Based on assuming pressure balance with individual CGM clouds at $P/k \approx 10~\mathrm{cm^{-3}\,K}$, an IGrM at these temperatures has physical densities of $n=10^{-5}$ to $10^{-6}~\mathrm{cm^{-3}}$, making its detection challenging \citep[although see][and M. Donahue et~al. 2019, in prep, for tentative detections]{nicastro18}.

With little evidence for gas escaping from these groups, systems with total halo masses of $10^{13.5}$ to $10^{14.5}~M_{\odot}$ should be considered ``closed boxes'' in the current epoch for galactic evolution and the baryon census. Since the currently detected baryons are insufficient to bring these groups up to the
cosmic ratio of baryons to dark matter, a massive IGrM is likely present. Indeed, the recent BAHAMAS numerical simulations predict that most of the groups studied here, specifically the most massive ones, have retained most of their baryons. If this is the case, a massive IGrM should be present in these groups to close the baryon accounting. Efforts should be made to extend these UV results downward in halo mass to systems the size of the Local Group to determine if these smaller groups show similar trends; e.g., presence of both cool and warm gas and few absorbers escaping. These lower mass groups are important because they are the sites where most of the star formation in the current Universe is taking place. Since different galaxy groups can have differing evolutionary and thus metallicity histories (e.g., only some have had major mergers early-on), the galaxy metallicity pattern in groups can vary. This may be observable as differences in the mass-metallicity relationship in different groups, contributing to the width observable in that relationship when galaxies from many groups are included (factor of $\sim4$ [95\% contour] in metallicity at fixed mass; \citealp{tremonti04}).

This study also clarifies that theoretically modeling the CGM of star-forming and passive galaxies must account for the IGrM of the groups in which the absorbers and galaxies are found. Specifically this means that the halo masses investigated in numerical simulations must concentrate on group-size halos ($10^{12.5}$ to $10^{14.5}~M_{\odot}$) to both fully account for the all the baryons in these systems and to predict CGM and IGrM properties successfully.

\acknowledgments
This research was supported by NASA/\hst\ grants 14277 (PI: JTS) and 14308 (PI: BDO). Special thanks to Alex Parker for stimulating discussions. Access to the MMTO through the University of Arizona is gratefully acknowledged.

\appendix
\section{Notes on Absorption toward Individual Sight Lines}
\label{sec:abs}

\autoref{tab:abs} lists the species, redshift, velocity offset with respect to the SDSS group redshift, significance level, equivalent width, Doppler $b$-value, and column density of all absorbers detected within $\pm1000~\kms$ of the SDSS group redshifts. \autoref{fig:abs} shows the portions of the COS/G130M spectra over approximately the same range. The absorption toward each sight line is discussed below.

\subsection{Group~12833; $z_0=0.14693$, RBS\,711}

A single, narrow \HI\ absorber is seen at $\Delta v_{\rm abs}=+129$~\kms\ with respect to the SDSS group redshift in both \lya\ and \lyb.  A CoG analysis of these lines is poorly constrained, but is consistent with the individual line fits of $\log{N_{\rm H\,I}}=14.6$-15.0 and $b=20$-25~\kms. Despite high S/N data, no \OVI\ 1032 absorption is seen corresponding to the \HI\ lines. A weak \SiIII\ line shows that metals are present in this system, but that the ionization state and temperature are likely low.  This is consistent with the temperature inferred from the narrow Lyman line profiles ($T\la80,000$~K).

\subsection{Group~16803; $z_0=0.14839$, SDSS\,J1540$-$0205}

This sight line probes one of the richest groups in our sample ($N=10$ SDSS galaxies; 56 galaxies identified by additional MOS; see \autoref{tab:groups} in \autoref{sec:groups:members}) at an impact parameter of $0.5\,\Rgrp$. The background AGN was predicted to have a flux of $\sim2\times10^{-15}~\mathrm{erg\,s^{-1}\,cm^{-2}\,\mbox{\AA}^{-1}}$ based on GALEX FUV and NUV fluxes and a detection by Swift/UVOT in the UVW1 channel.  However, the observed flux was only about half of this due to a previously unknown partial Lyman limit system at the AGN redshift ($z_{\rm em}=0.322$). This means that the continuum level blueward of $\sim1220$~\AA\ is only $\sim5\times10^{-17}~\mathrm{erg\,s^{-1}\,cm^{-2}\,\mbox{\AA}^{-1}}$.  The data quality around 1400~\AA\ (the region of \lya\ at the group redshift) is adequate ($\mathrm{S/N}\approx15$ per resel) for the \Lya\ and \SiIII\ regions.  However, $\mathrm{S/N} \sim 2$-3 for the \OVI\ and \Lyb\ regions of interest ($\lambda<1200$~\AA). The only absorption in the redshift range of the group is a single (presumed to be) \lya\ line at 1392~\AA\ ($\Delta v_{\rm abs}=-893$~\kms).  The low data quality at $\lambda<1200$~\AA\ precludes stringent limits to \OVI\ absorption, but no absorption is apparent in either line of the doublet.

\subsection{Group~19670; $z_0=0.13443$, SBS\,0956$+$510}

Strong \HI\ absorption at $z=0.13388$ is seen in \lya\ and \lyb.  A CoG solution gives $\log{N_{\rm H\,I}}=15.7:$ and $b=17:$~\kms, which fits line profiles reasonably well.  A weaker feature at $\lambda=1376.9$~\AA\ is identified as a possible \lya\ line at $\Delta v_{\rm abs}=+1069$~\kms\ ($z=0.13267$, $\log{N_{\rm H\,I}}=13.1\pm0.1$, $b=29\pm7$~\kms) just off the right-hand edge of \autoref{fig:abs}.  No \OVI\ is seen at the redshift of either \HI\ component.  \OVI\ 1032~\AA\ at the SDSS group redshift is blended with \Lyb\ at $z_{\rm abs}=0.1415$, but no \OVI\ 1038~\AA\ absorption is seen at any redshift.

\subsection{Group~25124; $z_0=0.18613$, B\,1612$+$266}

A strong \lya\ absorption line lies very close to the expected group redshift in this sight line ($z_{\rm abs}=0.1853$, $\Delta v_{\rm abs}=-212$~\kms).  Strong, broad \OVI\ 1038~\AA\ absorption is detected ($\log{N_{\rm O\,VI}}=14.49\pm0.05$, $b=54\pm6$~\kms) at a slightly higher redshift (see \autoref{sec:cos:BLAs}), but the stronger \OVI\ 1032~\AA\ line is blended with the \HI\ Ly10 line of a partial LLS at $z_{\rm abs}=0.3306$.  

We analyze the blended line in two ways: first, we use the unblended \HI\ lines of the partial LLS to model the column density and $b$-value of the system; Ly\,6-9 and 11-15 are clearly present in the data and show no obvious blending with other systems.  A curve-of-growth (CoG) analysis gives a solution of $\log{N_{\rm H\,I}}=16.60\pm0.02$, $b=27\pm2$~\kms.

With this information, we model the partial LLS absorption from the data and recover the \OVI\ 1032~\AA\ line, which appears consistent with the weaker 1038~\AA\ line. We check this for consistency with a second method that uses the observed \OVI\ 1038~\AA\ line as a model and subtracts it from the data near the \OVI+Ly10 blend at 1223~\AA, yielding a Ly10 profile consistent with the CoG solution.

In addition to \HI\ \lya\ and \OVI, corresponding absorption is also seen in \SiIII\ 1206.5 and \ion{Si}{2} 989, 1190, 1193 at $z_{\rm abs}=0.1853$.  Corresponding \lyb\ absorption is obscured in the Galactic \lya\ profile and is unrecoverable. This system is discussed  in detail in \autoref{sec:cos:BLAs}.

\clearpage
\input{./tab_abs.tex}
\onecolumngrid

\subsection{Group~32123; $z_0=0.15971$, SDSS\,J1333$+$4518}

Several strong absorbers are present in this sight line at $v_{\rm grp}\pm500$~\kms\ though none at $|\Delta v_{\rm abs}|<200$ \kms.  Strong \Lya\ systems are seen at $z=0.15874$ and $z=0.16081$, bracketing the SDSS group redshift.  The blue system shows two components in \Lyb\ while the red system shows a single \Lyb\ component along with weak \OVI\ absorption in both lines of the doublet.  Both red and blue absorbers show weak \SiIII\ absorption though the red \SiIII\ detection is marginal ($\sim2.8\sigma$). A possible metal-free BLA is in this group at $z$=0.16142 ($\Delta v_{\rm abs}=+442$~\kms).

\subsection{Group~36001; $z_0=0.18788$, SDSS\,J1028$+$2119}

While there is no absorption at the SDSS group redshift, three strong \lya\ lines are found at $\Delta v_{\rm abs}=-1100$ to $-500$~\kms. The first absorber at $z$=0.18429 ($\Delta v_{\rm abs}=-906$~\kms) is outside the relative velocity bounds established for association with this group. There is also weak \OVI\ 1031, 1038~\AA\ absorption associated with the $z$=0.18527 \lya\ absorber. Relative to the SDSS group centroid and virial radius, this sight line has the smallest impact parameter ($\rho_*$=295 kpc) in our sample.

\subsection{Groups~44565, 44564; $z_0=0.14925, 0.14506$, CSO\,1022}

This archival dataset probes two SDSS groups $\sim1200$~\kms\ apart, which were shown to be parts of a single, much more massive group (see \autoref{sec:indiv}). While a low-significance feature ($2.8\sigma$) is consistent with \OVI\ 1032~\AA\ absorption at the redshift of group~44564, it is identified as \OVI\ 1038\AA absorption from the unrelated, strong $z=0.1387$ system (confirmed via \HI\ and other metal lines at $z=0.1387$). No absorption is seen in \HI\ through either \lya\ or \lyb\ near the redshift of either group. The group characterization analysis discussed in \autoref{sec:groups:members} merges these two groups into one, with velocity centroid between the values above and sky position very close to the centroids of the two SDSS groups.

\subsection{Group~44726; $z_0=0.15208$, CSO\,1080}

The long IGM sight line ($z_{\rm em}=0.526$) toward this background AGN provides a complicated absorption-line population against which we must measure absorption from the group. Nevertheless, there is no \HI\ absorption within $|\Delta v_{\rm abs}|<1000$ \kms\ of the SDSS group redshift, though a strong \HI\ system is found at $\Delta v_{\rm abs}=+1300$~\kms. 

\subsection{Groups~44739, 44858; $z_0=0.11839, 0.12740$, FBQS\,J1519$+$2838}

At the redshift of group~44739 ($z_0=0.1184$), four strong \HI\ lines are seen in both \lya\ and \lyb\ between $-800$ and $-100$~\kms\ from the SDSS group redshift. Two marginal \OVI\ 1032 lines are seen which appear to correspond with the $-808$ and $-195$~\kms\ components. The potentially broad \OVI\ 1032~\AA\ detection is weak enough that it's consistent with a non-detection in the 1038~\AA\ line. There is \SiIII\ absorption seen corresponding to three of the \HI\ components as well.

For group~44858 ($z_0=0.1274$), a single, strong \HI\ absorber is best-fit as a pair of \lya\ lines at $\Delta v_{\rm abs}=-501  \& -460$~\kms\ with respect to the SDSS group redshift.  Moderate, narrow \OVI\ absorption is seen in both lines of the doublet at very nearly the $\Delta v_{\rm abs}= -460$~\kms\ velocity. \SiIII\ at this redshift is blended with another intervening system and no other commonly seen low-ionization species are present in these data.



\subsection{Group~50433; $z_0=0.13599$; FBQS\,J1030$+$3102}

High-S/N data at the expected \OVI, \Lya, and \Lyb\ line locations show no absorption features in \HI\ or \OVI\ within $\pm1500$~\kms\ of the group velocity.

\section{Notes on Properties of Individual Groups}
\label{sec:indiv}

\begin{figure*}
\epsscale{0.97}
\centering\plotone{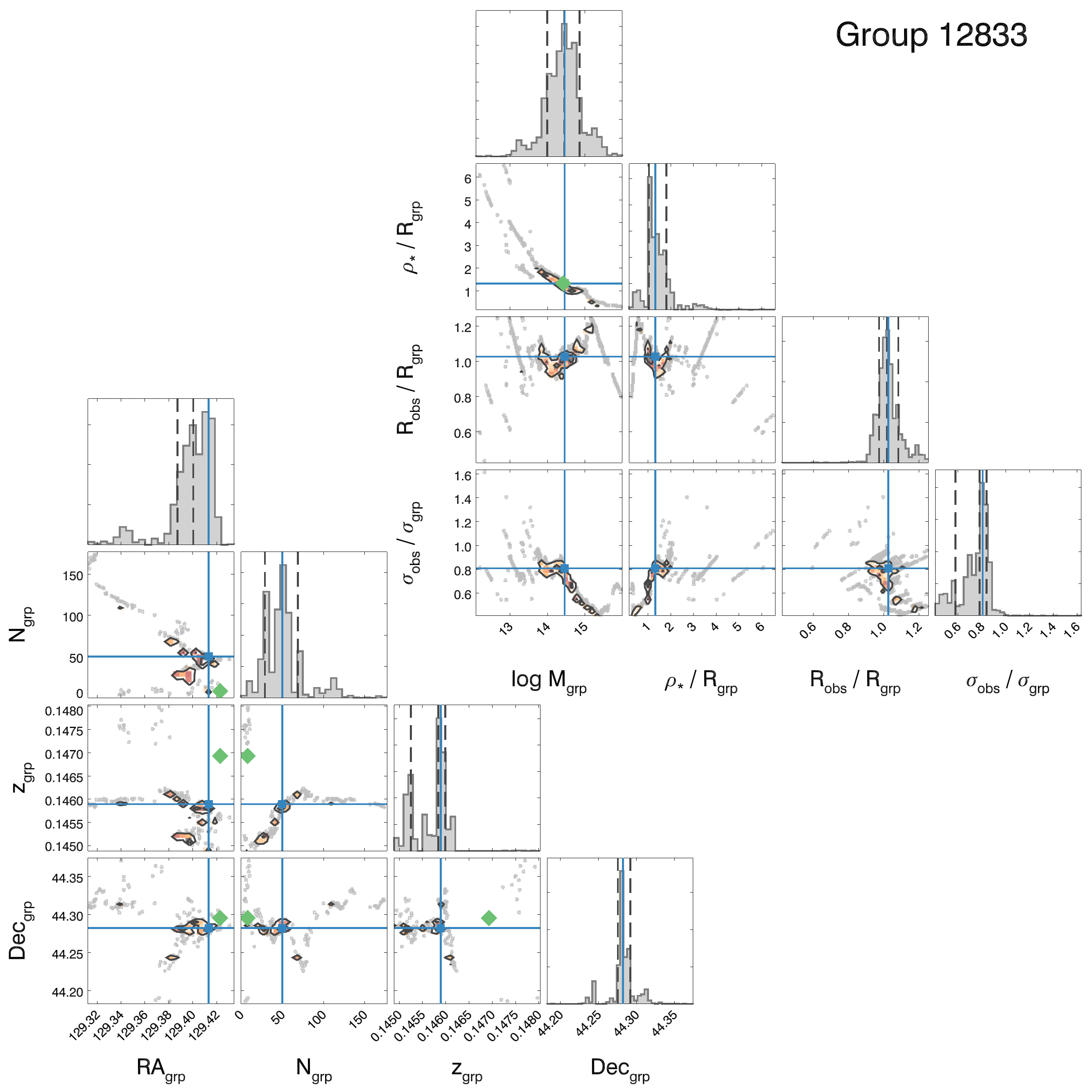}
\caption{Exploration of parameter covariance for group 12833. Contours are drawn at $1\sigma$ and $2\sigma$ significance. Inside the $2\sigma$ contour, data are color-coded according to density, with lighter colors indicating lower density. Outside the $2\sigma$ contour, data from individual Monte Carlo realizations are shown as gray circles. Dashed lines in the histograms show the 16th, 50th, and 84th percentile values (i.e., the ranges listed in \autoref{tab:params}). Solid blue lines show the adopted values from \autoref{tab:groups} (see \autoref{sec:groups:members}). Green diamonds show the values from the SDSS analysis in \autoref{tab:berlind}, where applicable.
\label{fig:corner}}
\end{figure*}

\autoref{fig:corner} examines how the derived properties of the galaxy groups are related. These properties show a high level of discretization for the sparsest groups ($N_{\rm grp} \lesssim 10$), as well as strong correlations with $M_{\rm grp}$ through \autoref{eqn:Rgrp}-\ref{eqn:sgrp}. Furthermore, there are often peaks for poor, low-mass and rich, high-mass configurations for a given group. Nevertheless, the group properties in \autoref{tab:groups} are almost always within the $1\sigma$ range of values derived from the individual Monte Carlo realizations (\autoref{tab:params}), indicating that the adopted group configurations are representative of the ensemble as a whole. This is reassuring but not surprising since we adopt the configuration that most closely matches the marginal distributions of these parameters (\autoref{sec:groups:members}). Below, we discuss the properties of individual groups.

\subsection{Group 12833}
\label{sec:indiv:12833}

The vast majority of the Monte Carlo realizations for this group have $<80$~members (\autoref{tab:params}), but there is a small tail up to $\sim150$~members (\autoref{fig:corner}). The group position is well-constrained in declination, but has two peaks in right ascension; the group redshift also shows two peaks separated by $\sim250~\kms$. The group mass shows a strong peak at $\sim10^{14.5}~M_{\Sun}$ and a weaker one at $\sim10^{13}~M_{\Sun}$, and the impact parameter of the QSO is well constrained at $\rho_\star \approx 1.3\,\Rgrp$. The distributions of $R_{\rm obs}/\Rgrp$ and $\sigma_{\rm obs}/\sigma_{\rm grp}$ are fairly broad, which is a general characteristic for all groups owing to the noisy nature of $R_{\rm obs}$ and $\sigma_{\rm obs}$. The adopted group properties (blue lines) all fall within the $1\sigma$ uncertainty.

\subsection{Group 16803}
\label{sec:indiv:16803}

The properties of this group are all reasonably well constrained (\autoref{fig:16803_corner}), except for the $R_{\rm obs}/\Rgrp$ and $\sigma_{\rm obs}/\sigma_{\rm grp}$ values, whose marginal likelihoods have multiple peaks. The adopted group properties are all near the peaks of the marginal distributions.

\subsection{Group 19670}
\label{sec:indiv:19670}

Most of the derived properties of this group show evidence for multiple peaks (\autoref{fig:19670_corner}), suggesting that a relatively low-mass, $\sim50$-member group and a higher-mass, $\sim90$-member group are both common outcomes. The adopted group properties favor the lower mass configuration, and all fall within the $1\sigma$ uncertainty.

\subsection{Group 25124}
\label{sec:indiv:25124}

This is perhaps the hardest group in our sample to characterize. The most common configurations (\autoref{fig:25124_corner}) are a $\sim30$-member, low-mass ($\sim10^{14.4}~M_{\Sun}$) group, and a $\sim40$-member, high-mass ($\sim10^{14.8}~M_{\Sun}$) group; however, the richness distribution for the high-mass configuration is broad and extends up to $\sim50$~members. The group redshift distribution shows two peaks separated by $\sim300~\mathrm{km\,s}^{-1}$, and the $R_{\rm obs}/\Rgrp$ and $\sigma_{\rm obs}/\sigma_{\rm grp}$ distributions are also multi-modal. Confoundingly, the two group configurations evident in the richness and mass distributions do not correspond to two clear positions on the sky; instead there are $\sim5$ fairly common group positions in the Monte Carlo realizations. Our adopted group membership favors the most common of these positions, which corresponds to one of the low-mass group configurations.

\subsection{Group 32123}
\label{sec:indiv:32123}

The Monte Carlo realizations for this group tend to find $\lesssim20$~members (\autoref{fig:32123_corner}), but there is a tail that extends up to $\sim50$~members. The group position on the sky shows a clearly favored position, but the group redshift is more widely distributed; the adopted group configuration matches the richness and favored position well. The mass and other derived properties also have multiple peaks in their distributions; the adopted group values are all in the high-density regions for these properties.

\subsection{Group 36001}
\label{sec:indiv:36001}

The Monte Carlo realizations all find a modest ($<15$-member) group, with a 6-member configuration being the most common (\autoref{fig:36001_corner}). Due to the small number of galaxies involved, the group position and redshift distributions tend to settle on a handful of discrete values. The mass and other derived properties show smoother distributions due to the uncertainty in $\Upsilon_{\rm grp}$ that is folded into the Monte Carlo process (see \autoref{sec:groups:algorithm}). The adopted group properties match the peaks in all of these quantities.

\subsection{Groups 44564 \& 44565}
\label{sec:indiv:44564_44565}

These two groups are both probed by the same QSO sight line (\autoref{tab:cos}), and the initial SDSS analysis found them to be separated by $1\farcm8$ on the sky and $1100~\kms$ in redshift (\autoref{tab:berlind}). We find many galaxies at intermediate redshifts in this region of the sky, so the two SDSS groups merge into a single, $\sim115$-member group by our analysis. The Monte Carlo realizations find positions and redshifts for the groups separated by $\sim10\arcsec$ on the sky and $\sim5~\kms$ in redshift (\autoref{tab:params}), and all other derived (and adopted) group quantities are also nearly identical. The adopted position of the merged group is closer to the SDSS position of group~44565, but the adopted redshift is closer to the SDSS redshift of group~44564 (\autoref{tab:groups}).

\subsection{Group 44726}
\label{sec:indiv:44726}

The vast majority of the Monte Carlo realizations find a modest group with $<30$~members, but there is a tail that extends all the way up to $>150$~members (\autoref{fig:44726_corner}). This leads to a bimodality in the inferred mass and other derived quantities. Our adopted group properties lie near the peaks of all of the parameter distributions, and prefer a low mass ($\sim10^{13.4}~M_{\Sun}$) group. The adopted 7-member group is also remarkably compact, with $R_{\rm obs} < 200$~kpc (\autoref{tab:groups}).

\subsection{Group 44739}
\label{sec:indiv:44739}

Groups~44739 and 44858 are probed by the same QSO sight line (\autoref{tab:cos}). However, the initial SDSS analysis found them to be much more separated ($30\arcmin$ on the sky and $2400~\mathrm{km\,s}^{-1}$ in redshift; \autoref{tab:berlind}) than Groups~44564 and 44565, so they do not merge into a single, larger group by our analysis. Thus, we continue to discuss their properties separately.

Most of the Monte Carlo realizations find groups with $\la100$~members, but occasionally they will merge this group with Group~44858, which allows the group to grow to several hundred members (\autoref{fig:44739_corner}). Ignoring this low-frequency tail, the group position, richness, and redshift are well-constrained. The inferred mass and other derived group properties are also well-behaved. The adopted group properties are all near the peaks of the marginal distributions.

\subsection{Group 44858}
\label{sec:indiv:44858}

As with Group~44739, most of the Monte Carlo realizations have $<100$~members, but there is a low-frequency tail that extends up to $\sim600$~members when the two groups occasionally merge (\autoref{fig:44858_corner}). Outside of these rare cases the group properties are well-constrained. The adopted group properties match the peaks of the marginal distributions, except for $R_{\rm obs}/\Rgrp$ and $\sigma_{\rm obs}/\sigma_{\rm grp}$, which lie outside the $1\sigma$ uncertainty.



\subsection{Group 50433}
\label{sec:indiv:50433}

The Monte Carlo realizations almost always find $<40$~members in this group (\autoref{fig:50433_corner}), with several discrete positions, richnesses, and redshifts. The distributions of inferred mass and other derived quantities are smooth and well-behaved, and the adopted group properties are near the peaks of all of these quantities.

\clearpage
\section{Supplemental Material}

\textbf{This is not a real section. It includes the full set of Figures and Tables that will be combined into the various Figure Sets and machine-readable Tables above after acceptance. It is included for completeness.}

\begin{figure*}[!h]
  \epsscale{0.48}
  \centering\plotone{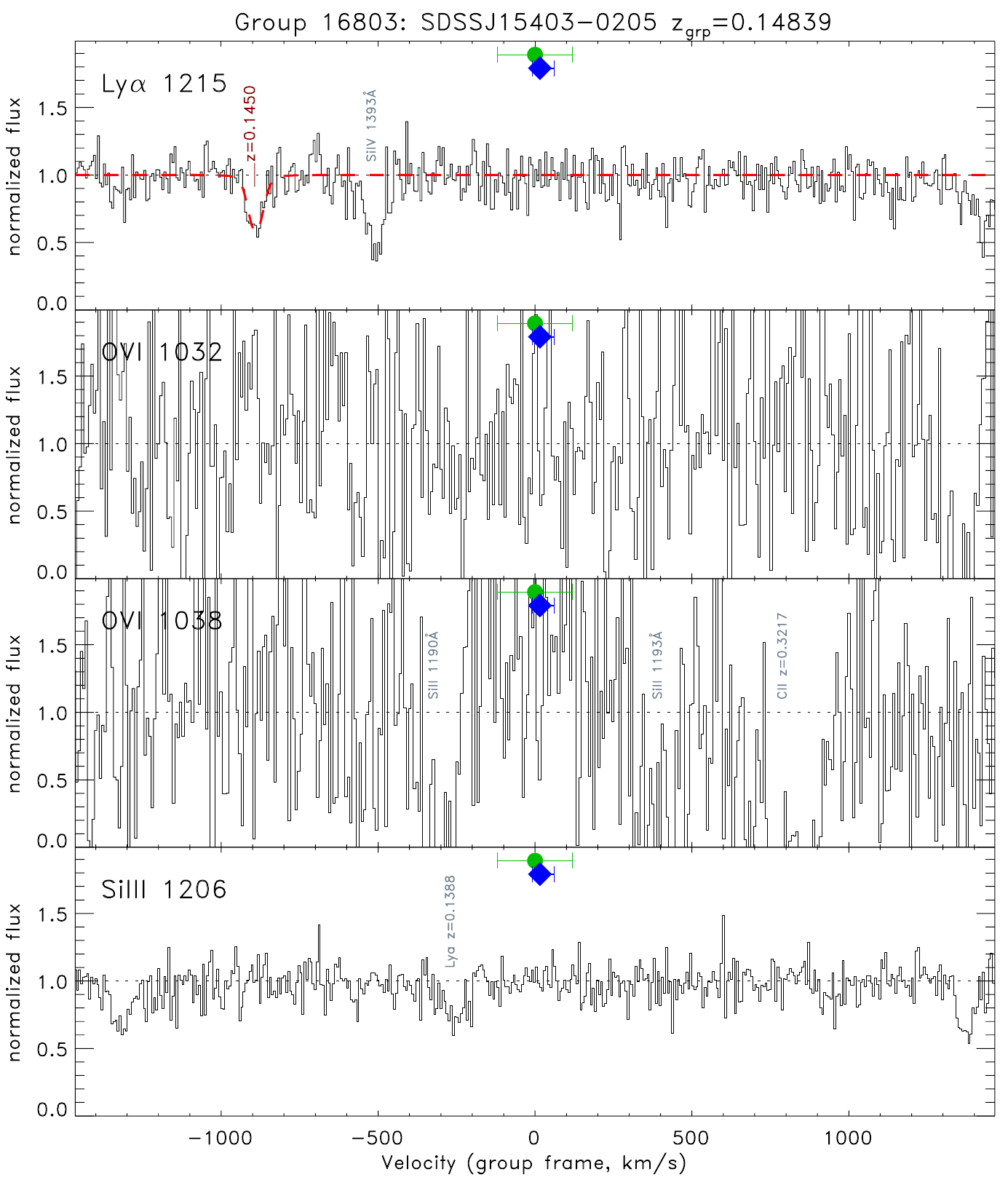}
  \caption{Same as \autoref{fig:abs}, but for group~16803.  A weak \lya\ line is seen, but no O\,VI is detected. The continuum is severely depressed due to a higher redshift LLS at the wavelength where OVI would appear at the group redshift. Therefore, this spectrum has an inconclusive result on the possibility of hot gas in this group.
  \label{fig:16803_abs}}
\end{figure*}

\begin{figure*}[!b]
  \epsscale{0.48}
  \centering\plotone{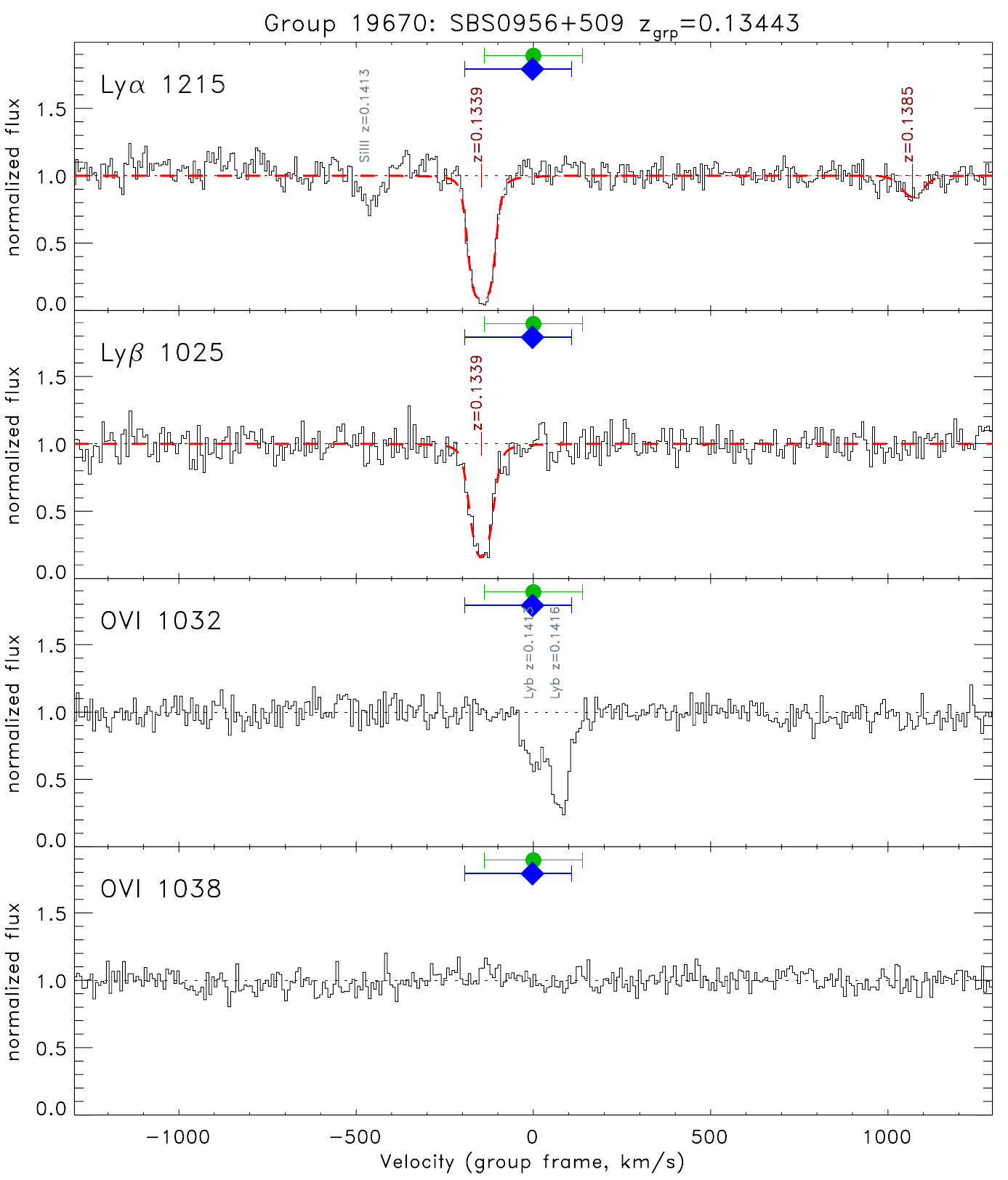}
  \caption{Same as \autoref{fig:abs}, but for group~19670. Strong \HI\ absorption is seen near the group velocity, but no corresponding \OVI\ absorption is seen.
  \label{fig:19670_abs}}
\end{figure*}

\clearpage
\begin{figure*}
  \epsscale{0.5}
  \centering\plotone{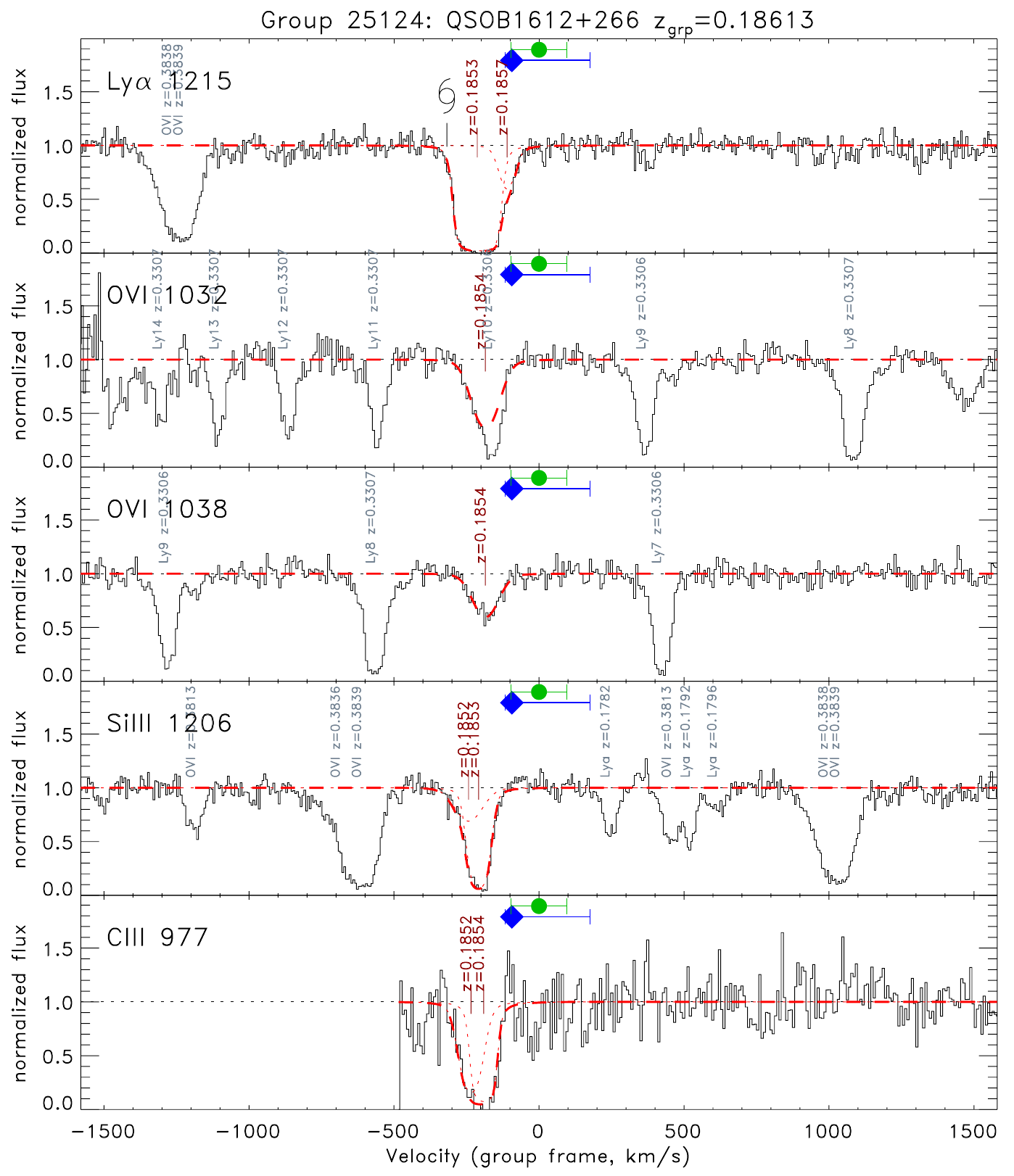}
  \caption{Same as \autoref{fig:abs}, but for group~25124.  Strong \lya\ absorption is seen at $\Delta v_{\rm abs}=-212$~\kms\ along with \OVI\ 1038 absorption. The corresponding \OVI\ 1032~\AA\ line is blended with a partial LLS Ly10 line, and attempts to recover this line using the observed Lyman series of the partial LLS successfully obtain a line strength and width consistent with the 1038~\AA\ detection. Because the \OVI\ lines are significantly offset in velocity from the narrow \HI\ lines in the fit displayed here and listed in \autoref{tab:abs}, the \lya\ complex was refit as discussed in \autoref{sec:cos:BLAs} with the result that a BLA is likely present coincident in velocity with the \OVI\ lines. The galaxy symbol marks the redshift of the nearest galaxy to this absorber 168~kpc ($0.6\,\Rvir$) away, to which these absorptions are likely associated. 
  \label{fig:25124_abs}}
\end{figure*}

\begin{figure}
  \epsscale{0.5}
  \centering\plotone{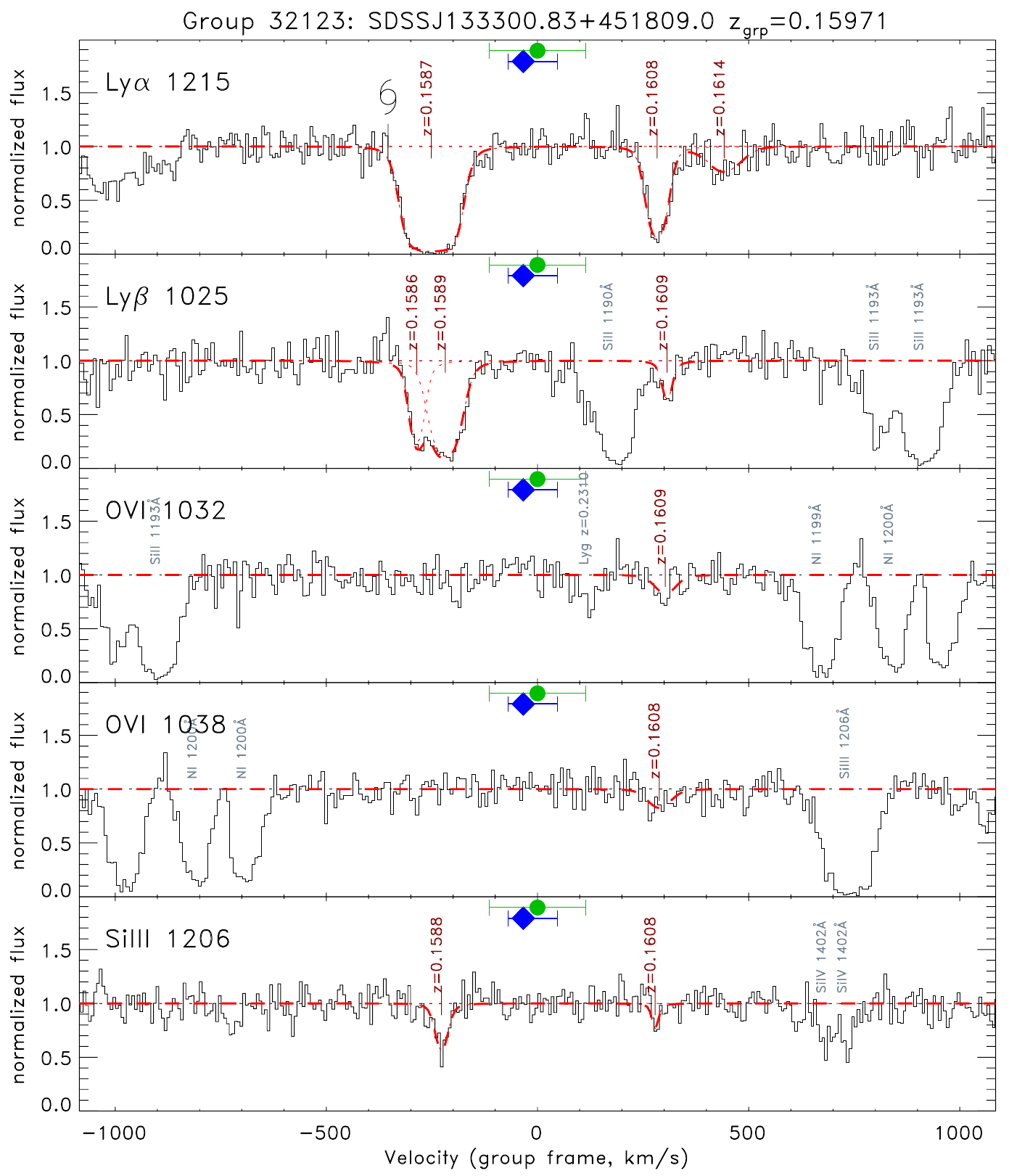}
  \caption{Same as \autoref{fig:abs}, but for group~32123. While no absorption is seen right at the SDSS group redshift, strong \lya\ features are seen at $\Delta v_{\rm abs} = -251$ and +283~\kms. A \edit1{possible,} metal-free BLA is associated with this group at $\Delta v_{\rm abs} = +442$~\kms\ 
  \label{fig:32123_abs}}
\end{figure}

\clearpage
\begin{figure*}
  \epsscale{0.5}
  \centering\plotone{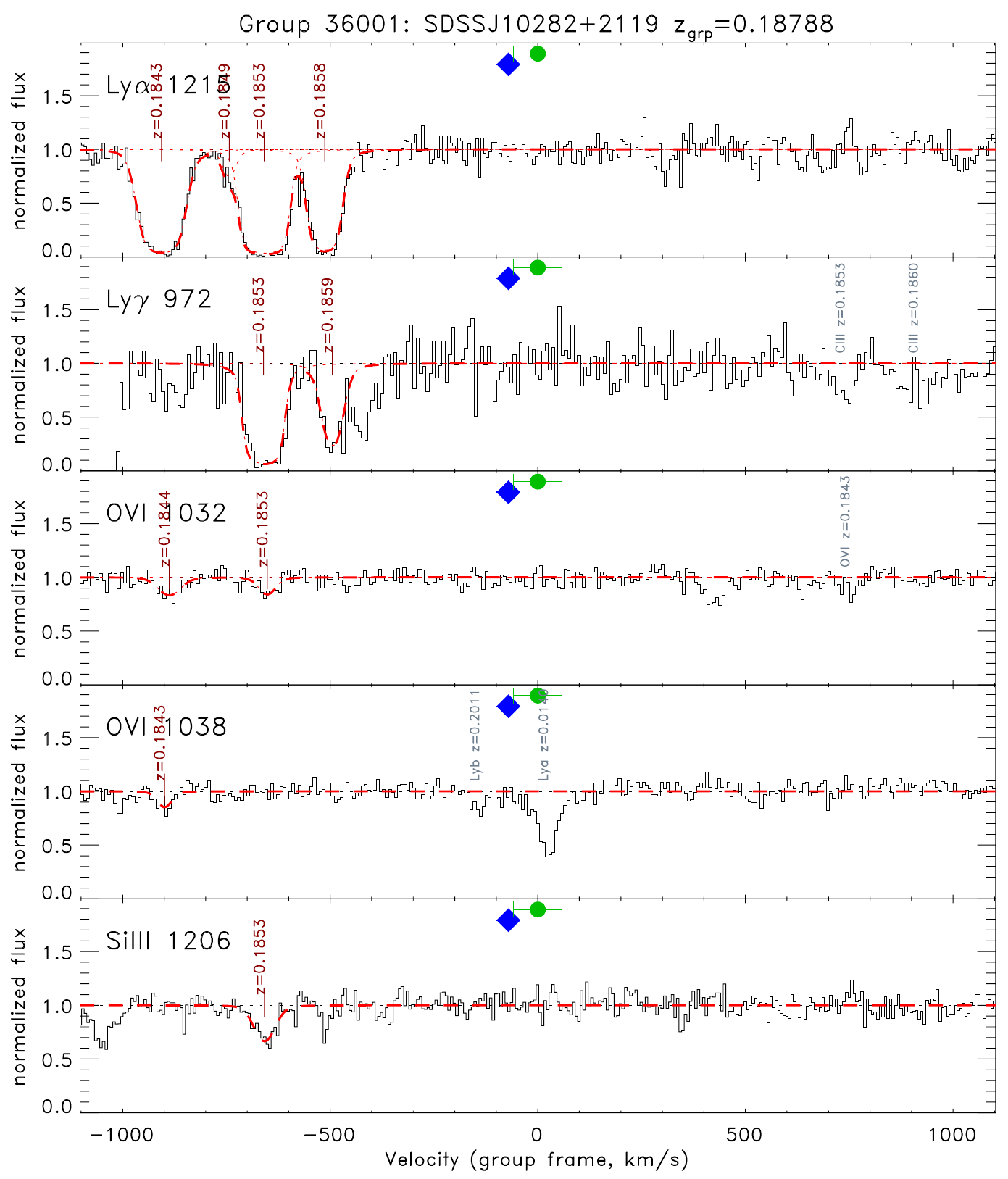}
  \caption{Same as \autoref{fig:abs}, but for group~36001. Weak \OVI\ 1032 and, \CIII\ \& \SiIII\ absorbers are aligned with an \HI\ system at $\Delta v_{\rm abs}=-659$~\kms. Two other \lya\ absorbers are present at higher negative relative velocities; the highest at $\Delta v_{\rm abs}=-906$~\kms is outside the velocity bounds we have considered for associated with this group.
  \label{fig:36001_abs}}
\end{figure*}

\begin{figure*}
  \epsscale{0.5}
  \centering\plotone{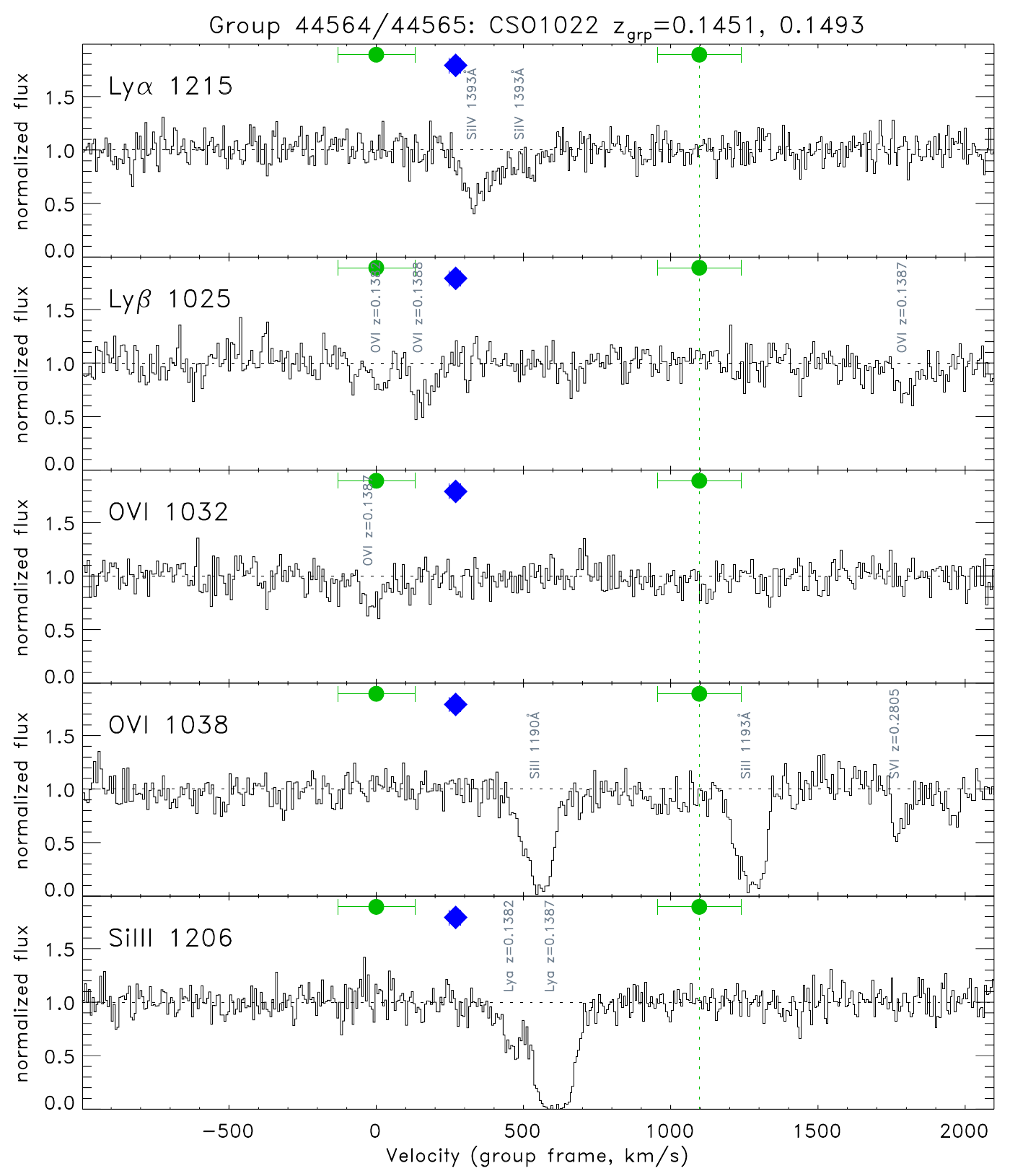}
  \caption{Same as \autoref{fig:abs}, but for groups~44564 and 44565, for which our detailed analysis finds only a single group (see \autoref{tab:groups}). The apparent \Lyb\ and \OVI\ absorption at $z\approx0.1451$ is coincidental and is, in fact, intervening \OVI\ from two lower-redshift systems with corroborating \HI\ \Lya\ and \Lyb\ absorption. A marginal ($<3\sigma$) detection is found consistent with very weak \OVI\ absorption at $z=0.1493$, but this absorber could also be plausibly identified as \HI\ Ly$\gamma$ at $z=0.2195$. No other \OVI\ or \HI\ absorption is seen associated with either group.
  \label{fig:44564_44565_abs}}
\end{figure*}

\clearpage
\begin{figure*}
  \epsscale{0.5}
  \centering\plotone{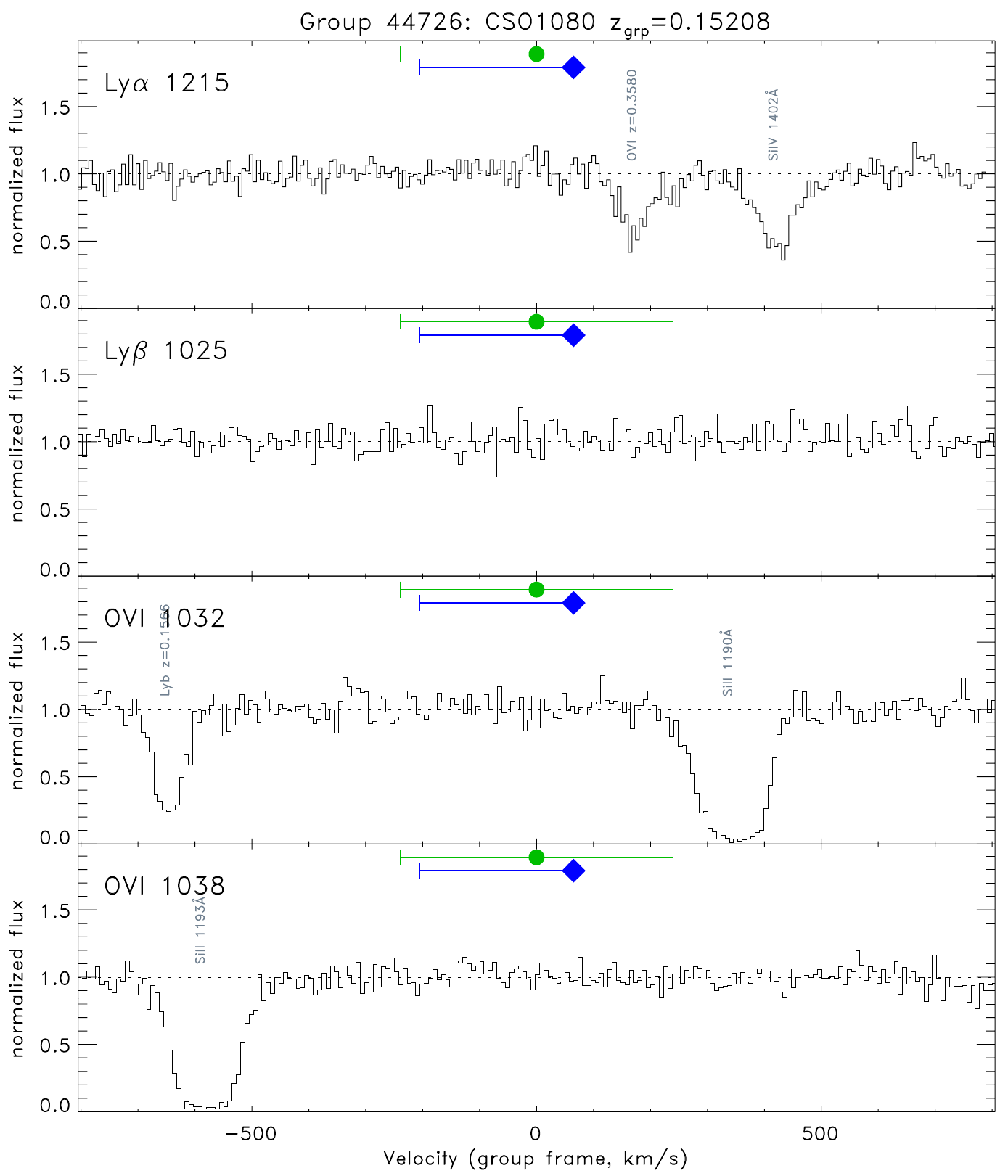}
  \caption{Same as \autoref{fig:abs}, but for group~44726. Strong \HI\ absorption is seen at $\Delta v_{\rm abs}=+1300$~\kms, but no absorption is seen at the SDSS group velocity. Numerous interstellar and unrelated intergalactic lines render the possiblity for blended lines high.
  \label{fig:44726_abs}}
\end{figure*}

\begin{figure*}
  \epsscale{0.5}
  \centering\plotone{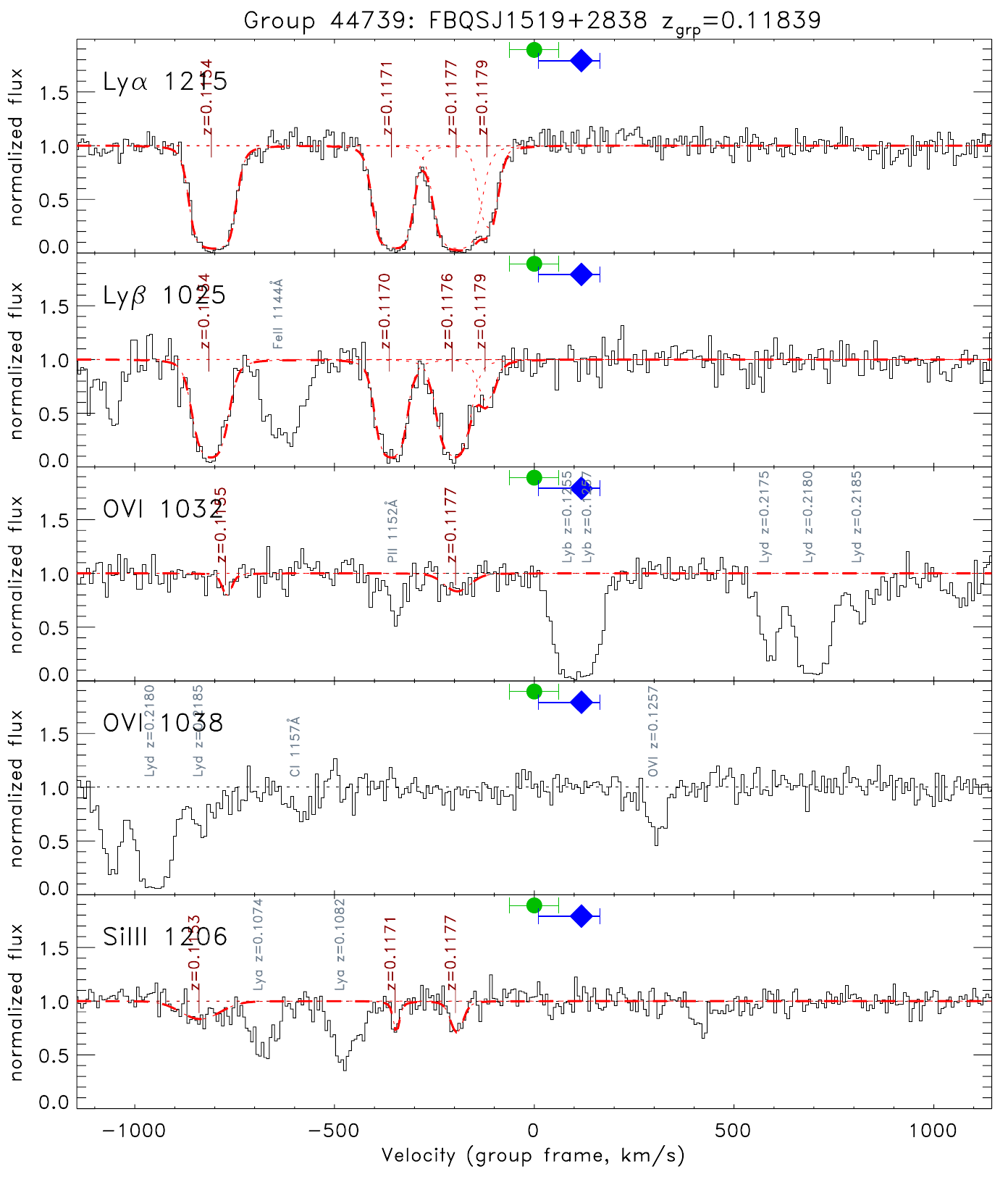}
  \caption{Same as \autoref{fig:abs}, but for group~44739. Strong \HI\ absorption is seen at $\Delta v_{\rm abs}\approx-800$ to $-120$~\kms.  Two weak features are consistent with \OVI\ 1032 absorption corresponding to two of the strong \HI\ components. Weak \SiIII\ absorption is also seen coincident with two \HI\ velocity components.
  \label{fig:44739_abs}}
\end{figure*}

\clearpage
\begin{figure*}
  \epsscale{0.5}
  \centering\plotone{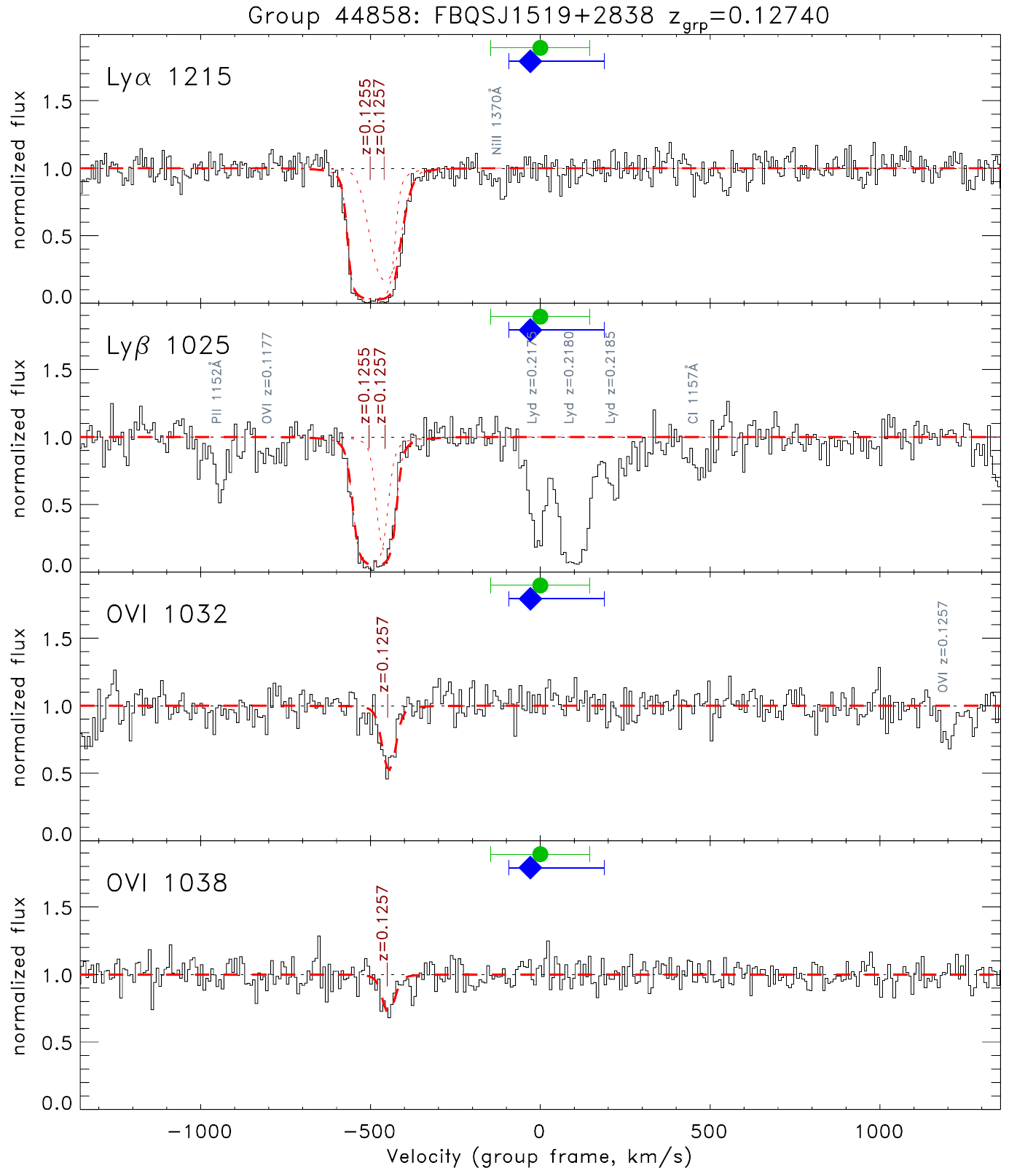}
  \caption{Same as \autoref{fig:abs}, but for group~44858. Strong \HI\ is aligned with moderate \OVI\ absorption at $\Delta v_{\rm abs}=-460$~\kms. A second \HI\ component at $\Delta v_{\rm abs}=-501$~\kms is metal-free.
  \label{fig:44858_abs}}
\end{figure*}


\begin{figure*}
  \epsscale{0.5}
  \centering\plotone{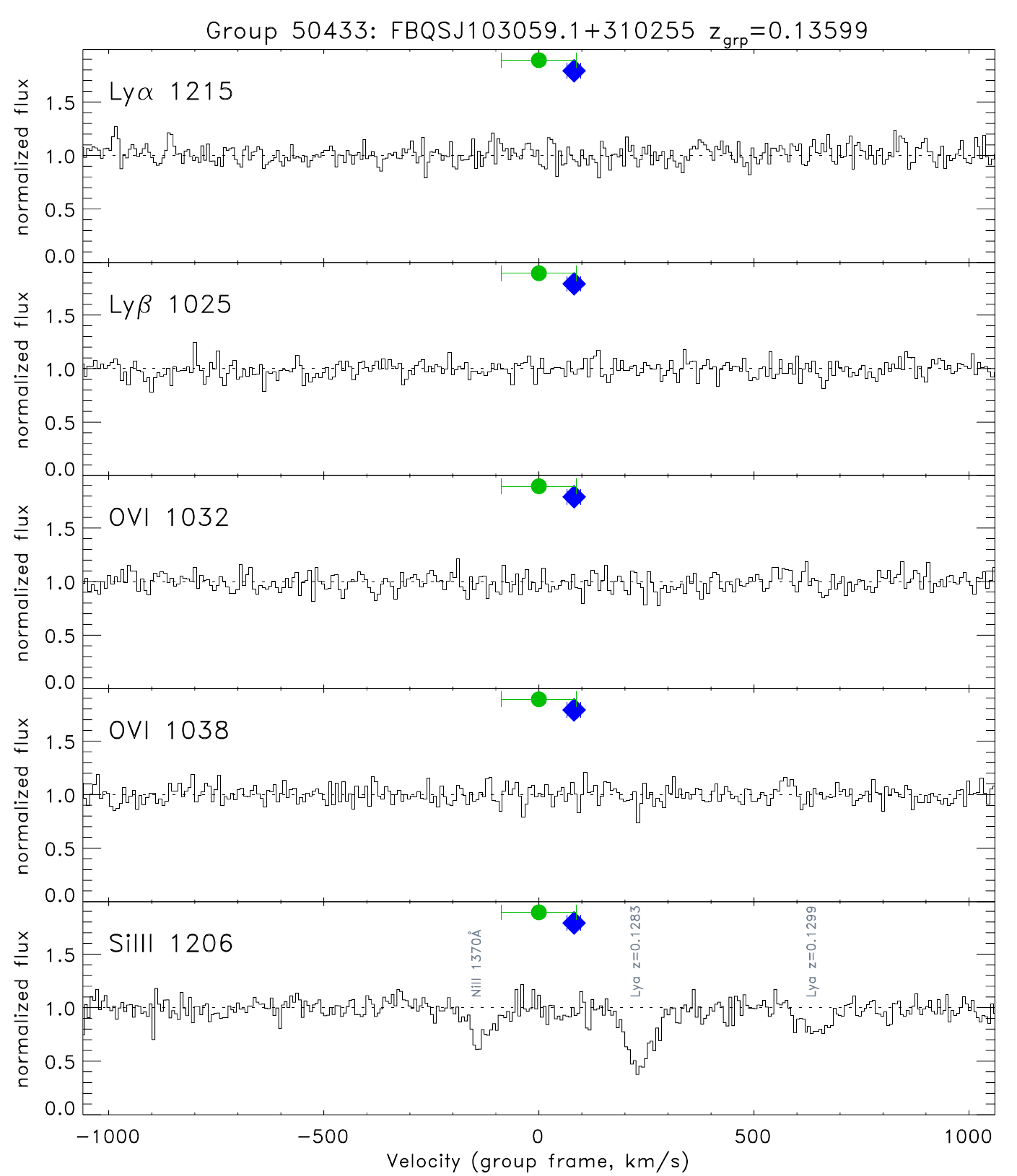}
  \caption{Same as \autoref{fig:abs}, but for group~50433.  No absorption is seen in any species near the group redshift.
  \label{fig:50433_abs}}
\end{figure*}

\clearpage
\begin{figure*}
\epsscale{0.95}
\centering\plotone{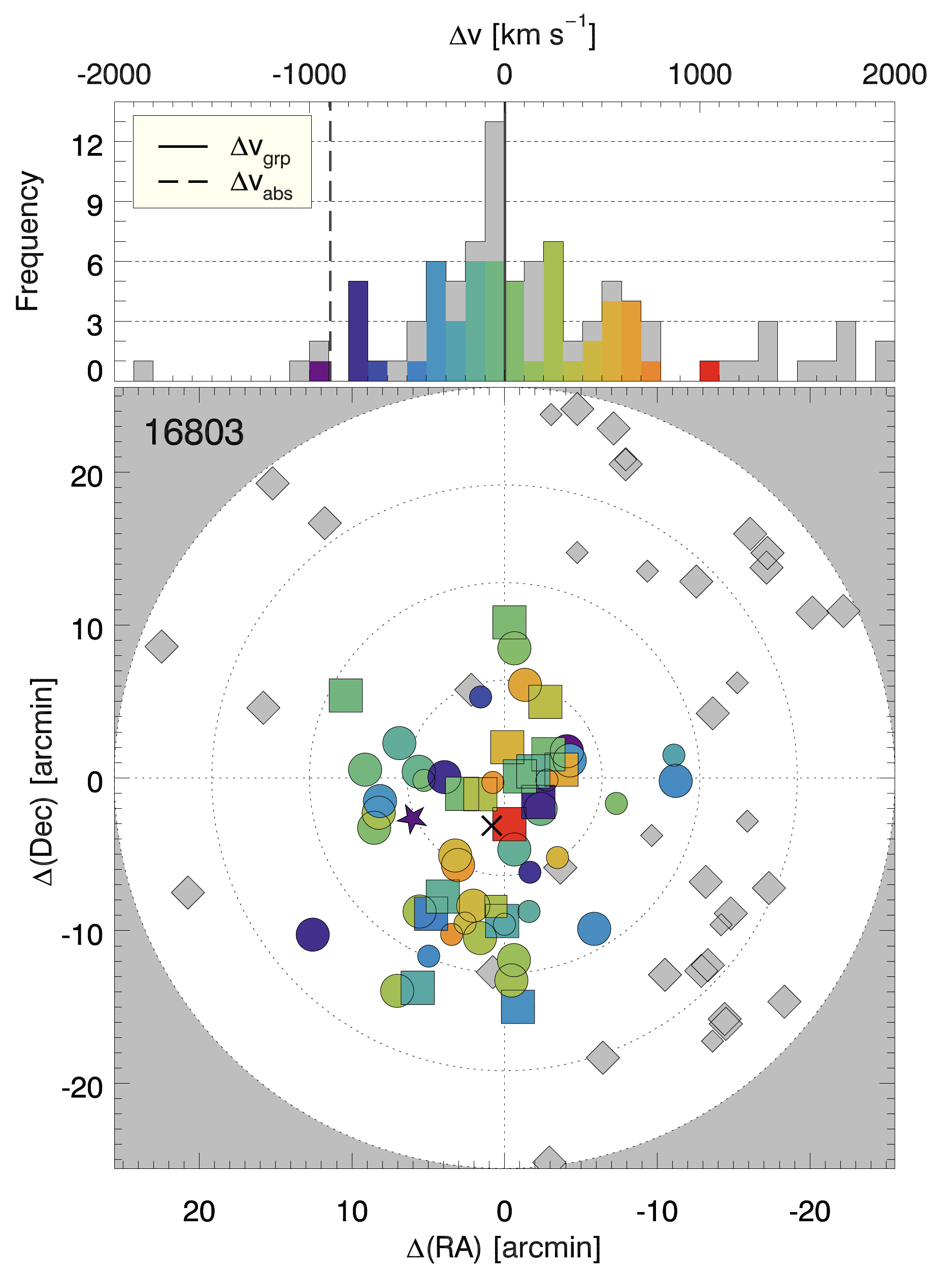}
\caption{Same as \autoref{fig:12833}, but for the galaxy group~16803.
\label{fig:16803}}
\end{figure*}
\clearpage
\input{./tab_16803.tex}

\clearpage
\begin{figure*}
\epsscale{0.95}
\centering\plotone{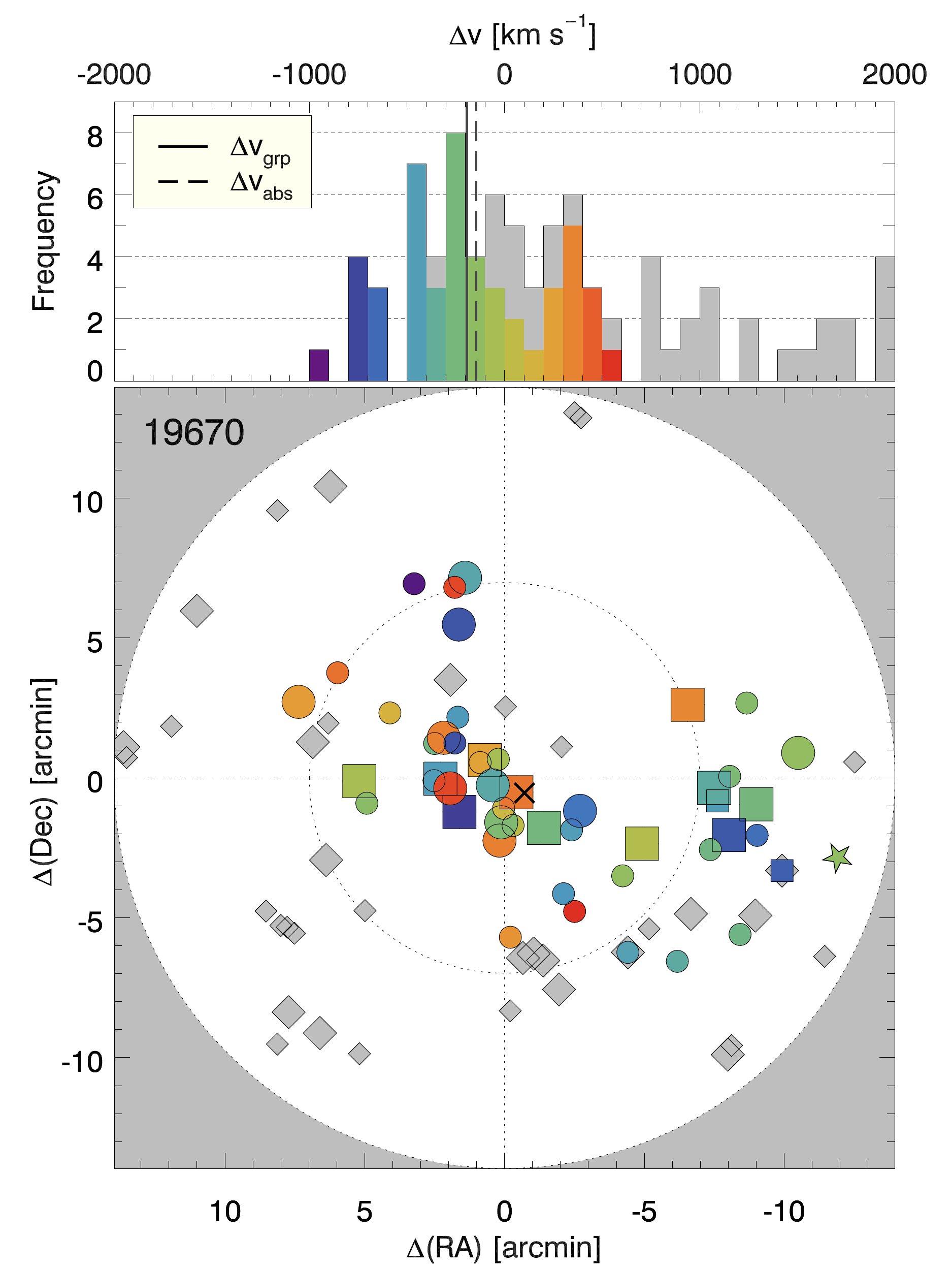}
\caption{Same as \autoref{fig:12833}, but for the galaxy group~19670.
\label{fig:19670}}
\end{figure*}
\clearpage
\input{./tab_19670.tex}

\clearpage
\begin{figure*}
\epsscale{0.95}
\centering\plotone{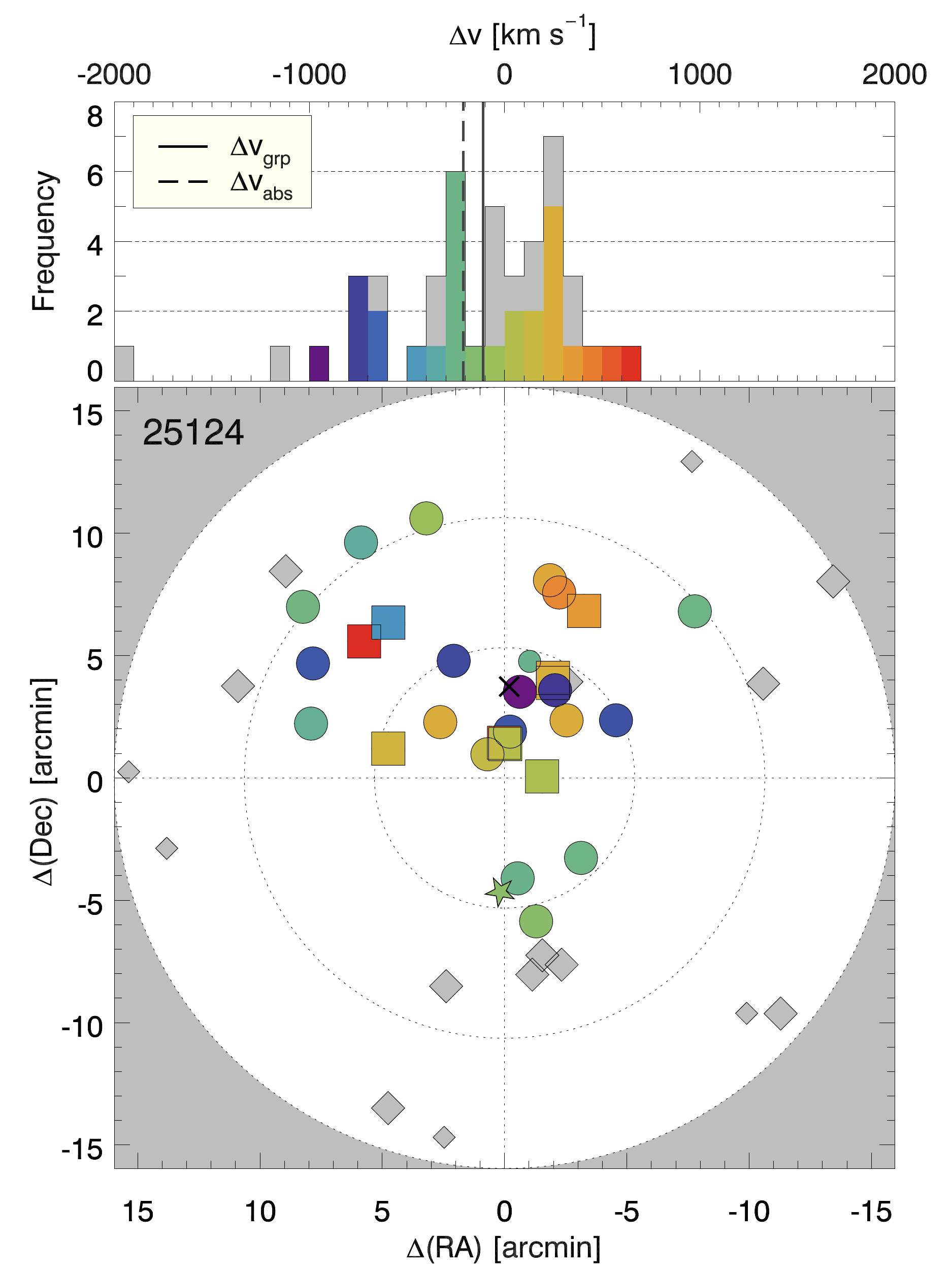}
\caption{Same as \autoref{fig:12833}, but for the galaxy group~25124.
\label{fig:25124}}
\end{figure*}
\newpage
\input{./tab_25124.tex}

\newpage
\begin{figure*}
\epsscale{0.95}
\centering\plotone{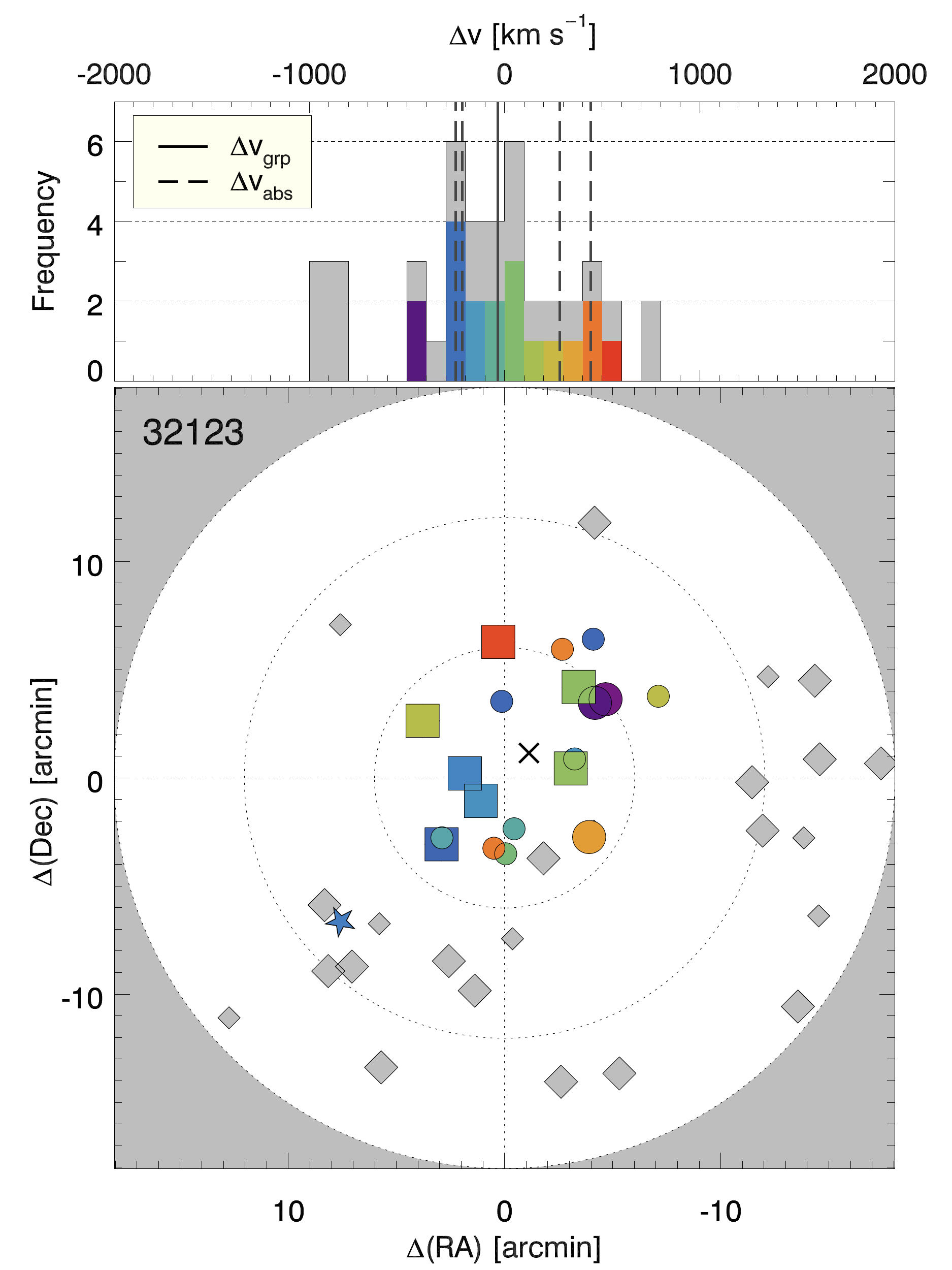}
\caption{Same as \autoref{fig:12833}, but for the galaxy group~32123.
\label{fig:32123}}
\end{figure*}
\newpage
\input{./tab_32123.tex}

\newpage
\begin{figure*}
\epsscale{0.95}
\centering\plotone{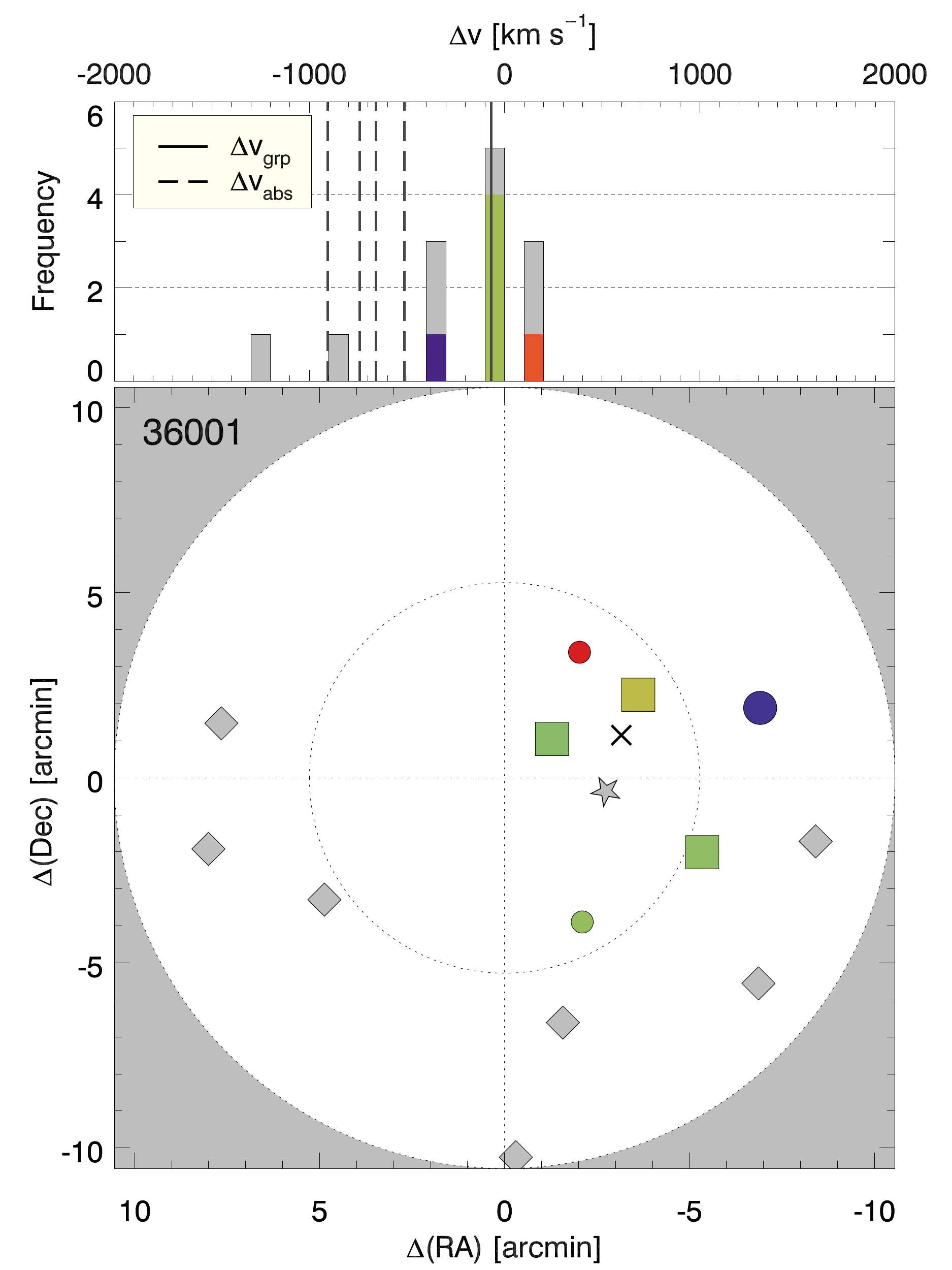}
\caption{Same as \autoref{fig:12833}, but for the galaxy group~36001.
\label{fig:36001}}
\end{figure*}
\input{./tab_36001.tex}

\newpage
\begin{figure*}
\epsscale{0.95}
\centering\plotone{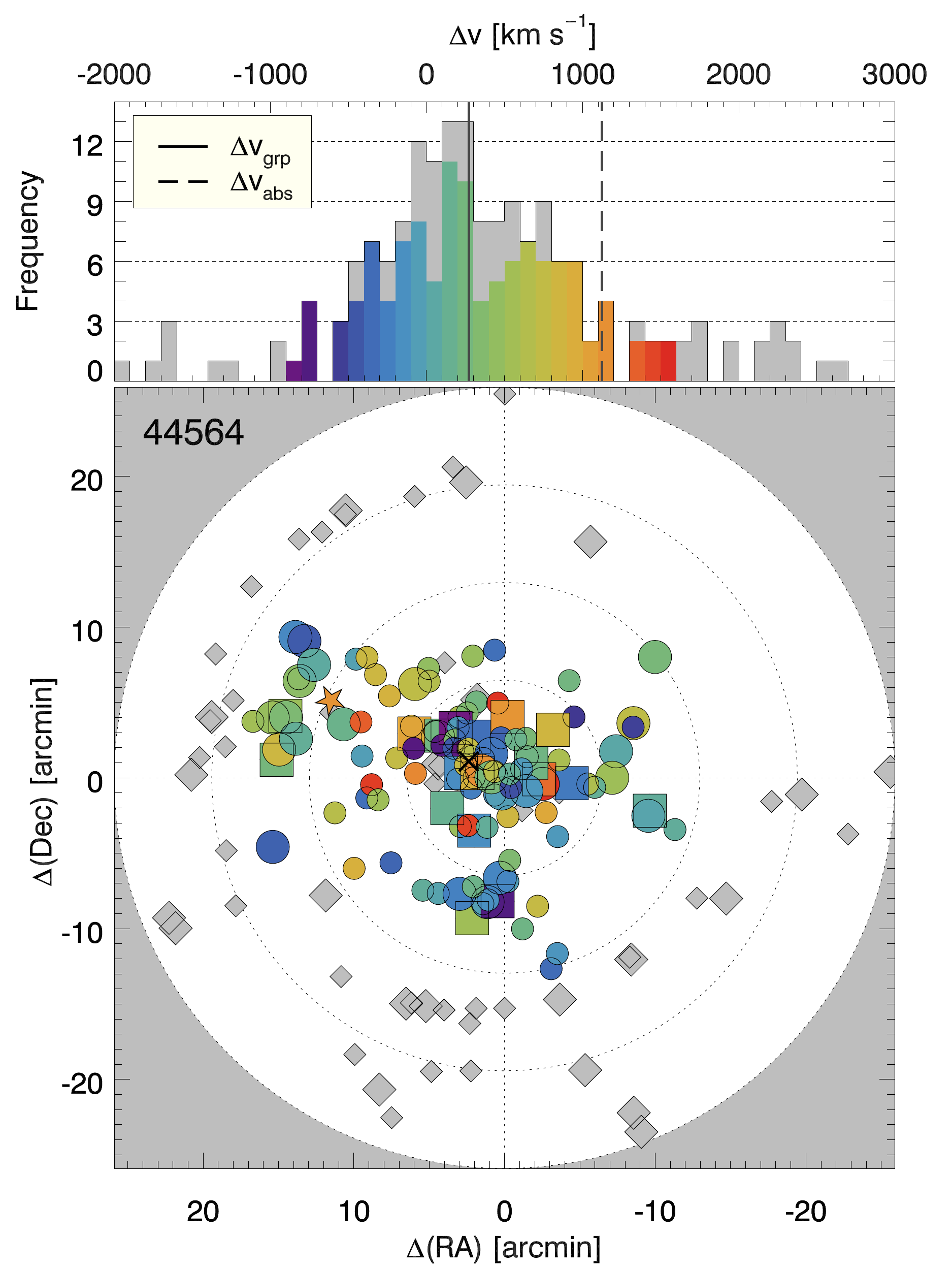}
\caption{Same as \autoref{fig:12833}, but for the galaxy group~44564.
\label{fig:44564}}
\end{figure*}
\clearpage
\input{./tab_44564.tex}

\clearpage
\begin{figure*}
\epsscale{0.95}
\centering\plotone{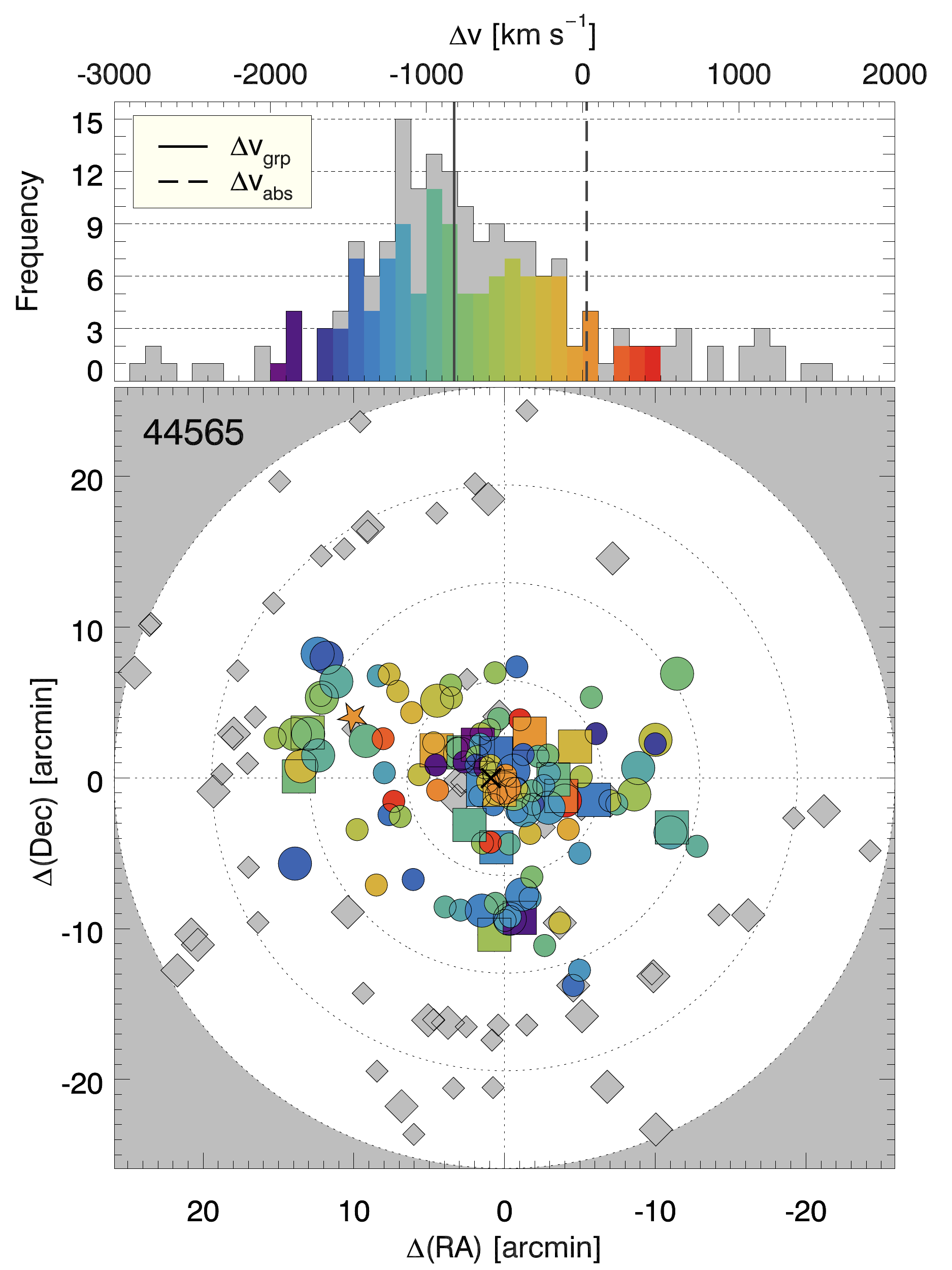}
\caption{Same as \autoref{fig:12833}, but for the galaxy group~44565.
\label{fig:44565}}
\end{figure*}
\clearpage
\input{./tab_44565.tex}

\clearpage
\begin{figure*}
\epsscale{0.95}
\centering\plotone{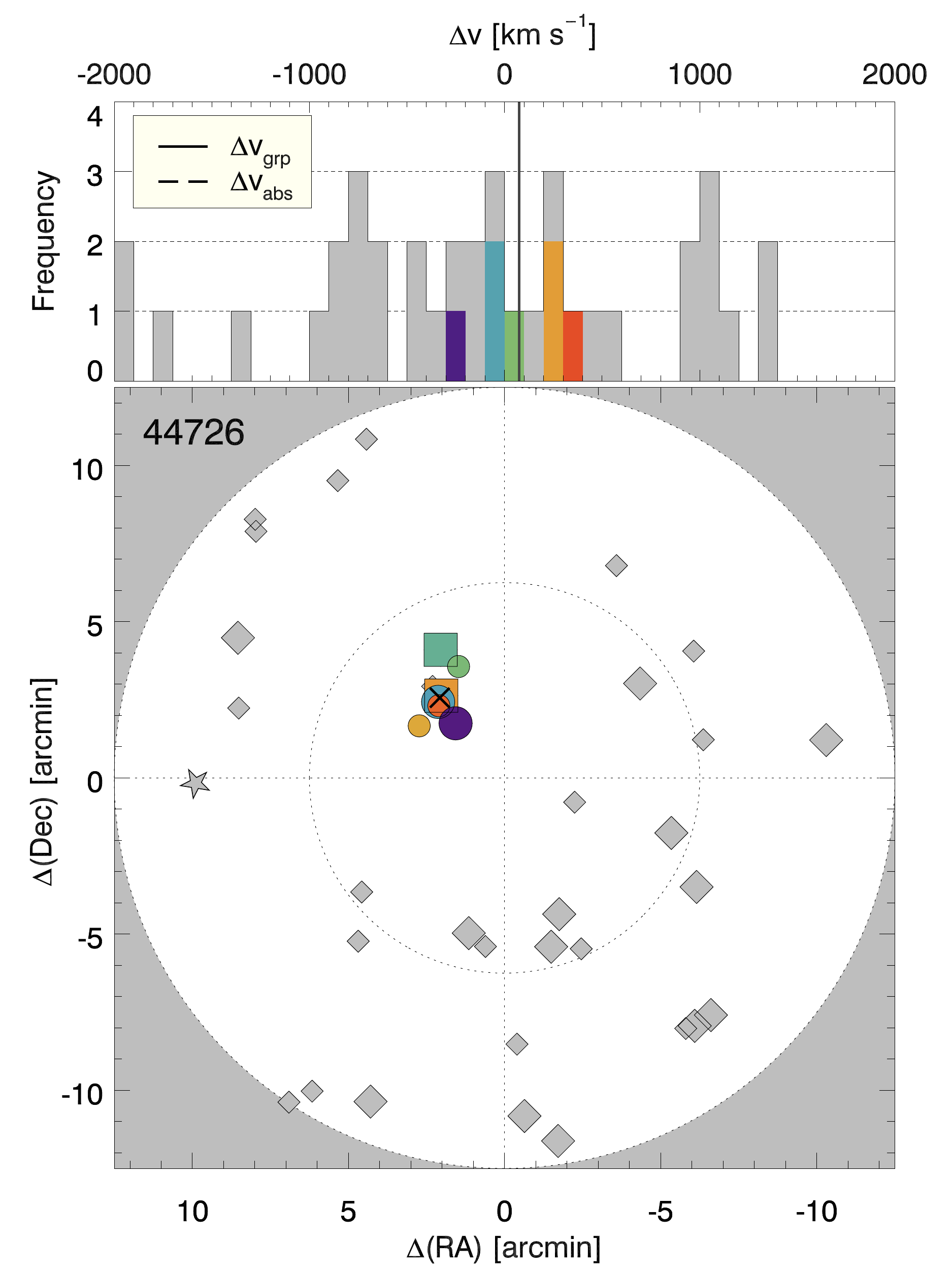}
\caption{Same as \autoref{fig:12833}, but for the galaxy group~44726.
\label{fig:44726}}
\end{figure*}
\newpage
\input{./tab_44726.tex}

\newpage
\begin{figure*}
\epsscale{0.95}
\centering\plotone{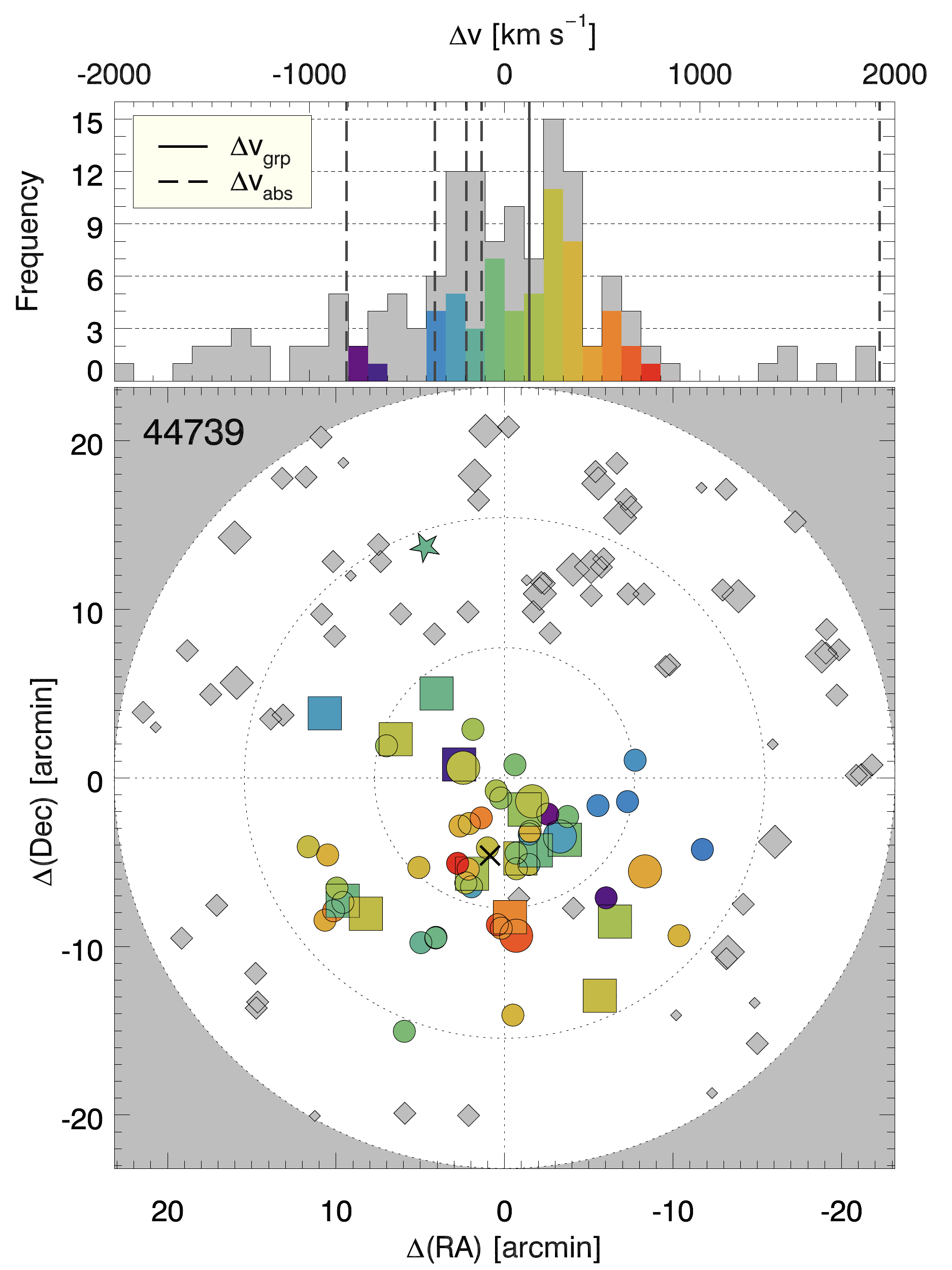}
\caption{Same as \autoref{fig:12833}, but for the galaxy group~44739.
\label{fig:44739}}
\end{figure*}
\clearpage
\input{./tab_44739.tex}

\clearpage
\begin{figure*}
\epsscale{0.95}
\centering\plotone{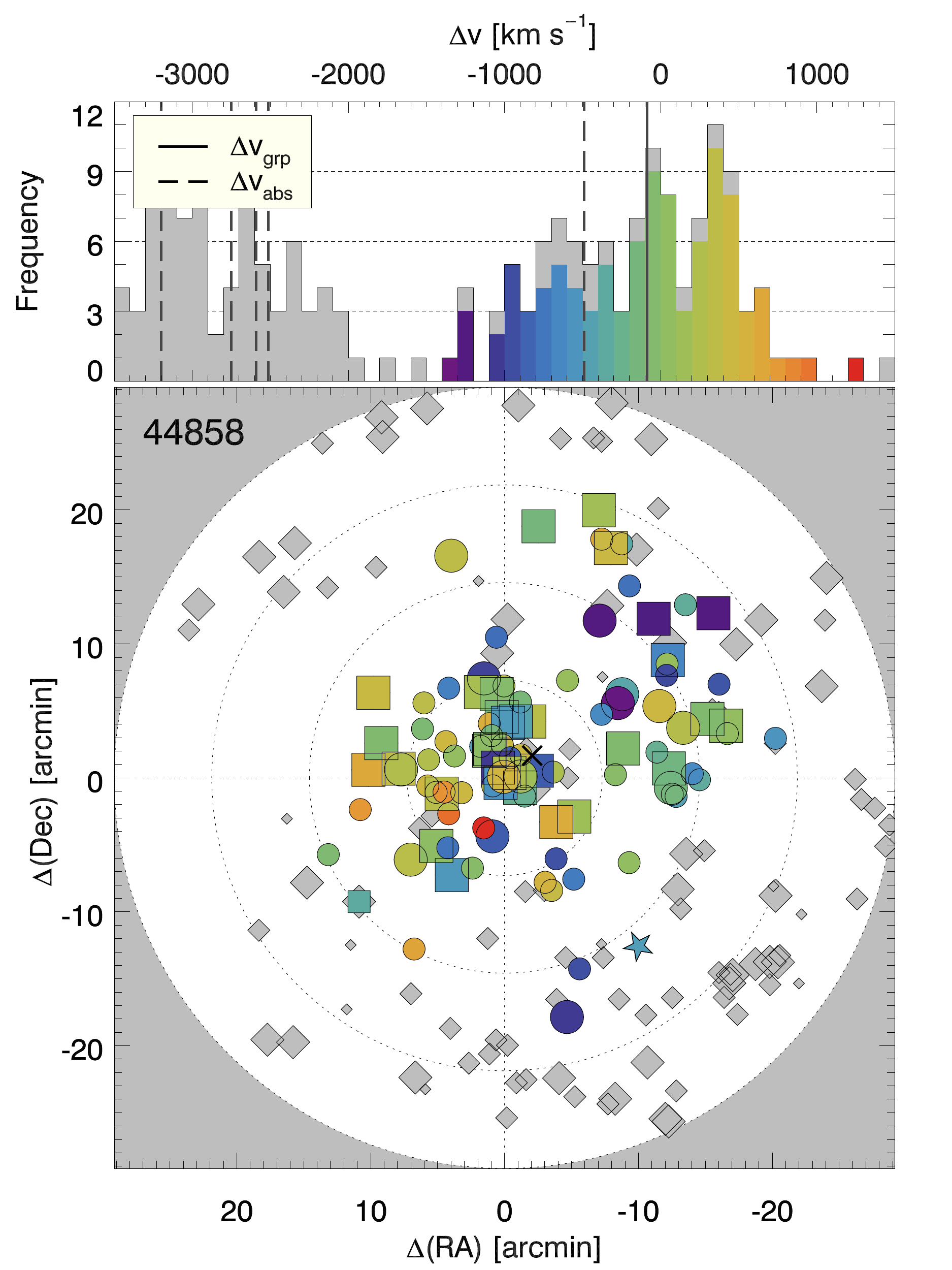}
\caption{Same as \autoref{fig:12833}, but for the galaxy group~44858.
\label{fig:44858}}
\end{figure*}
\clearpage
\input{tab_44858.tex}


\newpage
\begin{figure*}
\epsscale{0.95}
\centering\plotone{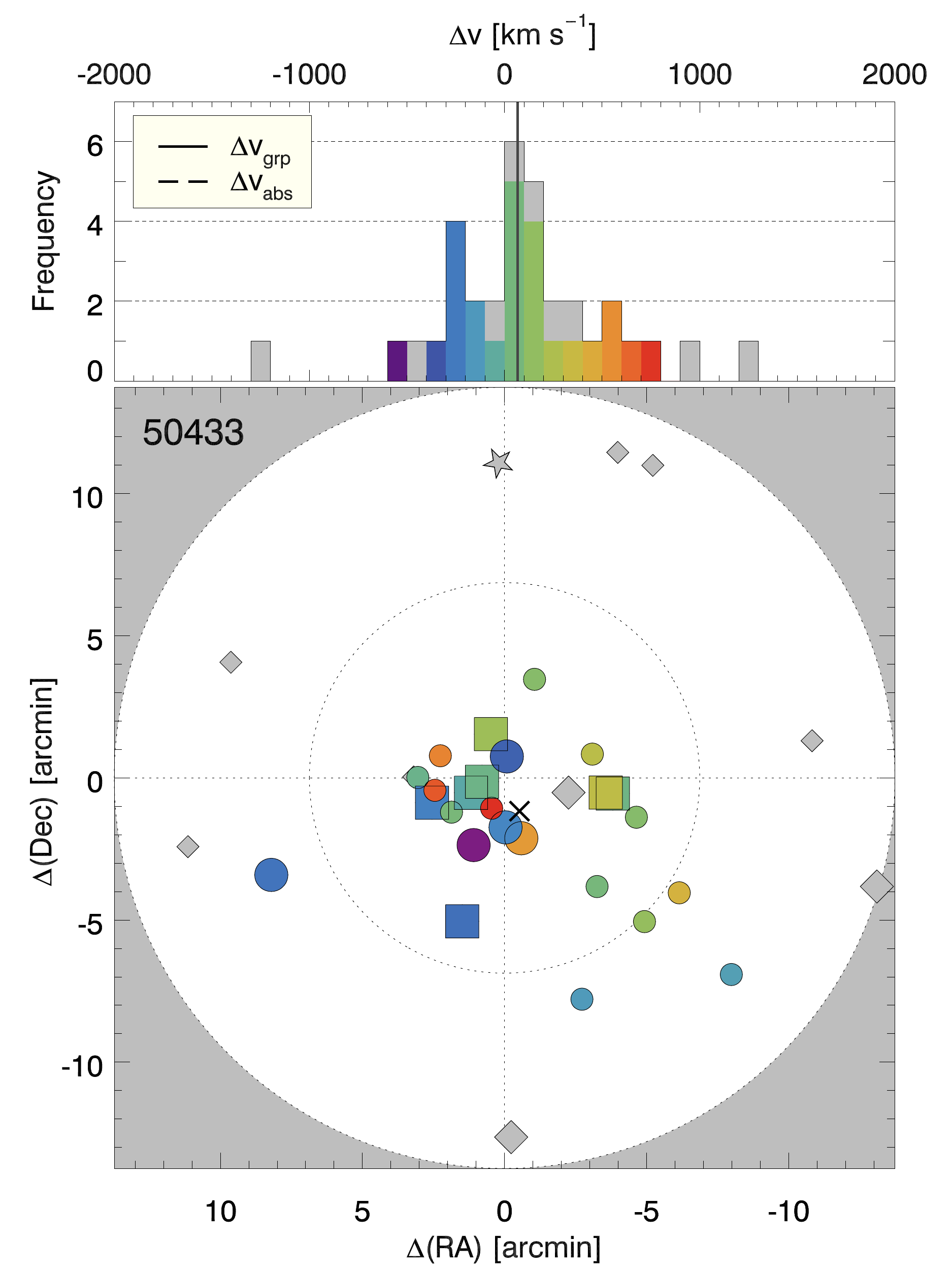}
\caption{Same as \autoref{fig:12833}, but for the galaxy group~50433.
\label{fig:50433}}
\end{figure*}
\newpage
\input{./tab_50433.tex}

\begin{figure*}
\epsscale{1.0}
\centering\plotone{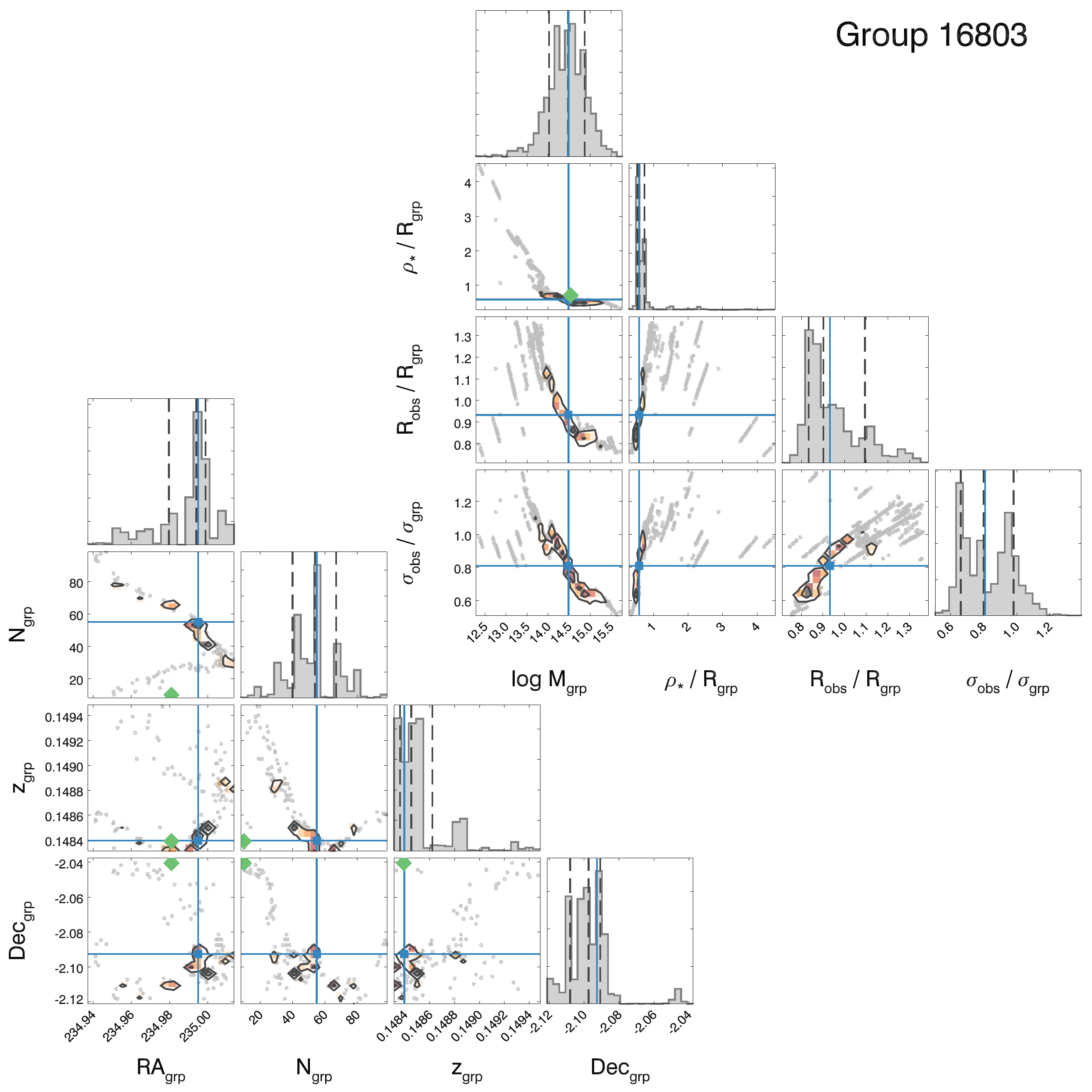}
\caption{Same as \autoref{fig:corner}, but for the galaxy group~16803.
\label{fig:16803_corner}}
\end{figure*}

\begin{figure*}
\epsscale{1.0}
\centering\plotone{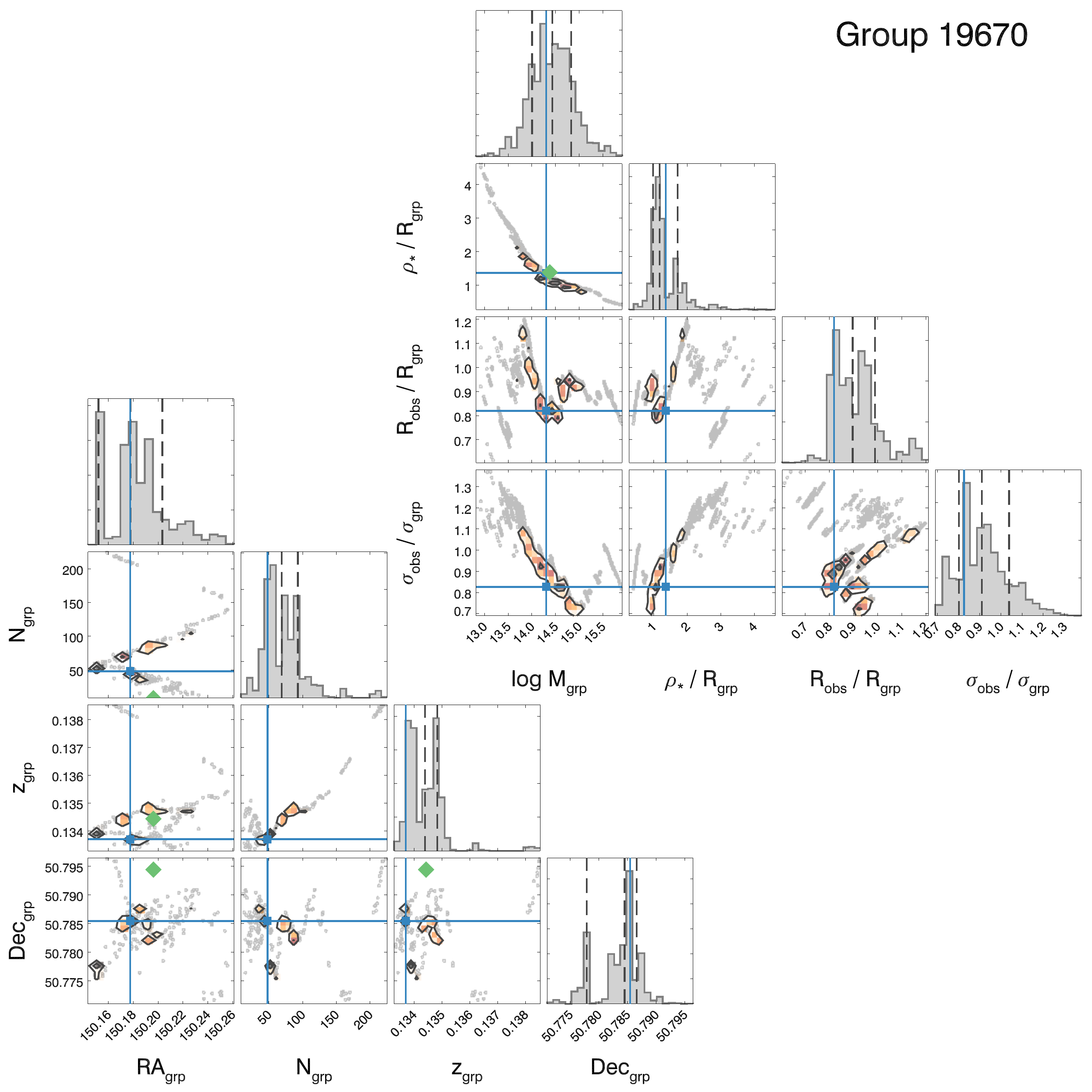}
\caption{Same as \autoref{fig:corner}, but for the galaxy group~19670.
\label{fig:19670_corner}}
\end{figure*}

\begin{figure*}
\epsscale{1.0}
\centering\plotone{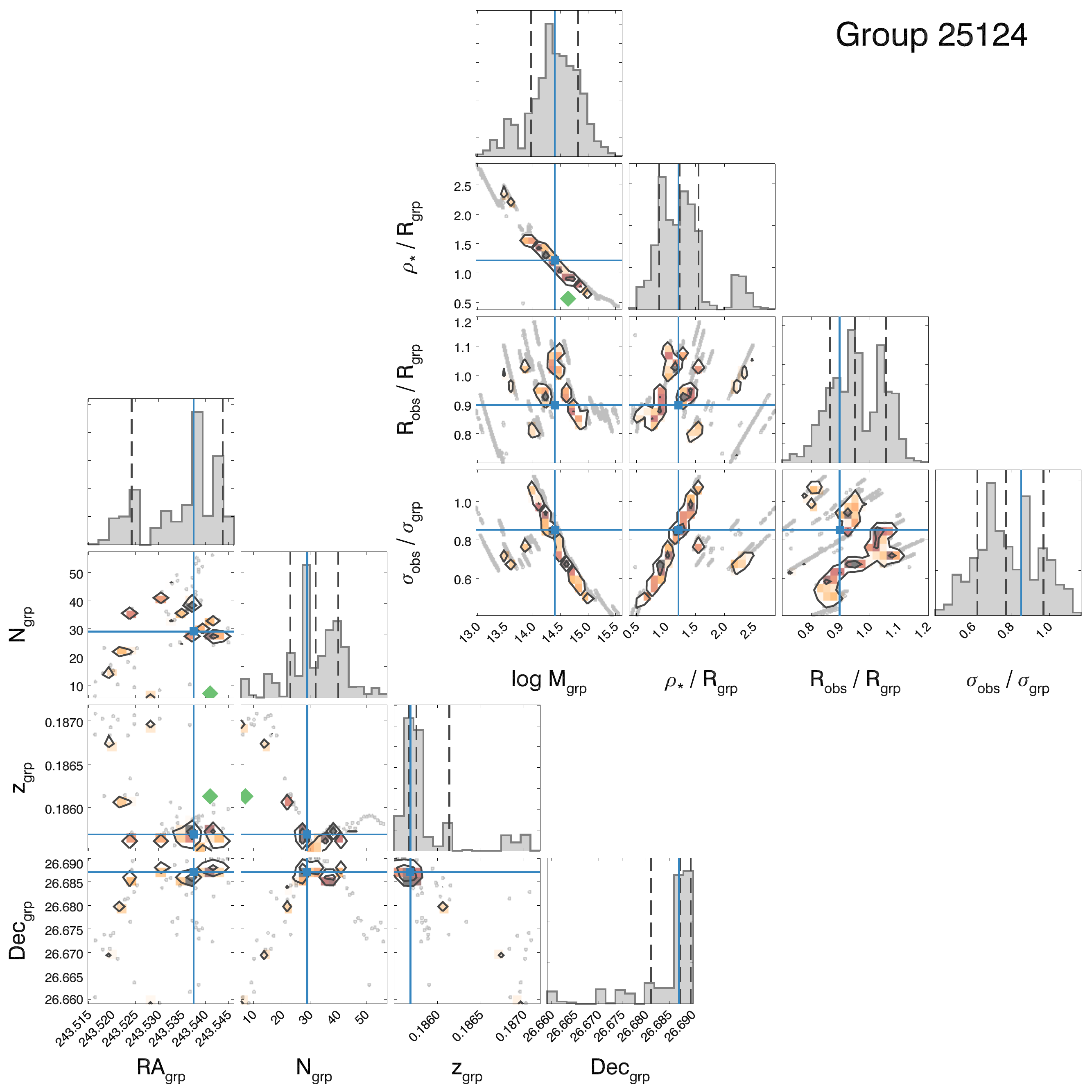}
\caption{Same as \autoref{fig:corner}, but for the galaxy group~25124.
\label{fig:25124_corner}}
\end{figure*}

\begin{figure*}
\epsscale{1.0}
\centering\plotone{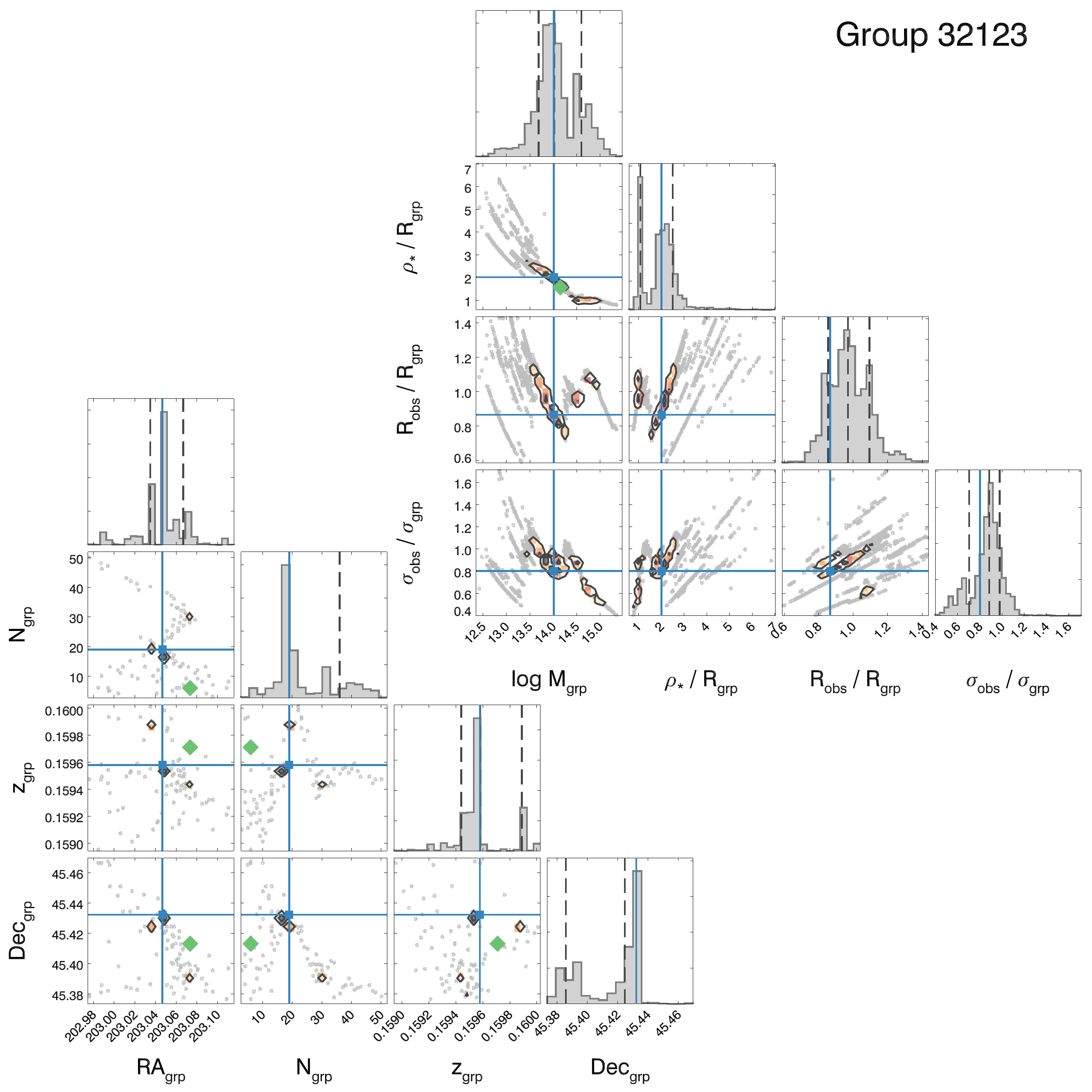}
\caption{Same as \autoref{fig:corner}, but for the galaxy group~32123.
\label{fig:32123_corner}}
\end{figure*}

\begin{figure*}
\epsscale{1.0}
\centering\plotone{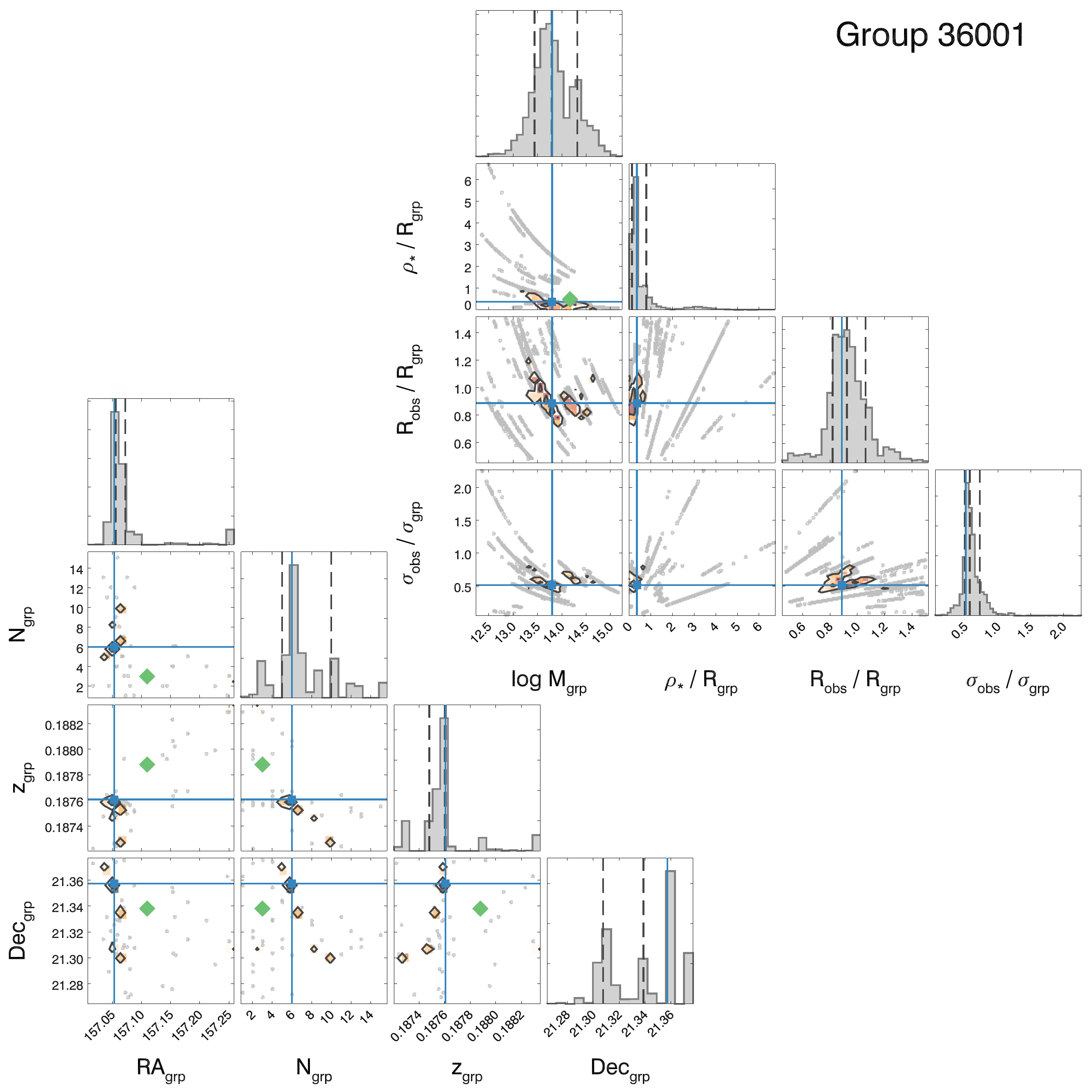}
\caption{Same as \autoref{fig:corner}, but for the galaxy group~36001.
\label{fig:36001_corner}}
\end{figure*}

\begin{figure*}
\epsscale{1.0}
\centering\plotone{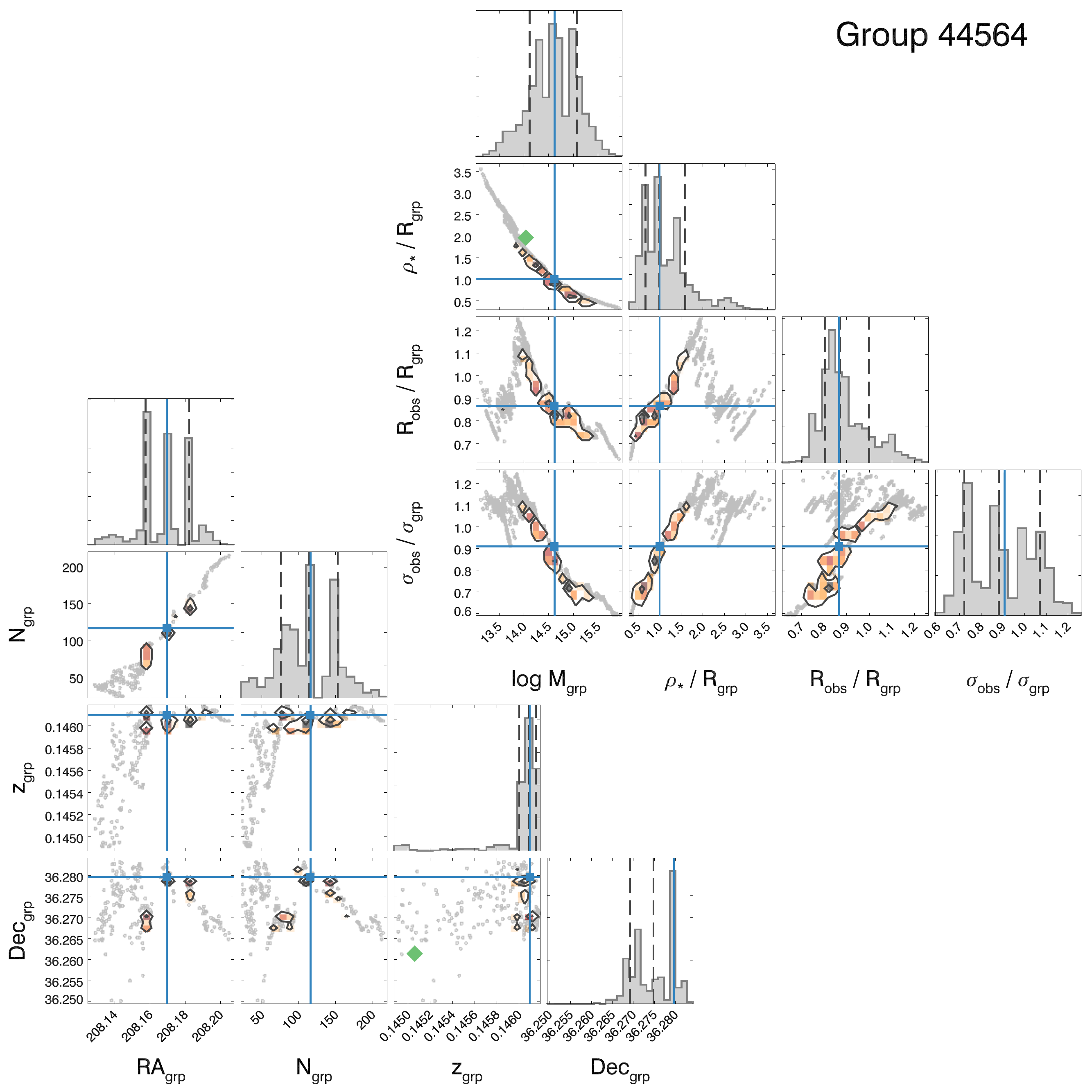}
\caption{Same as \autoref{fig:corner}, but for the galaxy group~44564.
\label{fig:44564_corner}}
\end{figure*}

\begin{figure*}
\epsscale{1.0}
\centering\plotone{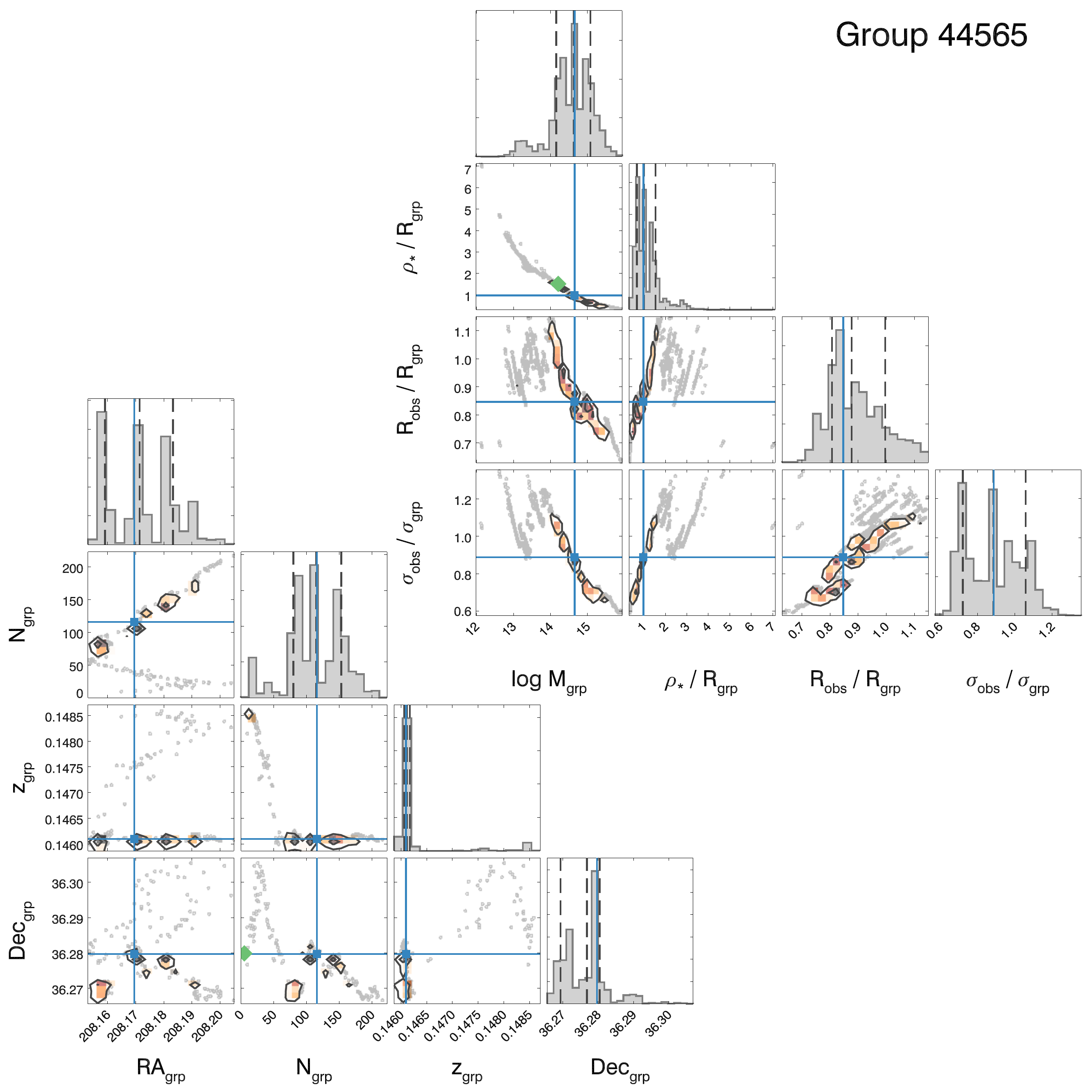}
\caption{Same as \autoref{fig:corner}, but for the galaxy group~44565.
\label{fig:44565_corner}}
\end{figure*}

\begin{figure*}
\epsscale{1.0}
\centering\plotone{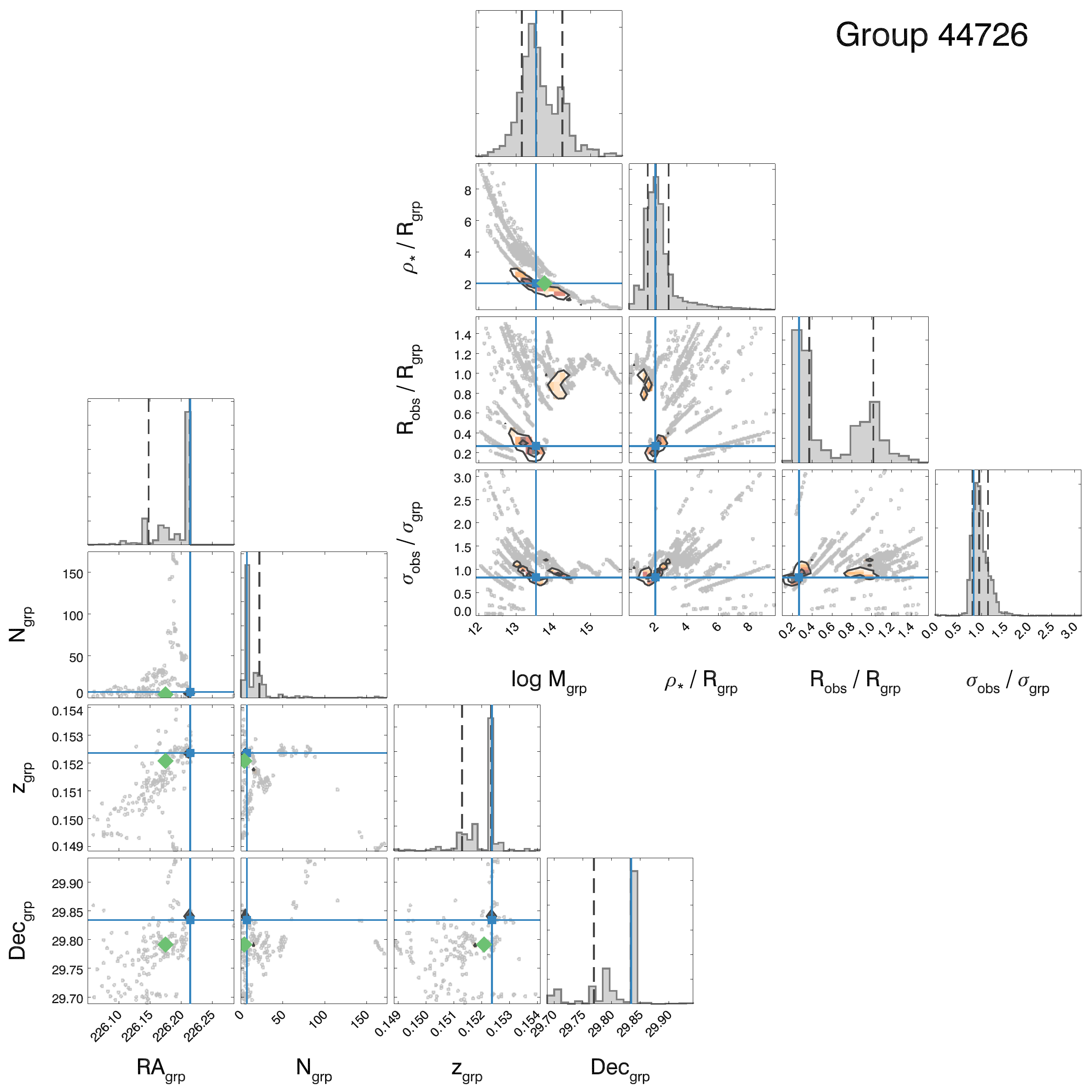}
\caption{Same as \autoref{fig:corner}, but for the galaxy group~44726.
\label{fig:44726_corner}}
\end{figure*}

\begin{figure*}
\epsscale{1.0}
\centering\plotone{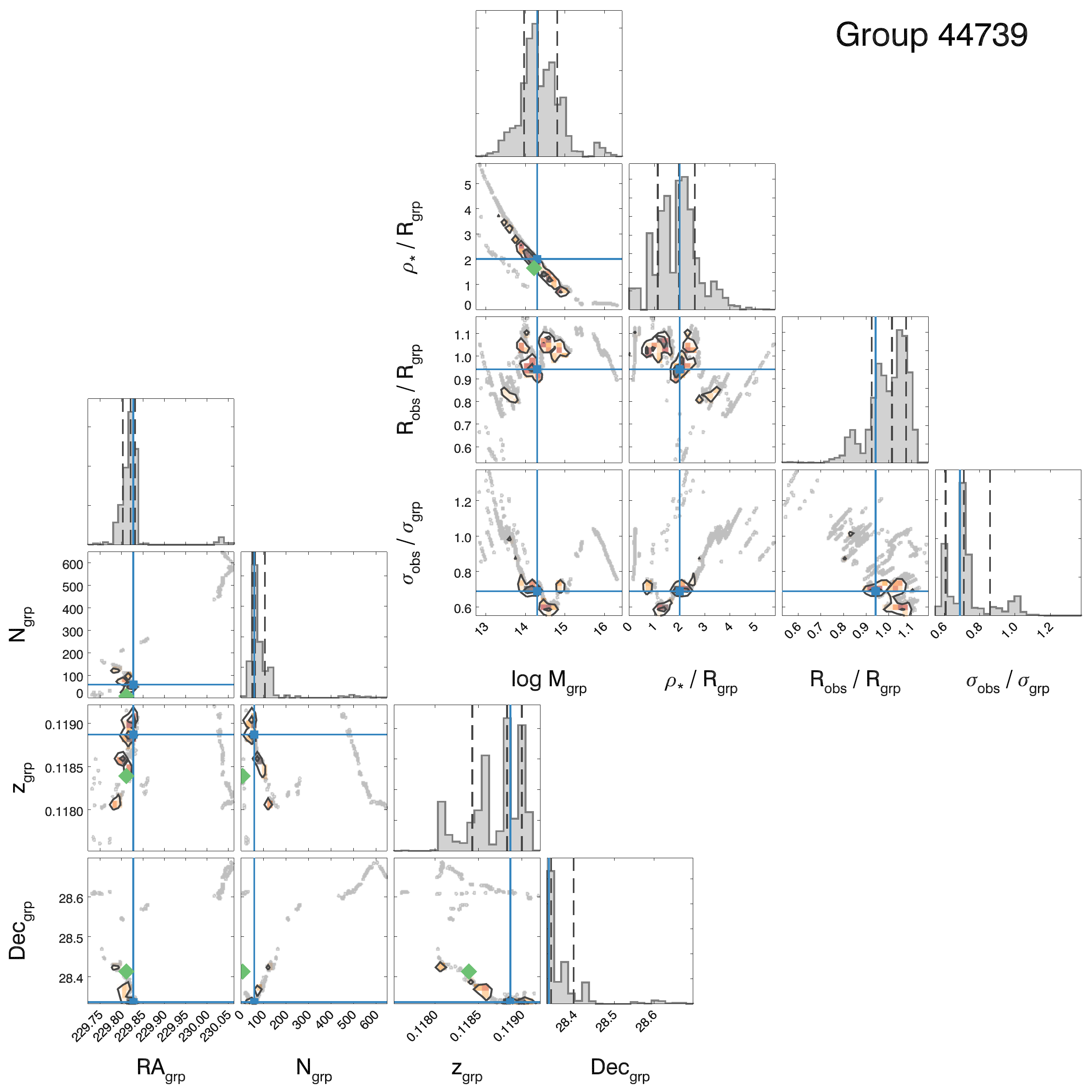}
\caption{Same as \autoref{fig:corner}, but for the galaxy group~44739.
\label{fig:44739_corner}}
\end{figure*}

\begin{figure*}
\epsscale{1.0}
\centering\plotone{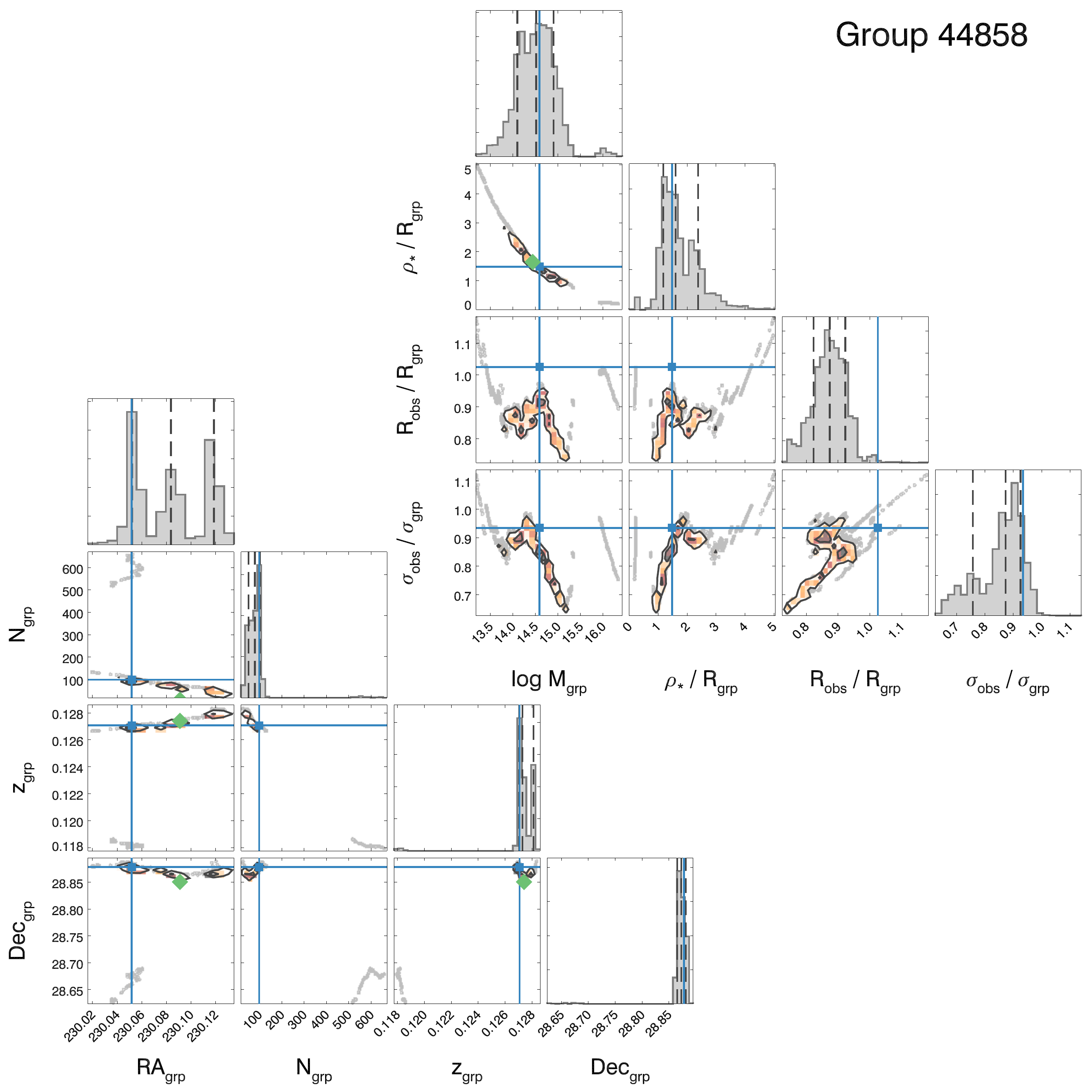}
\caption{Same as \autoref{fig:corner}, but for the galaxy group~44858.
\label{fig:44858_corner}}
\end{figure*}


\begin{figure*}
\epsscale{1.0}
\centering\plotone{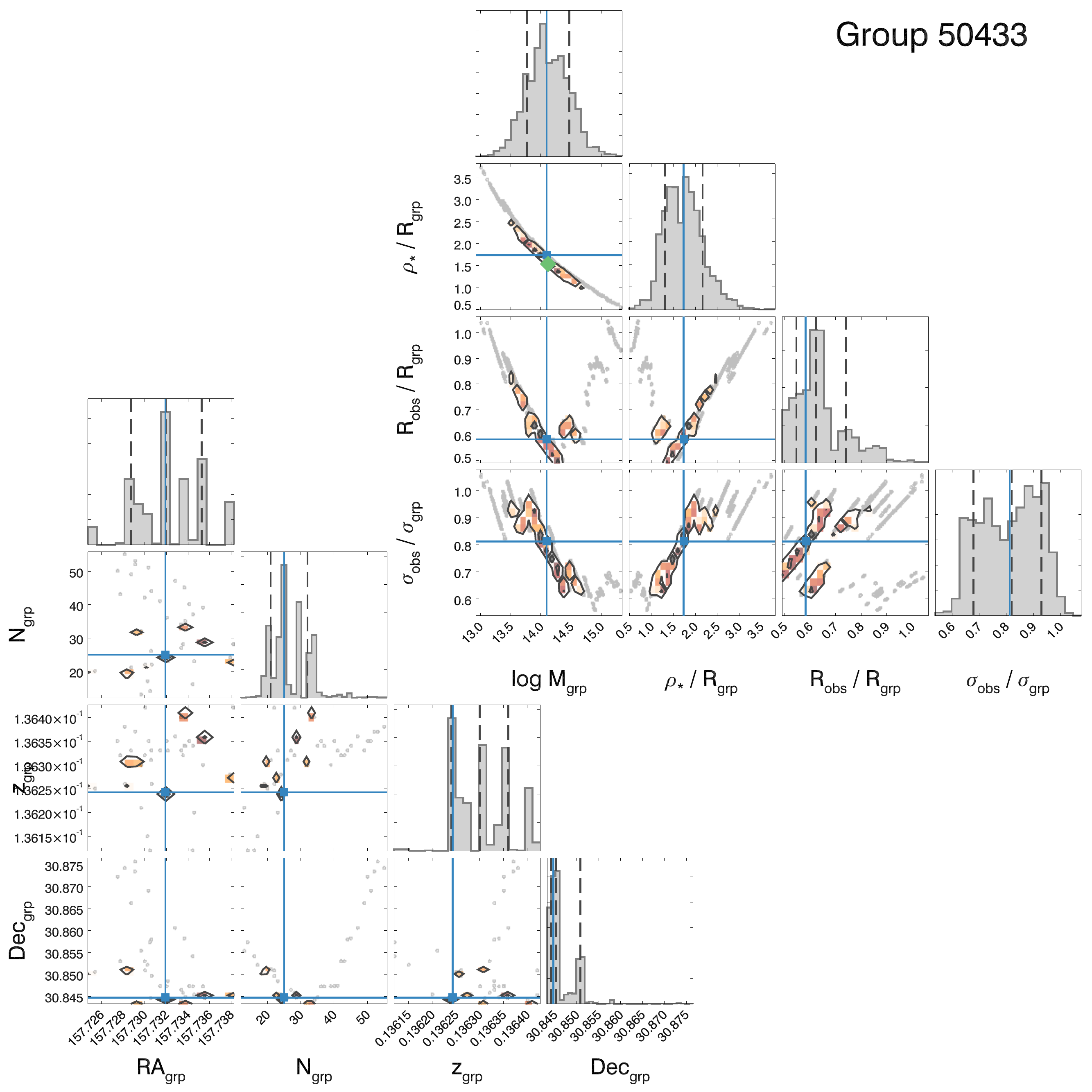}
\caption{Same as \autoref{fig:corner}, but for the galaxy group~50433.
\label{fig:50433_corner}}
\end{figure*}

\end{document}

%% file: tab_cos.tex
\begin{deluxetable*}{llcccccl}

\tablecaption{\hst/COS Observations
\label{tab:cos}}

\tablehead{
	\colhead{Group} &
	\colhead{AGN Target} &
	\colhead{$z_{\rm AGN}$} &
	\colhead{Obs. Date}&
	\colhead{$t_{\rm exp}$} &
	\colhead{$F_{\rm FUV}$} &
	\colhead{S/N} &
	\colhead{Notes} \\
	\colhead{ID} &
	\colhead{Name} &
	\colhead{} &
	\colhead{} &
	\colhead{(ksec)} &
	\colhead{($10^{-15}$)} &
	\colhead{} &
	\colhead{} 
	}	

\colnumbers
\startdata
12833 & RBS~711                 & 0.255 &  9/29/16       &  5   & 12 & 29     & 14277-07  \\		
16803 & SDSS~J1540--0205      & 0.321 &  7/3/16        & 25.3 &$<1$\tablenotemark{d} & 11 & 14277-05,-06  \\	
19670 & SBS~0956+510          & 0.143 &  6/23/16       & 8.0  &  8 & 21     & 14277-09  \\		
25124 & B~1612+266         & 0.395 & Jul/Aug '16\tablenotemark{c} & 25.4 &  2 & 20     & 14277-03,-04  \\	
32123 &SDSS~J1333+4518 & 0.320 & 4/26/17       &  11  &  2 & 16     & 14277-16 \\
36001 & SDSS~J1028+2119      & 0.374 &  4/2/16        & 15.9 &  4 & 26     & 14277-01,-02  \\	
44564 & CSO~1022\tablenotemark{a}           & 0.285 &  6/30/14       &  4.8 &  5 & 16     & 13444\tablenotemark{e} (Wakker) \\	
44565 & CSO~1022\tablenotemark{a} & 0.285 &  6/30/14       & 4.8  &  5 & 16     & 13444\tablenotemark{e} (Wakker) \\	
44726 & CSO~1080                & 0.526 &  7/7/16        & 10.1 &  5 & 19     & 14277-13,-15  \\
44739 & FBQS~J1519+2838\tablenotemark{b}  & 0.270 &  8/10/16       & 15.8 &  3 & 22     & 14277-11,-12  \\	
44858 & FBQS~J1519+2838\tablenotemark{b}  & 0.270 &  8/10/16       & 15.8 &  3 & 22     & 14277-11,-12  \\	
50433 & FBQS~J1030+3102 & 0.178 &  3/26/16       & 4.9  & 12 & 23     & 14277-10  \\		
\enddata

\tablenotetext{a}{Two groups are probed by the CSO~1022 sight line}
\tablenotetext{b}{Two groups are probed by the FBQS~J1519+2838 sight line}
\tablenotetext{c}{Two coadded visits 22~Jul and 6~Aug~2016}
\tablenotetext{d}{Partial LLS at source redshift mostly eliminates \OVI\ sensitivity.}
\tablenotetext{e}{Archival dataset}

\vspace{-2em}

\end{deluxetable*}

%% file: tab_berlind.tex
\begin{deluxetable*}{lcccccccc}

\tablecaption{Group Properties from SDSS Analysis
\label{tab:berlind}}

\tablehead{ \colhead{Group} & \colhead{$N_0$} & \colhead{$\mathrm{RA}_0$} & \colhead{$\mathrm{Dec}_0$} & \colhead{$z_0$} & \colhead{$\log{M_0}$} & \colhead{$\sigma_0$} & \colhead{$\mathrm{RA}_{\star}$} & \colhead{$\mathrm{Dec}_{\star}$} }

\colnumbers
\startdata
12833 &  8 & $129.423\pm0.040$ & $44.295\pm0.029$ & $0.14693\pm0.00077$ & $14.42\pm0.26$ & 534 & 129.245433 &  44.433969 \\
16803 & 10 & $234.981\pm0.015$ & $-2.040\pm0.040$ & $0.14839\pm0.00046$ & $14.54\pm0.25$ & 586 & 235.081530 & --2.084831 \\
19670 &  8 & $150.196\pm0.044$ & $50.794\pm0.028$ & $0.13443\pm0.00052$ & $14.38\pm0.26$ & 518 & 149.881960 &  50.746969 \\
25124 &  7 & $243.541\pm0.018$ & $26.625\pm0.036$ & $0.18613\pm0.00038$ & $14.64\pm0.24$ & 632 & 243.544250 &  26.547278 \\
32123 &  6 & $203.074\pm0.066$ & $45.413\pm0.024$ & $0.15971\pm0.00044$ & $14.15\pm0.29$ & 434 & 203.253458 &  45.302500 \\
36001 &  3 & $157.110\pm0.045$ & $21.338\pm0.028$ & $0.18788\pm0.00023$ & $14.17\pm0.28$ & 441 & 157.060646 &  21.331967 \\
44564 &  4 & $208.120\pm0.040$ & $36.261\pm0.022$ & $0.14506\pm0.00050$ & $14.04\pm0.30$ & 399 & 208.358845 &  36.347070 \\
44565 &  5 & $208.151\pm0.016$ & $36.280\pm0.013$ & $0.14925\pm0.00054$ & $14.21\pm0.28$ & 455 & 208.358845 &  36.347070 \\
44726 &  4 & $226.175\pm0.034$ & $29.791\pm0.035$ & $0.15208\pm0.00092$ & $13.76\pm0.35$ & 322 & 226.365002 &  29.788439 \\
44739 &  7 & $229.811\pm0.033$ & $28.413\pm0.051$ & $0.11839\pm0.00023$ & $14.22\pm0.28$ & 458 & 229.900610 &  28.641011 \\
44858 &  9 & $230.091\pm0.027$ & $28.850\pm0.013$ & $0.12740\pm0.00055$ & $14.44\pm0.26$ & 542 & 229.900610 &  28.641011 \\
50433 &  4 & $157.742\pm0.026$ & $30.864\pm0.009$ & $0.13599\pm0.00033$ & $14.12\pm0.29$ & 424 & 157.746220 &  31.048811 \\
\enddata

\tablecomments{All coordinates are J2000, $M_0$ has units of $M_{\Sun}$, and $\sigma_0$ has units of \kms. $\mathrm{RA}_{\star}$ and $\mathrm{Dec}_{\star}$ are the coordinates of the COS sight line.}

\vspace{-1em}

\end{deluxetable*}

%% file: tab_abssummary.tex
\begin{deluxetable*}{ccll}

\tablecaption{Detections/non-detections in group sight lines
\label{tab:abssummary}}

\tablecolumns{5}

\tablehead{
	\colhead{Group} &
	\colhead{$\rho/R_0$} &
	\colhead{AGN} &
	\colhead{Absorption component notes}
	}

\colnumbers
\startdata 
12833 &  0.9 &  RBS711   &  strong \HI; no \OVI\\
16803 &  0.5 & SDSS15403 &  weak \HI\ at $-1036$~\kms; partial LLS blocks \OVI \\
19670 &  0.9 &  SBS0956  &  strong \HI; no \OVI\\
25124 &  0.4 & B1612     &  2 strong \HI, 1 possible BLA; \OVI\ aligned with BLA \\
32123 &  1.1 & SDSS1333  &  3 strong \HI, 1 weak BLA; 1 weak \OVI; 1 \ion{Si}{3} \\
36001 &  0.3 & SDSS1028  &  3 strong \HI; 1 weak \OVI; 1 \ion{Si}{3}; probably aligned \\
44564 &  1.3 & CSO1022   &  nothing \\
44565 &  1.0 & CSO1022   &  nothing \\
44726 &  1.4 & CSO1080   &  strong \HI\ at +1200~\kms; no \OVI\\
44739 &  1.1 & FBQS1519  &  4 strong \HI; 2 weak \OVI, 1 with \ion{Si}{3}; probably aligned \\
44858 &  1.1 & FBQS1519  &  strong \HI; moderate \OVI; slight misalignment\\
50433 &  1.1 & FBQS1030  &  nothing\\
\enddata

\tablecomments{Column~2 lists the impact parameter between the SDSS group barycenter (\autoref{tab:berlind}) and the QSO sight line, normalized by the group virial radius.}

\vspace{-2em}

\end{deluxetable*}

%% file: tab_bla.tex
\begin{deluxetable}{cccccc}

\tablecaption{The Probable BLA in Group~25124
\label{tab:bla}}

\tablehead{
	\colhead{} & 
	\colhead{$\lambda$} & 
	\colhead{$z$} & 
	\colhead{$b$} & 
	\colhead{EW} & 
	\colhead{$\log\,N$} \\ 
	\colhead{} & 
	\colhead{(\AA)} & 
	\colhead{} & 
	\colhead{(\kms)} & 
	\colhead{(m\AA)} & 
	\colhead{($N$ in cm$^{-2}$)} 
	}
	\startdata
	\#1 & 1440.89 & 0.18526 &  $37\pm4$  &  708 & $15.89\pm0.41$ \\
	\#2 & 1441.05 & 0.18540 & $128\pm57$ &  140 & $13.43\pm0.38$ \\
	\#3 & 1441.41 & 0.18569 &  $18\pm7$  &   63 & $13.15\pm0.19$ \\
\enddata 

\vspace{-2em}

\end{deluxetable}

%% file: tab_ng.tex
\begin{deluxetable*}{lccccccccc}

\tablecaption{Nearest Galaxy Properties
\label{tab:ng}}

\tablehead{ \colhead{Group} & \colhead{$z_{\rm abs}$} & \colhead{RA} & \colhead{Dec} & \colhead{$z_{\rm gal}$} & \colhead{$L_{\rm gal}$} & \colhead{$R_{\rm vir}$} & \colhead{$\rho$} & \colhead{$\Delta v_{\rm abs}$} & \colhead{$D/R_{\rm vir}$} \\ & & \colhead{($\degr$)} & \colhead{($\degr$)} & & \colhead{($L^*$)} & \colhead{(kpc)} & \colhead{(kpc)} & \colhead{(km\,s$^{-1}$)} }

\colnumbers
\startdata
12833       & 0.14743 & 129.343638 &  44.390149 & 0.14844 & 2.6     & 248     &  772    &     263 &  3.1 \\ [1ex]
16803       & 0.14497 & 235.046611 & --2.039641 & 0.14549 & 1.2     & 191     &  525    &     136 &  2.7 \\ [1ex]
19670       & 0.13388 & 149.958823 &  50.778650 & 0.13346 & 1.1     & 185     &  499    &   --111 &  2.7 \\ [1ex]
25124       & 0.18529 & 243.531018 &  26.556435 & 0.18507 & 1.9     & 223     &  168    &    --55 &  0.8 \\
            & 0.18569 &            &            &         &         &         &         &   --159 &      \\ [1ex]
32123       & 0.15874 & 203.271113 &  45.315150 & 0.15853 & 5.0     & 311     &  175    &    --54 &  0.6 \\
            & 0.16081 & 203.211036 &  45.300789 & 0.16102 & 0.78    & 167     &  299    &      54 &  1.8 \\
            & 0.16142 &            &            &         &         &         &         &   --103 &      \\ [1ex]
36001       & 0.18429 & 156.986890 &  21.245411 & 0.18432 & 2.7     & 252     & 1239    &       7 &  4.9 \\
            & 0.18494 & 156.986060 &  21.369490 & 0.18658 & 2.0     & 228     &  887    &     414 &  4.0 \\
            & 0.18527 &            &            &         &         &         &         &   --323 &  3.9 \\
            & 0.18585 & 157.086777 &  21.355513 & 0.18752 & 4.7     & 304     &  382    &     422 &  1.6 \\ [1ex]
44564/44565 & \nodata & \nodata    & \nodata    & \nodata & \nodata & \nodata & \nodata & \nodata & \nodata \\ [1ex]
44726       & \nodata & \nodata    & \nodata    & \nodata & \nodata & \nodata & \nodata & \nodata & \nodata \\ [1ex]
44739       & 0.11537 & 229.950843 &  28.626721 & 0.11502 & 1.1     & 185     &  350    &    --94 &  1.9 \\
            & 0.11706 & 229.844996 &  28.711684 & 0.11801 & 1.4     & 203     &  661    &     254 &  3.3 \\
            & 0.11766 &            &            &         &         &         &         &      93 &      \\
            & 0.11795 &            &            &         &         &         &         &      16 &      \\
44858       & 0.12551 & 230.069021 &  28.838277 & 0.12638 & 4.0     & 287     & 2006    &     231 &  7.0 \\
            & 0.12567 &            &            &         &         &         &         &     189 &      \\ [1ex]
50433       & \nodata & \nodata    & \nodata    & \nodata & \nodata & \nodata & \nodata & \nodata & \nodata \\ [1ex]
\tableline
49980       & 0.11351 & 152.472930 &  30.033939 & 0.11345 & 2.7     & 252     &  254    &    --16 &  1.0 \\
            & 0.11381 &            &            &         &         &         &         &    --96 &      \\
            & 0.11466 &            &            &         &         &         &         &   --325 &      \\
            & 0.11675 & 152.426382 &  29.942452 & 0.11551 & 3.5     & 274     & 1005    &   --332 &  3.7 \\
\enddata

\tablecomments{An individual galaxy can be identified as the nearest galaxy to several \lya\ absorbers; when this occurs the galaxy properties are listed once and only the galaxy-absorber velocity difference is updated for subsequent absorbers. Similarly, some groups have no absorption within 1000~\kms\ of the group centroid (see \autoref{tab:abs}), as indicated by the ``\nodata'' symbol in each column.}

\end{deluxetable*}

%% file: tab_params.tex
\movetabledown=1.5in

\begin{rotatetable}
\begin{deluxetable*}{lcccccccccccc}

\tablecaption{Range of Group Properties
\label{tab:params}}

\tablehead{ \colhead{Group} & \colhead{RA} & \colhead{Dec} & \colhead{$z_{\rm grp}$} & \colhead{$N_{\rm grp}$} & \colhead{$L_{\rm grp}$} & \colhead{$\log{M_{\rm grp}}$} & \colhead{$R_{\rm obs}$} & \colhead{$\sigma_{\rm obs}$} & \colhead{$\rho_{\star}$} & \colhead{$R_{\rm obs}/R_{\rm grp}$} & \colhead{$\sigma_{\rm obs}/\sigma_{\rm grp}$} & \colhead{$\rho_{\star}/R_{\rm grp}$} \\ & \colhead{($\degr$)} & \colhead{($\degr$)} & & & \colhead{($L^*$)} & & \colhead{(kpc)} & \colhead{(km\,s$^{-1}$)} & \colhead{(kpc)} }

\colnumbers
\startdata
12833 & $129.400^{+0.013}_{-0.013}$ & $44.282^{+0.010}_{-0.007}$ & $0.14584^{+0.00015}_{-0.00062}$ &  $50^{+19}_{-21}$ & $88.2^{+25.8}_{-27.0}$ & $14.44^{+0.41}_{-0.45}$ & $1395^{+637}_{-436}$ & $434^{+ 15}_{-119}$ & $1741^{+ 69}_{- 48}$ & $1.02^{+0.06}_{-0.05}$ & $0.78^{+0.06}_{-0.20}$ & $1.33^{+0.47}_{-0.29}$ \\
16803 & $234.994^{+0.005}_{-0.026}$ & $-2.098^{+0.009}_{-0.011}$ & $0.14845^{+0.00017}_{-0.00009}$ &  $54^{+13}_{-18}$ & $94.9^{+24.1}_{-29.6}$ & $14.48^{+0.40}_{-0.45}$ & $1254^{+333}_{-175}$ & $445^{+ 57}_{- 76}$ &  $824^{+297}_{- 35}$ & $0.90^{+0.19}_{-0.07}$ & $0.80^{+0.18}_{-0.14}$ & $0.59^{+0.14}_{-0.06}$ \\
19670 & $150.178^{+0.028}_{-0.026}$ & $50.784^{+0.002}_{-0.007}$ & $0.13442^{+0.00042}_{-0.00072}$ &  $69^{+24}_{-21}$ & $87.0^{+22.0}_{-24.8}$ & $14.43^{+0.40}_{-0.43}$ & $1144^{+606}_{-154}$ & $494^{+ 76}_{- 91}$ & $1651^{+136}_{-158}$ & $0.90^{+0.09}_{-0.08}$ & $0.91^{+0.13}_{-0.11}$ & $1.18^{+0.54}_{-0.20}$ \\
25124 & $243.538^{+0.005}_{-0.015}$ & $26.687^{+0.002}_{-0.006}$ & $0.18576^{+0.00107}_{-0.00009}$ &  $30^{+10}_{-14}$ & $80.5^{+24.5}_{-19.1}$ & $14.40^{+0.42}_{-0.43}$ & $1380^{+242}_{-521}$ & $451^{+ 13}_{-143}$ & $1586^{+ 19}_{- 62}$ & $0.95^{+0.11}_{-0.09}$ & $0.77^{+0.20}_{-0.15}$ & $1.24^{+0.32}_{-0.35}$ \\
32123 & $203.047^{+0.020}_{-0.012}$ & $45.425^{+0.008}_{-0.039}$ & $0.15958^{+0.00031}_{-0.00014}$ &  $19^{+17}_{- 0}$ & $32.8^{+31.7}_{- 0.7}$ & $14.22^{+0.59}_{-0.33}$ &  $837^{+667}_{-  0}$ & $316^{+129}_{-  3}$ & $1942^{+  1}_{-366}$ & $0.97^{+0.12}_{-0.12}$ & $0.88^{+0.10}_{-0.18}$ & $2.00^{+0.50}_{-0.92}$ \\
36001 & $157.056^{+0.019}_{-0.003}$ & $21.339^{+0.018}_{-0.031}$ & $0.18760^{+0.00001}_{-0.00012}$ &   $6^{+ 4}_{- 1}$ & $20.0^{+12.3}_{- 2.1}$ & $13.79^{+0.53}_{-0.36}$ &  $728^{+334}_{- 74}$ & $194^{+110}_{- 24}$ &  $297^{+179}_{-150}$ & $0.92^{+0.14}_{-0.11}$ & $0.57^{+0.15}_{-0.08}$ & $0.36^{+0.46}_{-0.23}$ \\
44564 & $208.170^{+0.013}_{-0.012}$ & $36.275^{+0.005}_{-0.006}$ & $0.14609^{+0.00006}_{-0.00009}$ & $114^{+39}_{-38}$ &  $132^{+58.0}_{-49.3}$ & $14.62^{+0.46}_{-0.50}$ & $1336^{+471}_{-200}$ & $547^{+ 84}_{- 91}$ & $1545^{+126}_{- 82}$ & $0.87^{+0.13}_{-0.06}$ & $0.88^{+0.19}_{-0.16}$ & $0.99^{+0.60}_{-0.32}$ \\
44565 & $208.171^{+0.012}_{-0.012}$ & $36.277^{+0.004}_{-0.007}$ & $0.14611^{+0.00006}_{-0.00006}$ & $115^{+38}_{-35}$ &  $134^{+57.0}_{-44.2}$ & $14.62^{+0.45}_{-0.47}$ & $1336^{+471}_{-183}$ & $557^{+ 74}_{- 82}$ & $1535^{+115}_{- 75}$ & $0.88^{+0.12}_{-0.07}$ & $0.89^{+0.17}_{-0.16}$ & $1.00^{+0.55}_{-0.32}$ \\
44726 & $226.214^{+0.001}_{-0.066}$ & $29.833^{+0.007}_{-0.064}$ & $0.15233^{+0.00003}_{-0.00104}$ &   $7^{+15}_{- 0}$ & $11.0^{+18.1}_{- 0.4}$ & $13.53^{+0.70}_{-0.40}$ &  $188^{+852}_{- 10}$ & $226^{+213}_{-  4}$ & $1331^{+493}_{-  6}$ & $0.37^{+0.65}_{-0.10}$ & $0.94^{+0.19}_{-0.14}$ & $2.03^{+0.83}_{-0.54}$ \\
44739 & $229.822^{+0.011}_{-0.019}$ & $28.342^{+0.058}_{-0.006}$ & $0.11883^{+0.00017}_{-0.00040}$ &  $62^{+44}_{-11}$ & $65.0^{+37.0}_{- 6.7}$ & $14.31^{+0.49}_{-0.35}$ & $1144^{+658}_{-151}$ & $348^{+ 96}_{- 63}$ & $2394^{+ 28}_{-416}$ & $1.01^{+0.06}_{-0.09}$ & $0.71^{+0.15}_{-0.10}$ & $1.98^{+0.62}_{-0.85}$ \\
44858 & $230.084^{+0.035}_{-0.032}$ & $28.873^{+0.008}_{-0.008}$ & $0.12729^{+0.00082}_{-0.00024}$ &  $80^{+23}_{-28}$ &  $103^{+22.0}_{-23.0}$ & $14.52^{+0.39}_{-0.42}$ & $1313^{+248}_{-419}$ & $507^{+ 74}_{-134}$ & $2289^{+172}_{- 31}$ & $0.87^{+0.05}_{-0.05}$ & $0.87^{+0.05}_{-0.12}$ & $1.61^{+0.78}_{-0.43}$ \\
50433 & $157.732^{+0.003}_{-0.003}$ & $30.845^{+0.006}_{-0.001}$ & $0.13630^{+0.00006}_{-0.00006}$ &  $25^{+ 7}_{- 4}$ & $39.7^{+ 8.1}_{- 3.0}$ & $14.10^{+0.38}_{-0.33}$ &  $600^{+234}_{- 38}$ & $343^{+ 27}_{- 57}$ & $1781^{+ 11}_{- 47}$ & $0.62^{+0.12}_{-0.08}$ & $0.82^{+0.11}_{-0.14}$ & $1.73^{+0.43}_{-0.42}$ \\
\enddata

\tablecomments{Tabulated values are the median from the Monte Carlo realizations, and ranges are the 16th and 84th percentile values. $M_{\rm grp}$ has units of $M_{\Sun}$, and $\rho_{\star}$ is the impact parameter between the group center and the COS sight line.}

\end{deluxetable*}
\end{rotatetable}

%% file: tab_12833.tex
\startlongtable
\begin{deluxetable*}{lcccccccc}

\tablecaption{Members of Group 12833
\label{tab:12833} }

\tablehead{ \colhead{Name} & 
            \colhead{RA} & 
            \colhead{Dec} & 
            \colhead{$z_{\rm gal}$} & 
            \colhead{$L_{\rm gal}$} & 
            \colhead{Type} & 
            \colhead{$\Delta\rho$} & 
            \colhead{$\Delta v$} & 
            \colhead{$f_{\rm mem}$} \\ 
            & 
            & 
            & 
            & 
            \colhead{($L^*$)} & 
            &
            \colhead{(kpc)} & 
            \colhead{($\mathrm{km\,s}^{-1}$)} & 
            \colhead{(\%)} } 

\colnumbers
\startdata
    rbs711\_126\_554 & 129.398389 & 44.325363 & 0.14572 &    0.575 & A &  323 & $ -316$ &  99.6 \\
    rbs711\_135\_607 & 129.383140 & 44.296795 & 0.14519 &    0.998 & A &  265 & $ -455$ &  99.5 \\
 1237654652565651620 & 129.394131 & 44.314455 & 0.14505 &     1.66 & A &  261 & $ -492$ &  99.5 \\
    rbs711\_123\_564 & 129.410462 & 44.330803 & 0.14518 &     1.05 & C &  340 & $ -457$ &  99.5 \\
    rbs711\_143\_470 & 129.334488 & 44.319729 & 0.14485 &     1.44 & A &  631 & $ -544$ &  99.4 \\
    rbs711\_136\_895 & 129.444666 & 44.229919 & 0.14614 &    0.519 & C &  627 & $ -207$ &  99.3 \\
    rbs711\_132\_770 & 129.432936 & 44.267044 & 0.14492 &     1.20 & A &  272 & $ -525$ &  99.3 \\
    rbs711\_117\_767 & 129.491115 & 44.313179 & 0.14454 &    0.757 & A &  484 & $ -625$ &  99.3 \\
    rbs711\_124\_864 & 129.492903 & 44.271641 & 0.14459 &     1.20 & A &  517 & $ -612$ &  99.3 \\
    rbs711\_146\_793 & 129.382281 & 44.236610 & 0.14369 &    0.584 & C &  610 & $ -847$ &  99.3 \\
 1237654652565586007 & 129.396584 & 44.231755 & 0.14514 &     4.72 & A &  618 & $ -468$ &  99.3 \\
 1237654652565651637 & 129.497242 & 44.248864 & 0.14458 &     8.14 & A &  658 & $ -613$ &  99.2 \\
 1237654652565651551 & 129.520232 & 44.348909 & 0.14386 &     2.15 & C &  818 & $ -802$ &  96.8 \\
 1237654652565651712 & 129.446442 & 44.379164 & 0.14691 &     6.55 & A &  796 & $   -5$ &  95.9 \\
    rbs711\_100\_604 & 129.474092 & 44.395351 & 0.14496 &    0.669 & C &  991 & $ -515$ &  95.7 \\
    rbs711\_154\_921 & 129.364843 & 44.192634 & 0.14592 &    0.482 & C & 1032 & $ -264$ &  95.0 \\
    rbs711\_170\_421 & 129.266596 & 44.317944 & 0.14545 &     5.19 & C & 1062 & $ -387$ &  95.0 \\
    rbs711\_172\_401 & 129.262476 & 44.323036 & 0.14467 &    0.803 & A & 1099 & $ -591$ &  94.9 \\
    rbs711\_140\_550 & 129.357548 & 44.303642 & 0.14728 &    0.714 & C &  442 & $   91$ &  94.4 \\
    rbs711\_121\_315 & 129.340653 & 44.378765 & 0.14738 &     1.03 & C &  950 & $  118$ &  94.3 \\
 1237655108362108987 & 129.285682 & 44.320084 & 0.14751 &     6.03 & A &  942 & $  151$ &  94.3 \\
 1237654652565586095 & 129.344588 & 44.180009 & 0.14761 &     3.71 & A & 1194 & $  177$ &  93.4 \\
 1237654652565586098 & 129.366472 & 44.166001 & 0.14542 &     2.68 & C & 1262 & $ -396$ &  92.6 \\
   rbs711\_121\_1082 & 129.569020 & 44.242108 & 0.14424 &     1.46 & C & 1093 & $ -703$ &  92.6 \\
    rbs711\_165\_990 & 129.321227 & 44.164215 & 0.14478 &     1.15 & A & 1397 & $ -562$ &  92.1 \\
 1237654652565586126 & 129.430232 & 44.151193 & 0.14501 &     2.17 & C & 1343 & $ -501$ &  92.1 \\
 1237654652565586160 & 129.341944 & 44.282553 & 0.14802 &     2.03 & C &  552 & $  285$ &  88.2 \\
    rbs711\_151\_952 & 129.384227 & 44.188805 & 0.14800 &     1.15 & A & 1025 & $  280$ &  88.1 \\
    rbs711\_203\_553 & 129.183812 & 44.286568 & 0.14586 &    0.919 & C & 1595 & $ -280$ &  87.4 \\
   rbs711\_122\_1207 & 129.603338 & 44.217339 & 0.14451 &    0.691 & A & 1406 & $ -633$ &  81.5 \\
 1237655107825238285 & 129.602299 & 44.217301 & 0.14477 &     1.81 & A & 1400 & $ -564$ &  81.5 \\
 1237655108362109124 & 129.343638 & 44.390149 & 0.14844 &     2.53 & C & 1028 & $  394$ &  77.4 \\
     rbs711\_97\_567 & 129.463434 & 44.409596 & 0.14857 &    0.831 & C & 1097 & $  429$ &  73.2 \\
   rbs711\_157\_1246 & 129.386401 & 44.102791 & 0.14504 &     1.30 & C & 1810 & $ -494$ &  71.1 \\
   rbs711\_152\_1014 & 129.385972 & 44.170589 & 0.14877 &     1.09 & A & 1188 & $  481$ &  70.2 \\
 1237655107825303599 & 129.649314 & 44.232407 & 0.14367 &     3.64 & E & 1620 & $ -851$ &  69.2 \\
 1237654652565717293 & 129.560989 & 44.425002 & 0.14736 &     2.03 & C & 1517 & $  112$ &  69.2 \\
   rbs711\_143\_1352 & 129.499970 & 44.105369 & 0.14697 &     1.26 & A & 1842 & $   10$ &  68.9 \\
   rbs711\_126\_1339 & 129.614897 & 44.172196 & 0.14399 &    0.580 & E & 1719 & $ -768$ &  68.9 \\
 1237655108362174725 & 129.466743 & 44.459596 & 0.14737 &     1.98 & A & 1556 & $  114$ &  68.9 \\
   rbs711\_109\_1207 & 129.668913 & 44.289364 & 0.14334 &    0.459 & C & 1641 & $ -938$ &  67.4 \\
     rbs711\_73\_243 & 129.331584 & 44.461864 & 0.14702 &    0.580 & E & 1665 & $   24$ &  67.2 \\
 1237654652565651690 & 129.426343 & 44.374180 & 0.14927 &     2.69 & A &  734 & $  612$ &  63.3 \\
 1237654652565520483 & 129.338620 & 44.089159 & 0.14882 &    0.160 & C & 2001 & $  493$ &  61.2 \\
     rbs711\_91\_910 & 129.599690 & 44.429485 & 0.14513 &    0.529 & C & 1716 & $ -471$ &  61.1 \\
    rbs711\_214\_724 & 129.125490 & 44.251896 & 0.14891 &    0.783 & A & 2023 & $  518$ &  55.4 \\
    rbs711\_152\_955 & 129.377389 & 44.186016 & 0.14973 &    0.869 & A & 1062 & $  732$ &  55.3 \\
     rbs711\_87\_866 & 129.581308 & 44.454063 & 0.14640 &     1.13 & C & 1815 & $ -139$ &  54.7 \\
   rbs711\_117\_1407 & 129.693074 & 44.209873 & 0.14552 &    0.714 & A & 1970 & $ -369$ &  52.9 \\
 1237655108362174700 & 129.354902 & 44.492543 & 0.14700 &     1.93 & C & 1890 & $   18$ &  52.1 \\
\enddata

\tablecomments{Column~6 lists the galaxy's spectral type: ``E'' for emission-line galaxies, ``A'' for absorption-line galaxies, and ``C'' for composite galaxies that show both emission and absorption. $\Delta\rho$ and $\Delta v$ are calculated with respect to the original SDSS group centers in \autoref{tab:berlind}; $f_{\rm mem}$ is the percentage of the Monte Carlo trials in which a galaxy was identified as a group member.}

\end{deluxetable*}

%% file: tab_groups.tex
\begin{deluxetable*}{lccccccccccccc}

\tablecaption{Adopted Group Properties
\label{tab:groups}}

\tablehead{ \colhead{Group} & \colhead{RA} & \colhead{Dec} & \colhead{$z_{\rm grp}$} & \colhead{$N_{\rm grp}$} & \colhead{$\log{M_{\rm grp}}$} & \colhead{\Rgrp} & \colhead{$\sigma_{\rm grp}$} &  \colhead{$R_{\rm obs}$} & \colhead{$\sigma_{\rm obs}$} & \colhead{$\rho_{\star}/R_{\rm grp}$} & \colhead{$\Delta\rho$} & \colhead{$\Delta v$} & \colhead{$f_{\rm spiral}$} \\ & \colhead{($\degr$)} & \colhead{($\degr$)} & & & & \colhead{(kpc)} & \colhead{(\kms)}  & \colhead{(kpc)} & \colhead{(\kms)} & & \colhead{(kpc)}  & \colhead{(\kms)} & \colhead{(\%)}}

\colnumbers
\startdata
12833 & 129.413 &  44.282 & 0.14589 &  50 & 14.45 & 1352 & 547 & 1396 & 444 & 1.32 & 137 & --271 & 43.9 \\
16803 & 234.995 & --2.093 & 0.14839 &  55 & 14.49 & 1394 & 563 & 1297 & 455 & 0.59 & 507 &     1 & 49.1 \\
19670 & 150.178 &  50.785 & 0.13370 &  48 & 14.30 & 1205 & 487 &  991 & 404 & 1.36 & 128 & --193 & 50.0 \\
25124 & 243.538 &  26.687 & 0.18569 &  29 & 14.40 & 1301 & 526 & 1167 & 448 & 1.22 & 704 & --111 & 44.8 \\
32123 & 203.047 &  45.432 & 0.15958 &  19 & 14.01 &  965 & 390 &  837 & 313 & 2.01 & 268 &  --34 & 50.0 \\
36001 & 157.053 &  21.357 & 0.18761 &   6 & 13.80 &  821 & 332 &  729 & 170 & 0.36 & 638 &  --69 & 66.7 \\
44564 & 208.170 &  36.280 & 0.14610 & 116 & 14.63 & 1553 & 628 & 1337 & 567 & 1.00 & 402 &   271 & 37.5 \\
44565 & 208.170 &  36.280 & 0.14610 & 116 & 14.63 & 1553 & 628 & 1337 & 567 & 0.99 & 142 & --823 & 37.5 \\
44726 & 226.215 &  29.834 & 0.15236 &   7 & 13.53 &  667 & 270 &  178 & 223 & 1.98 & 527 &    74 & 40.0 \\
44739 & 229.828 &  28.336 & 0.11887 &  59 & 14.30 & 1205 & 487 & 1135 & 336 & 2.00 & 604 &   128 & 68.0 \\
44858 & 230.052 &  28.878 & 0.12707 &  99 & 14.59 & 1506 & 609 & 1546 & 569 & 1.49 & 364 &  --88 & 50.0 \\
50433 & 157.732 &  30.845 & 0.13624 &  25 & 14.10 & 1034 & 418 &  600 & 338 & 1.73 & 187 &    67 & 31.3 \\
\enddata

\tablecomments{$M_{\rm grp}$ has units of $M_{\Sun}$, and is calculated assuming $\Upsilon_{\rm grp}=2.5$. $\rho_{\star}$ is the impact parameter between the adopted group center and the COS sight line. $\Delta\rho$ and $\Delta v$ are offsets from the SDSS group centroids in \autoref{tab:berlind}. The final column lists an estimate of the group's spiral fraction derived from the spectral types of the adopted group members (see text for details).}

\vspace{-2em}

\end{deluxetable*}

%% file: tab_cf.tex
\begin{deluxetable}{ccl}

\tablecaption{Covering Fraction of Group Absorption
\label{tab:cf}}

\tablehead{
	\colhead{$\rho_\star/\Rgrp$} & 
	\colhead{Cov. Frac.} & 
	\colhead{Notes}
	}
	\startdata
	0.0-0.5 & $1\,/\,1$ & All have \OVI \\
        0.5-1.0 & $1\,/\,2$ & Metal-free \\
        1.0-1.5 & $2\,/\,2$ & One metal-free, one low-ions only \\
        1.5-2.0 & $3\,/\,5$ & All have \OVI \\
\enddata 

\vspace{-2em}

\end{deluxetable}

%% file: tab_abs.tex
\startlongtable
\begin{deluxetable}{ccccccc}

\tablecolumns{7}

\tablecaption{Absorption lines measured within $2.5\sigma_0$ of SDSS group velocities
\label{tab:abs}}

\tablehead{
	\colhead{Species} &
	\colhead{$z_{\rm abs}$} &
	\colhead{$\Delta v_{\rm abs}$} &
	\colhead{S.L.} &
	\colhead{${\rm EW}_r$} &
	\colhead{$b$} &
	\colhead{$\log N$} \\
	\colhead{} &
	\colhead{} &
	\colhead{(\kms)} &
	\colhead{($\sigma$)} &
	\colhead{(m\AA)} &
	\colhead{(\kms)} &
	\colhead{($N$ in cm$^{-2}$)} 
	}
\startdata 
\sidehead{RBS711, Group \#12833; $z_0= 0.14693$}
  Ly$\alpha$ 1215 &  0.14743 &$    129 $& 42.8&$  333\pm 33$&$  23\pm  2$&$  14.65\pm0.19$ \\ [-1ex]
  Ly$\beta$ 1025  &  0.14743 &$    131 $& 31.7&$  241\pm 14$&$  24\pm  1$&$  15.03\pm0.07$ \\ [-1ex]
  Si\,III 1206    &  0.14742 &$    127 $&  6.9&$   42\pm 10$&$  15\pm  4$&$  12.37\pm0.09$ \\
\sidehead{SDSS\,J15403-0205, Group \#16803; $z_0= 0.14839$}
  Ly$\alpha$ 1215 &  0.14497 &$   -893 $&  7.1&$  110\pm 17$&$  27\pm  5$&$  13.41\pm0.07$ \\
\sidehead{SBS0956+509, Group \#19670; $z_0= 0.13443$}
  Ly$\alpha$ 1215 &  0.13388 &$   -146 $& 30.7&$  335\pm 26$&$  25\pm  2$&$  14.49\pm0.12$ \\ [-1ex]
  Ly$\alpha$ 1215 &  0.13848 &$   1069 $&  5.3&$   51\pm  9$&$  38\pm  8$&$  13.01\pm0.07$ \\ [-1ex]
  Ly$\beta$ 1025  &  0.13387 &$   -146 $& 20.8&$  233\pm 15$&$  26\pm  2$&$  14.90\pm0.05$ \\
\sidehead{QSOB1612+266, Group \#25124; $z_0= 0.18613$}
  Ly$\alpha$ 1215 &  0.18529 &$   -212 $& 49.7&$  705\pm 77$&$  37\pm  3$&$  15.81\pm0.33$ \\ [-1ex]
  Ly$\alpha$ 1215 &  0.18569 &$   -110 $&  9.0&$  120\pm 10$&$  32\pm  4$&$  13.44\pm0.05$ \\ [-1ex]
  O\,VI 1032      &  0.18540 &$   -184 $& 13.0&$  248\pm 90$&$  53\pm  4$&$  14.49\pm0.05$ \\ [-1ex]
  O\,VI 1038      &  0.18540 &$   -184 $& 13.0&$  160\pm 21$&$  53\pm  4$&$  14.49\pm0.05$ \\ [-1ex]
  Si\,III 1206    &  0.18517 &$   -242 $&  9.7&$  150\pm 25$&$  59\pm 10$&$  12.92\pm0.08$ \\ [-1ex]
  Si\,III 1206    &  0.18531 &$   -208 $& 32.7&$  364\pm 22$&$  30\pm  2$&$  13.80\pm0.07$ \\ [-1ex]
  C\,III 977      &  0.18520 &$   -234 $&  9.5&$  289\pm129$&$  38\pm 11$&$  13.98\pm0.45$ \\ [-1ex]
  C\,III 977      &  0.18537 &$   -191 $& 13.7&$  323\pm 83$&$  21\pm  4$&$  15.32\pm0.72$ \\ [-1ex]
  N\,III 989      &  0.18543 &$   -176 $& 14.8&$  215\pm 21$&$  34\pm  3$&$  14.54\pm0.05$ \\ [-1ex]
  Si\,II 1190     &  0.18529 &$   -212 $& 13.9&$  166\pm 15$&$  33\pm  2$&$  13.79\pm0.04$ \\ [-1ex]
  Si\,II 1193     &  0.18531 &$   -207 $& 16.5&$  199\pm103$&$  28\pm  1$&$  13.65\pm0.02$ \\
\sidehead{SDSSJ133300.83+451809.0, Group \#32123; $z_0= 0.15971$}
  Ly$\alpha$ 1215 &  0.15874 &$   -251 $& 37.1&$  643\pm 67$&$  41\pm  4$&$  15.10\pm0.22$ \\ [-1ex]
  Ly$\alpha$ 1215 &  0.16081 &$    283 $& 15.7&$  234\pm 20$&$  22\pm  2$&$  14.02\pm0.07$ \\ [-1ex]
  Ly$\alpha$ 1215 &  0.16142 &$    442 $&  4.5&$   95\pm 30$&$  49\pm 13$&$  13.29\pm0.11$ \\ [-1ex]
  Ly$\beta$ 1025  &  0.15860 &$   -285 $& 10.1&$  136\pm 32$&$  15\pm  4$&$  14.67\pm0.15$ \\ [-1ex]
  Ly$\beta$ 1025  &  0.15887 &$   -218 $& 16.2&$  295\pm 33$&$  32\pm  4$&$  15.03\pm0.07$ \\ [-1ex]
  Ly$\beta$ 1025  &  0.16090 &$    306 $&  4.8&$   46\pm 15$&$  11\pm  6$&$  13.94\pm0.13$ \\ [-1ex]
  O\,VI 1032      &  0.16088 &$    302 $&  2.3&$   36\pm 15$&$  29\pm 13$&$  13.55\pm0.25$ \\ [-1ex]
  O\,VI 1038      &  0.16082 &$    288 $&  3.0&$   44\pm 35$&$  33\pm 16$&$  13.89\pm0.28$ \\ [-1ex]
  Si\,III 1206    &  0.15883 &$   -227 $&  6.0&$   77\pm 16$&$  15\pm  4$&$  12.69\pm0.09$ \\ [-1ex]
  Si\,III 1206    &  0.16079 &$    278 $&  2.9&$   24\pm 11$&$   5\pm  0$&$  12.19\pm0.21$ \\
\tablebreak
\sidehead{SDSSJ10282+2119, Group \#36001; $z_0= 0.18788$}
  Ly$\alpha$ 1215 &  0.18429 &$   -906 $& 35.3&$  507\pm 32$&$  38\pm  2$&$  14.68\pm0.09$ \\ [-1ex]
  Ly$\alpha$ 1215 &  0.18494 &$   -743 $&  5.5&$   46\pm 25$&$  17\pm 11$&$  13.00\pm0.22$ \\ [-1ex]
  Ly$\alpha$ 1215 &  0.18527 &$   -659 $& 44.8&$  523\pm106$&$  28\pm  5$&$  15.56\pm0.62$ \\ [-1ex]
  Ly$\alpha$ 1215 &  0.18585 &$   -513 $& 27.9&$  367\pm 43$&$  26\pm  2$&$  14.63\pm0.18$ \\ [-1ex]
  Ly$\gamma$ 972  &  0.18526 &$   -660 $& 18.7&$  344\pm 12$&$  18\pm  0$&$  18.00\pm0.00$ \\ [-1ex]
  Ly$\gamma$ 972  &  0.18592 &$   -494 $&  8.5&$  172\pm 28$&$  23\pm  5$&$  15.16\pm0.09$ \\ [-1ex]
  O\,VI 1032      &  0.18436 &$   -887 $&  4.1&$   42\pm 36$&$  32\pm 10$&$  13.56\pm0.43$ \\ [-1ex]
  O\,VI 1032      &  0.18530 &$   -651 $&  3.6&$   30\pm  1$&$  23\pm  9$&$  13.43\pm0.13$ \\ [-1ex]
  O\,VI 1038      &  0.18432 &$   -898 $&  3.6&$   24\pm  2$&$  18\pm  8$&$  13.63\pm0.17$ \\ [-1ex]
  Si\,III 1206    &  0.18527 &$   -658 $&  8.1&$   90\pm 21$&$  29\pm  4$&$  12.71\pm0.09$ \\ [-1ex]
  C\,III 977      &  0.18532 &$   -646 $&  3.8&$   65\pm 24$&$  22\pm  7$&$  13.10\pm0.14$ \\ [-1ex]
  C\,III 977      &  0.18604 &$   -464 $&  4.4&$  111\pm 38$&$  66\pm 14$&$  13.29\pm0.07$ \\
\sidehead{CSO1022, Groups \#44564, 44565; $z_0=0.14506,0.14925$ [no absorbers]}
\sidehead{CSO1080, Group \#44726; $z_0= 0.15208$ [no absorbers]}
\sidehead{FBQSJ1519+2838, Group \#44739; $z_0= 0.11839$}
  Ly$\alpha$ 1215 &  0.11537 &$   -808 $& 46.6&$  497\pm 40$&$  32\pm  2$&$  14.98\pm0.21$ \\ [-1ex]
  Ly$\alpha$ 1215 &  0.11706 &$   -357 $& 36.7&$  436\pm 39$&$  30\pm  2$&$  14.78\pm0.17$ \\ [-1ex]
  Ly$\alpha$ 1215 &  0.11766 &$   -195 $& 35.6&$  509\pm 48$&$  37\pm  3$&$  14.72\pm0.11$ \\ [-1ex]
  Ly$\alpha$ 1215 &  0.11795 &$   -118 $& 17.2&$  191\pm 26$&$  21\pm  3$&$  13.84\pm0.09$ \\ [-1ex]
  Ly$\beta$ 1025  &  0.11535 &$   -814 $& 19.9&$  325\pm 30$&$  32\pm  3$&$  15.20\pm0.12$ \\ [-1ex]
  Ly$\beta$ 1025  &  0.11704 &$   -362 $& 26.2&$  290\pm 54$&$  25\pm  4$&$  15.34\pm0.28$ \\ [-1ex]
  Ly$\beta$ 1025  &  0.11762 &$   -205 $& 20.8&$  304\pm 30$&$  31\pm  3$&$  15.11\pm0.09$ \\ [-1ex]
  Ly$\beta$ 1025  &  0.11793 &$   -123 $&  6.1&$   77\pm 21$&$  22\pm  7$&$  14.13\pm0.11$ \\ [-1ex]
  O\,VI 1032      &  0.11551 &$   -773 $&  3.2&$   26\pm 19$&$  10\pm 10$&$  13.39\pm0.33$ \\ [-1ex]
  O\,VI 1032      &  0.11766 &$   -197 $&  3.6&$   49\pm 27$&$  40\pm 15$&$  13.63\pm0.18$ \\ [-1ex]
  Si\,III 1206    &  0.11526 &$   -839 $&  4.9&$   85\pm 31$&$  64\pm 15$&$  12.64\pm0.12$ \\ [-1ex]
  Si\,III 1206    &  0.11709 &$   -348 $&  4.6&$   30\pm 29$&$   5\pm  0$&$  12.32\pm0.12$ \\ [-1ex]
  Si\,III 1206    &  0.11765 &$   -197 $&  5.2&$   53\pm 12$&$  17\pm  4$&$  12.48\pm0.09$ \\
\sidehead{FBQSJ1519+2838, Group \#44858; $z_0= 0.12740$}
  Ly$\alpha$ 1215 &  0.12551 &$   -501 $& 61.2&$  581\pm166$&$  29\pm  6$&$  15.94\pm0.96$ \\ [-1ex]
  Ly$\alpha$ 1215 &  0.12567 &$   -460 $& 32.4&$  372\pm189$&$  43\pm 13$&$  14.12\pm0.49$ \\ [-1ex]
  Ly$\beta$ 1025  &  0.12550 &$   -504 $& 30.8&$  365\pm 75$&$  30\pm  5$&$  15.57\pm0.37$ \\ [-1ex]
  Ly$\beta$ 1025  &  0.12568 &$   -457 $& 20.6&$  211\pm147$&$  21\pm 28$&$  15.00\pm2.36$ \\ [-1ex]
  O\,VI 1032      &  0.12571 &$   -449 $&  9.8&$   93\pm 12$&$  21\pm  3$&$  14.01\pm0.06$ \\ [-1ex]
  O\,VI 1038      &  0.12571 &$   -450 $&  5.1&$   52\pm 15$&$  20\pm  6$&$  14.00\pm0.10$ \\
\sidehead{FBQSJ103059.1+310255, Group \#50433; $z_0= 0.13599$ [no absorbers]}
\enddata 

\tablecomments{Column~3 lists the velocity offset between the absorber and the SDSS group redshift in \autoref{tab:berlind}, and Column~4 lists the significance level of the detected absorption.}

\tablenotetext{a}{\OVI\ in this sight line is obscured by a partial LLS at $z=0.322$.}

\end{deluxetable}

%% file: tab_16803.tex
\startlongtable
\begin{deluxetable*}{lcccccccc}

\tablecaption{Members of Group 16803
\label{tab:16803} }

\tablehead{ \colhead{Name} & 
            \colhead{RA} & 
            \colhead{Dec} & 
            \colhead{$z_{\rm gal}$} & 
            \colhead{$L_{\rm gal}$} & 
            \colhead{Type} & 
            \colhead{$\Delta\rho$} & 
            \colhead{$\Delta v$} & 
            \colhead{$f_{\rm mem}$} \\ 
            & 
            & 
            & 
            & 
            \colhead{($L^*$)} & 
            &
            \colhead{(kpc)} & 
            \colhead{($\mathrm{km\,s}^{-1}$)} & 
            \colhead{(\%)} } 

\colnumbers
\startdata
 1237655498141270296 & 234.964227 & --2.038841 & 0.14815 &     6.80 & A &  158 & $  -63$ &  99.7 \\
  sdss1540\_275\_503 & 234.941540 & --2.074340 & 0.14812 &     1.03 & A &  488 & $  -70$ &  99.7 \\
 1237655498141270269 & 234.949579 & --2.033421 & 0.14790 &     2.01 & C &  302 & $ -127$ &  99.6 \\
 1237655691945836795 & 234.933090 & --2.014718 & 0.14841 &     2.31 & C &  510 & $    5$ &  99.6 \\
  sdss1540\_254\_416 & 234.970307 & --2.118609 & 0.14785 &     1.03 & A &  740 & $ -141$ &  99.3 \\
 1237655498141270424 & 235.027343 & --2.057876 & 0.14837 &     3.46 & C &  465 & $   -4$ &  99.3 \\
 1237655498141270391 & 235.007329 & --2.058280 & 0.14952 &     4.89 & A &  299 & $  295$ &  99.3 \\
  sdss1540\_291\_657 & 234.909754 & --2.021402 & 0.14704 &     1.12 & A &  691 & $ -352$ &  97.8 \\
\enddata

\tablecomments{Column~6 lists the galaxy's spectral type: ``E'' for emission-line galaxies, ``A'' for absorption-line galaxies, and ``C'' for composite galaxies that show both emission and absorption. $\Delta\rho$ and $\Delta v$ are calculated with respect to the original SDSS group centers in \autoref{tab:berlind}; $f_{\rm mem}$ is the percentage of the Monte Carlo trials in which a galaxy was identified as a group member.}

\end{deluxetable*}

%% file: tab_19670.tex
\startlongtable
\begin{deluxetable*}{lcccccccc}

\tablecaption{Members of Group 19670
\label{tab:19670} }

\tablehead{ \colhead{Name} & 
            \colhead{RA} & 
            \colhead{Dec} & 
            \colhead{$z_{\rm gal}$} & 
            \colhead{$L_{\rm gal}$} & 
            \colhead{Type} & 
            \colhead{$\Delta\rho$} & 
            \colhead{$\Delta v$} & 
            \colhead{$f_{\rm mem}$} \\ 
            & 
            & 
            & 
            & 
            \colhead{($L^*$)} &
            &
            \colhead{(kpc)} & 
            \colhead{($\mathrm{km\,s}^{-1}$)} & 
            \colhead{(\%)} } 

\colnumbers
\startdata
    sbs0956\_87\_726 & 150.199427 & 50.767933 & 0.13361 &     2.71 & A &  229 & $ -217$ & 100.0 \\
    sbs0956\_84\_882 & 150.262799 & 50.792747 & 0.13341 &    0.743 & C &  363 & $ -270$ & 100.0 \\
    sbs0956\_80\_899 & 150.262213 & 50.814877 & 0.13337 &    0.601 & C &  401 & $ -280$ & 100.0 \\
 1237657632190627883 & 150.159170 & 50.764552 & 0.13344 &     4.70 & C &  328 & $ -263$ & 100.0 \\
    sbs0956\_80\_758 & 150.202131 & 50.805466 & 0.13421 &    0.408 & C &  101 & $  -58$ & 100.0 \\
    sbs0956\_87\_700 & 150.188212 & 50.766033 & 0.13455 &    0.312 & A &  249 & $   32$ & 100.0 \\
    sbs0956\_85\_725 & 150.197110 & 50.776314 & 0.13479 &    0.748 & A &  156 & $   95$ &  99.9 \\
   sbs0956\_86\_1018 & 150.326143 & 50.779293 & 0.13377 &    0.392 & A &  721 & $ -174$ &  99.9 \\
\enddata

\tablecomments{Column~6 lists the galaxy's spectral type: ``E'' for emission-line galaxies, ``A'' for absorption-line galaxies, and ``C'' for composite galaxies that show both emission and absorption. $\Delta\rho$ and $\Delta v$ are calculated with respect to the original SDSS group centers in \autoref{tab:berlind}; $f_{\rm mem}$ is the percentage of the Monte Carlo trials in which a galaxy was identified as a group member.}

\end{deluxetable*}

%% file: tab_25124.tex
\begin{deluxetable*}{lcccccccc}

\tablecaption{Members of Group 25124
\label{tab:25124} }

\tablehead{ \colhead{Name} & 
            \colhead{RA} & 
            \colhead{Dec} & 
            \colhead{$z_{\rm gal}$} & 
            \colhead{$L_{\rm gal}$} & 
            \colhead{Type} & 
            \colhead{$\Delta\rho$} & 
            \colhead{$\Delta v$} & 
            \colhead{$f_{\rm mem}$} \\ 
            & 
            & 
            & 
            & 
            \colhead{($L^*$)} &
            &
            \colhead{(kpc)} & 
            \colhead{($\mathrm{km\,s}^{-1}$)} & 
            \colhead{(\%)} } 

\colnumbers
\startdata
 1237662619189838119 & 243.512392 & 26.625859 & 0.18623 &     3.71 & A &  290 & $   26$ &  99.1 \\
       b1612\_6\_338 & 243.554077 & 26.640946 & 0.18664 &     1.84 & C &  224 & $  129$ &  99.1 \\
 1237662619189837900 & 243.540465 & 26.647870 & 0.18640 &     2.21 & C &  260 & $   69$ &  99.1 \\
     b1612\_337\_450 & 243.493767 & 26.664057 & 0.18719 &     2.84 & A &  651 & $  268$ &  99.1 \\
 1237662619189837901 & 243.541445 & 26.648761 & 0.18812 &     5.51 & A &  270 & $  504$ &  99.0 \\
 1237662619189837844 & 243.504132 & 26.692949 & 0.18728 &     8.19 & C &  855 & $  290$ &  99.0 \\
 1237662619189837842 & 243.504250 & 26.689543 & 0.18710 &     1.12 & A &  820 & $  244$ &  99.0 \\
      b1612\_22\_441 & 243.590040 & 26.662842 & 0.18718 &     1.24 & C &  655 & $  265$ &  98.8 \\
\enddata

\tablecomments{Column~6 lists the galaxy's spectral type: ``E'' for emission-line galaxies, ``A'' for absorption-line galaxies, and ``C'' for composite galaxies that show both emission and absorption. $\Delta\rho$ and $\Delta v$ are calculated with respect to the original SDSS group centers in \autoref{tab:berlind}; $f_{\rm mem}$ is the percentage of the Monte Carlo trials in which a galaxy was identified as a group member.}

\end{deluxetable*}

%% file: tab_32123.tex
\begin{deluxetable*}{lcccccccc}

\tablecaption{Members of Group 32123
\label{tab:32123} }

\tablehead{ \colhead{Name} & 
            \colhead{RA} & 
            \colhead{Dec} & 
            \colhead{$z_{\rm gal}$} & 
            \colhead{$L_{\rm gal}$} & 
            \colhead{Type} & 
            \colhead{$\Delta\rho$} & 
            \colhead{$\Delta v$} & 
            \colhead{$f_{\rm mem}$} \\ 
            & 
            & 
            & 
            & 
            \colhead{($L^*$)} &
            &
            \colhead{(kpc)} & 
            \colhead{($\mathrm{km\,s}^{-1}$)} & 
            \colhead{(\%)} } 

\colnumbers
\startdata
  sdss1333\_291\_546 & 203.063049 & 45.373932 & 0.15941 &    0.628 & C &  398 & $  -78$ &  99.3 \\
 1237661360761471098 & 203.099533 & 45.395529 & 0.15906 &     2.54 & A &  253 & $ -169$ &  99.3 \\
 1237661433777029221 & 203.117247 & 45.416476 & 0.15892 &     2.83 & C &  308 & $ -203$ &  99.2 \\
  sdss1333\_287\_495 & 203.072233 & 45.354637 & 0.15978 &    0.968 & A &  584 & $   18$ &  98.4 \\
 1237661433776963784 & 203.000758 & 45.420384 & 0.16007 &     4.51 & C &  515 & $   94$ &  97.9 \\
  sdss1333\_296\_790 & 202.996616 & 45.427727 & 0.15909 &    0.516 & C &  558 & $ -160$ &  97.9 \\
  sdss1333\_301\_364 & 203.142242 & 45.366959 & 0.15934 &    0.712 & A &  667 & $  -96$ &  97.2 \\
 1237661360761471128 & 203.142678 & 45.361986 & 0.15867 &     2.27 & C &  704 & $ -269$ &  97.2 \\
\enddata

\tablecomments{Column~6 lists the galaxy's spectral type: ``E'' for emission-line galaxies, ``A'' for absorption-line galaxies, and ``C'' for composite galaxies that show both emission and absorption. $\Delta\rho$ and $\Delta v$ are calculated with respect to the original SDSS group centers in \autoref{tab:berlind}; $f_{\rm mem}$ is the percentage of the Monte Carlo trials in which a galaxy was identified as a group member.}

\end{deluxetable*}

%% file: tab_36001.tex
\begin{deluxetable*}{lcccccccc}

\tablecaption{Members of Group 36001
\label{tab:36001} }

\tablehead{ \colhead{Name} & 
            \colhead{RA} & 
            \colhead{Dec} & 
            \colhead{$z_{\rm gal}$} & 
            \colhead{$L_{\rm gal}$} & 
            \colhead{Type} & 
            \colhead{$\Delta\rho$} & 
            \colhead{$\Delta v$} & 
            \colhead{$f_{\rm mem}$} \\ 
            & 
            & 
            & 
            & 
            \colhead{($L^*$)} &
            &
            \colhead{(kpc)} &            
            \colhead{($\mathrm{km\,s}^{-1}$)} & 
            \colhead{(\%)} } 

\colnumbers
\startdata
 1237667549268148342 & 157.086777 & 21.355513 & 0.18752 &     4.69 & C &  315 & $  -92$ &  92.8 \\
   sdss1028\_12\_229 & 157.073436 & 21.394539 & 0.18867 &    0.678 & E &  750 & $  199$ &  89.4 \\
 1237667549268082878 & 157.045047 & 21.375382 & 0.18783 &     8.49 & A &  807 & $  -12$ &  89.2 \\
 1237667536399564944 & 157.014096 & 21.304513 & 0.18756 &     3.29 & A & 1083 & $  -80$ &  85.9 \\
  sdss1028\_297\_284 & 156.986060 & 21.369490 & 0.18658 &     2.01 & C & 1359 & $ -328$ &  82.6 \\
  sdss1028\_169\_215 & 157.072120 & 21.273083 & 0.18759 &    0.816 & C &  839 & $  -73$ &  79.7 \\
\enddata

\tablecomments{Column~6 lists the galaxy's spectral type: ``E'' for emission-line galaxies, ``A'' for absorption-line galaxies, and ``C'' for composite galaxies that show both emission and absorption. $\Delta\rho$ and $\Delta v$ are calculated with respect to the original SDSS group centers in \autoref{tab:berlind}; $f_{\rm mem}$ is the percentage of the Monte Carlo trials in which a galaxy was identified as a group member.}

\end{deluxetable*}

%% file: tab_44564.tex
\startlongtable
\begin{deluxetable*}{lcccccccc}

\tablecaption{Members of Group 44564
\label{tab:44564} }

\tablehead{ \colhead{Name} & 
            \colhead{RA} & 
            \colhead{Dec} & 
            \colhead{$z_{\rm gal}$} & 
            \colhead{$L_{\rm gal}$} & 
            \colhead{Type} & 
            \colhead{$\Delta\rho$} & 
            \colhead{$\Delta v$} & 
            \colhead{$f_{\rm mem}$} \\ 
            & 
            & 
            & 
            & 
            \colhead{($L^*$)} &
            &
            \colhead{(kpc)} & 
            \colhead{($\mathrm{km\,s}^{-1}$)} & 
            \colhead{(\%)} } 

\colnumbers
\startdata
   cso1022\_252\_769 & 208.113456 & 36.265675 & 0.14554 &    0.399 & A &   65 & $  126$ & 100.0 \\
   cso1022\_249\_708 & 208.138490 & 36.262070 & 0.14610 &     9.66 & A &  134 & $  272$ & 100.0 \\
   cso1022\_247\_774 & 208.124285 & 36.244125 & 0.14490 &     1.46 & A &  162 & $  -42$ & 100.0 \\
   cso1022\_254\_807 & 208.096947 & 36.271324 & 0.14479 &    0.493 & A &  197 & $  -71$ & 100.0 \\
   cso1022\_246\_753 & 208.133912 & 36.242416 & 0.14443 &    0.327 & C &  202 & $ -165$ & 100.0 \\
   cso1022\_250\_858 & 208.090482 & 36.246872 & 0.14467 &     1.06 & A &  260 & $ -102$ & 100.0 \\
   cso1022\_262\_790 & 208.091225 & 36.305073 & 0.14620 &    0.311 & A &  457 & $  298$ & 100.0 \\
 1237662685753311429 & 208.083703 & 36.278130 & 0.14589 &     3.30 & A &  313 & $  218$ & 100.0 \\
\enddata

\tablecomments{Column~6 lists the galaxy's spectral type: ``E'' for emission-line galaxies, ``A'' for absorption-line galaxies, and ``C'' for composite galaxies that show both emission and absorption. $\Delta\rho$ and $\Delta v$ are calculated with respect to the original SDSS group centers in \autoref{tab:berlind}; $f_{\rm mem}$ is the percentage of the Monte Carlo trials in which a galaxy was identified as a group member.}

\end{deluxetable*}

%% file: tab_44565.tex
\startlongtable
\begin{deluxetable*}{lcccccccc}

\tablecaption{Members of Group 44565
\label{tab:44565} }

\tablehead{ \colhead{Name} & 
            \colhead{RA} & 
            \colhead{Dec} & 
            \colhead{$z_{\rm gal}$} & 
            \colhead{$L_{\rm gal}$} & 
            \colhead{Type} & 
            \colhead{$\Delta\rho$} & 
            \colhead{$\Delta v$} & 
            \colhead{$f_{\rm mem}$} \\ 
            & 
            & 
            & 
            & 
            \colhead{($L^*$)} &
            &
            \colhead{(kpc)} & 
            \colhead{($\mathrm{km\,s}^{-1}$)} & 
            \colhead{(\%)} } 

\colnumbers
\startdata
   cso1022\_253\_602 & 208.165812 & 36.285011 & 0.14765 &    0.387 & A &  124 & $ -417$ & 100.0 \\
   cso1022\_251\_710 & 208.134227 & 36.268360 & 0.14768 &    0.231 & A &  167 & $ -410$ & 100.0 \\
   cso1022\_249\_594 & 208.174067 & 36.275593 & 0.14799 &    0.488 & A &  181 & $ -329$ & 100.0 \\
   cso1022\_255\_581 & 208.169818 & 36.293491 & 0.14794 &    0.783 & A &  193 & $ -342$ & 100.0 \\
 1237662685753376885 & 208.157511 & 36.266795 & 0.14855 &     1.85 & A &  133 & $ -184$ & 100.0 \\
   cso1022\_255\_450 & 208.211760 & 36.306549 & 0.14868 &     1.71 & C &  527 & $ -149$ &  99.8 \\
   cso1022\_257\_431 & 208.216181 & 36.312786 & 0.14862 &    0.231 & A &  586 & $ -164$ &  99.8 \\
   cso1022\_265\_514 & 208.182921 & 36.328602 & 0.14738 &    0.674 & A &  520 & $ -488$ &  99.8 \\
\enddata

\tablecomments{Column~6 lists the galaxy's spectral type: ``E'' for emission-line galaxies, ``A'' for absorption-line galaxies, and ``C'' for composite galaxies that show both emission and absorption. $\Delta\rho$ and $\Delta v$ are calculated with respect to the original SDSS group centers in \autoref{tab:berlind}; $f_{\rm mem}$ is the percentage of the Monte Carlo trials in which a galaxy was identified as a group member.}

\end{deluxetable*}

%% file: tab_44726.tex
\begin{deluxetable*}{lcccccccc}

\tablecaption{Members of Group 44726
\label{tab:44726} }

\tablehead{ \colhead{Name} & 
            \colhead{RA} & 
            \colhead{Dec} & 
            \colhead{$z_{\rm gal}$} & 
            \colhead{$L_{\rm gal}$} & 
            \colhead{Type} & 
            \colhead{$\Delta\rho$} & 
            \colhead{$\Delta v$} & 
            \colhead{$f_{\rm mem}$} \\ 
            & 
            & 
            & 
            & 
            \colhead{($L^*$)} &
            &
            \colhead{(kpc)} & 
            \colhead{($\mathrm{km\,s}^{-1}$)} & 
            \colhead{(\%)} } 

\colnumbers
\startdata
   cso1080\_287\_491 & 226.215806 & 29.831860 & 0.15185 &     1.15 & A &  516 & $  -60$ &  85.9 \\
   cso1080\_292\_552 & 226.203217 & 29.850737 & 0.15222 &    0.525 & C &  616 & $   36$ &  85.8 \\
 1237664853111603254 & 226.214239 & 29.859650 & 0.15205 &     3.80 & A &  732 & $   -8$ &  85.7 \\
   cso1080\_283\_443 & 226.227436 & 29.819086 & 0.15296 &    0.786 & C &  511 & $  229$ &  85.6 \\
 1237664853111603398 & 226.213860 & 29.835203 & 0.15309 &     2.31 & A &  531 & $  262$ &  85.6 \\
   cso1080\_286\_490 & 226.215362 & 29.829822 & 0.15333 &    0.510 & E &  499 & $  325$ &  85.6 \\
   cso1080\_282\_512 & 226.205034 & 29.820446 & 0.15108 &     1.70 & C &  375 & $ -260$ &  83.0 \\
\enddata

\tablecomments{Column~6 lists the galaxy's spectral type: ``E'' for emission-line galaxies, ``A'' for absorption-line galaxies, and ``C'' for composite galaxies that show both emission and absorption. $\Delta\rho$ and $\Delta v$ are calculated with respect to the original SDSS group centers in \autoref{tab:berlind}; $f_{\rm mem}$ is the percentage of the Monte Carlo trials in which a galaxy was identified as a group member.}

\end{deluxetable*}

%% file: tab_44739.tex
\startlongtable
\begin{deluxetable*}{lcccccccc}

\tablecaption{Members of Group 44739
\label{tab:44739} }

\tablehead{ \colhead{Name} & 
            \colhead{RA} & 
            \colhead{Dec} & 
            \colhead{$z_{\rm gal}$} & 
            \colhead{$L_{\rm gal}$} & 
            \colhead{Type} & 
            \colhead{$\Delta\rho$} & 
            \colhead{$\Delta v$} & 
            \colhead{$f_{\rm mem}$} \\ 
            & 
            & 
            & 
            & 
            \colhead{($L^*$)} & 
            &
            \colhead{(kpc)} & 
            \colhead{($\mathrm{km\,s}^{-1}$)} & 
            \colhead{(\%)} } 

\colnumbers
\startdata
  fbqs1519\_206\_837 & 229.799752 & 28.425806 & 0.11834 &    0.216 & C &  128 & $  -13$ &  99.5 \\
  fbqs1519\_192\_796 & 229.857945 & 28.422901 & 0.11924 &     1.25 & A &  326 & $  228$ &  99.5 \\
  fbqs1519\_199\_903 & 229.820809 & 28.400061 & 0.11904 &    0.228 & C &  118 & $  174$ &  99.5 \\
  fbqs1519\_199\_932 & 229.815774 & 28.392901 & 0.11864 &    0.829 & C &  158 & $   67$ &  99.5 \\
  fbqs1519\_206\_981 & 229.780355 & 28.389717 & 0.11918 &     2.00 & C &  278 & $  212$ &  99.4 \\
 1237664853113110825 & 229.788788 & 28.381629 & 0.11870 &     6.48 & C &  287 & $   84$ &  99.4 \\
 fbqs1519\_189\_1155 & 229.852309 & 28.322714 & 0.11976 &    0.435 & A &  752 & $  367$ &  99.4 \\
 1237664853113176111 & 229.847981 & 28.318204 & 0.11892 &     4.44 & A &  774 & $  141$ &  99.4 \\
\enddata

\tablecomments{Column~6 lists the galaxy's spectral type: ``E'' for emission-line galaxies, ``A'' for absorption-line galaxies, and ``C'' for composite galaxies that show both emission and absorption. $\Delta\rho$ and $\Delta v$ are calculated with respect to the original SDSS group centers in \autoref{tab:berlind}; $f_{\rm mem}$ is the percentage of the Monte Carlo trials in which a galaxy was identified as a group member.}

\end{deluxetable*}

%% file: tab_44858.tex
\startlongtable
\begin{deluxetable*}{lcccccccc}

\tablecaption{Members of Group 44858
\label{tab:44858} }

\tablehead{ \colhead{Name} & 
            \colhead{RA} & 
            \colhead{Dec} & 
            \colhead{$z_{\rm gal}$} & 
            \colhead{$L_{\rm gal}$} & 
            \colhead{Type} & 
            \colhead{$\Delta\rho$} & 
            \colhead{$\Delta v$} & 
            \colhead{$f_{\rm mem}$} \\ 
            & 
            & 
            & 
            & 
            \colhead{($L^*$)} & 
            &
            \colhead{(kpc)} & 
            \colhead{($\mathrm{km\,s}^{-1}$)} & 
            \colhead{(\%)} } 

\colnumbers
\startdata
  fbqs1519\_36\_1006 & 230.067973 & 28.879044 & 0.12915 &    0.487 & A &  289 & $  465$ & 100.0 \\
   fbqs1519\_47\_971 & 230.107970 & 28.840452 & 0.12854 &    0.588 & A &  147 & $  303$ & 100.0 \\
 1237664853649981733 & 230.084212 & 28.856470 & 0.12827 &     2.39 & C &   71 & $  231$ & 100.0 \\
   fbqs1519\_43\_969 & 230.092135 & 28.851719 & 0.12928 &     2.13 & A &   13 & $  500$ & 100.0 \\
  fbqs1519\_38\_1088 & 230.094152 & 28.891317 & 0.12903 &    0.813 & A &  338 & $  433$ & 100.0 \\
   fbqs1519\_39\_926 & 230.068302 & 28.852137 & 0.12794 &     2.34 & C &  165 & $  144$ & 100.0 \\
          A1518+2903 & 230.112920 & 28.884167 & 0.12748 &     9.00 & \nodata &  320 & $   22$ & 100.0 \\
 1237664853649981766 & 230.111703 & 28.886344 & 0.12717 &     7.80 & A &  332 & $  -61$ & 100.0 \\
\enddata

\tablecomments{Column~6 lists the galaxy's spectral type: ``E'' for emission-line galaxies, ``A'' for absorption-line galaxies, and ``C'' for composite galaxies that show both emission and absorption. $\Delta\rho$ and $\Delta v$ are calculated with respect to the original SDSS group centers in \autoref{tab:berlind}; $f_{\rm mem}$ is the percentage of the Monte Carlo trials in which a galaxy was identified as a group member.}

\end{deluxetable*}

%% file: tab_50433.tex
\begin{deluxetable*}{lcccccccc}

\tablecaption{Members of Group 50433
\label{tab:50433} }

\tablehead{ \colhead{Name} & 
            \colhead{RA} & 
            \colhead{Dec} & 
            \colhead{$z_{\rm gal}$} & 
            \colhead{$L_{\rm gal}$} & 
            \colhead{Type} & 
            \colhead{$\Delta\rho$} & 
            \colhead{$\Delta v$} & 
            \colhead{$f_{\rm mem}$} \\ 
            & 
            & 
            & 
            & 
            \colhead{($L^*$)} & 
            &
            \colhead{(kpc)} & 
            \colhead{($\mathrm{km\,s}^{-1}$)} & 
            \colhead{(\%)} } 

\colnumbers
\startdata
 1237665015235870817 & 157.765069 & 30.855472 & 0.13571 &     2.36 & A &  188 & $  -74$ &  99.9 \\
 1237665015235870836 & 157.751425 & 30.889867 & 0.13673 &     3.51 & A &  234 & $  195$ &  99.9 \\
  fbqs1030\_182\_769 & 157.741456 & 30.835203 & 0.13518 &     1.03 & A &  253 & $ -214$ &  99.9 \\
  fbqs1030\_172\_743 & 157.778378 & 30.844095 & 0.13621 &    0.207 & A &  323 & $   58$ &  99.9 \\
 1237665015235870848 & 157.791725 & 30.849692 & 0.13512 &     3.26 & A &  392 & $ -229$ &  99.9 \\
 1237665015235870816 & 157.757518 & 30.861972 & 0.13605 &     11.4 & A &  117 & $   16$ &  99.9 \\
  fbqs1030\_164\_685 & 157.801409 & 30.864449 & 0.13600 &    0.542 & A &  444 & $    3$ &  99.9 \\
 1237665015235805294 & 157.668127 & 30.855027 & 0.13606 &     3.82 & A &  560 & $   19$ &  99.9 \\
\enddata

\tablecomments{Column~6 lists the galaxy's spectral type: ``E'' for emission-line galaxies, ``A'' for absorption-line galaxies, and ``C'' for composite galaxies that show both emission and absorption. $\Delta\rho$ and $\Delta v$ are calculated with respect to the original SDSS group centers in \autoref{tab:berlind}; $f_{\rm mem}$ is the percentage of the Monte Carlo trials in which a galaxy was identified as a group member.}

\end{deluxetable*}